\documentclass[a4paper,12pt]{book}

\usepackage{graphicx}
\usepackage{float}
\usepackage{longtable}
\usepackage{mathdots}
\usepackage{eucal}
\usepackage{ stmaryrd }
\usepackage{amssymb,amsmath,cite,amsthm}
\usepackage{mathtools}
\usepackage[pdftex,colorlinks,bookmarks=true,pdfauthor={Fazlul Hoque}]{hyperref}

\usepackage{array}

\usepackage[final]{pdfpages}
\usepackage{ stmaryrd }
\usepackage[top=1.2in, bottom=1in, left=1in, right=1in]{geometry}
\usepackage{tikz}
\usepackage{amsthm}
\usepackage{pifont}
\usepackage{lipsum}
\usepackage{enumerate}
\usepackage[all]{xy}

\usepackage[utf8]{inputenc}
\usepackage{fancyhdr}
\usepackage{blindtext}
 
\usepackage{sectsty}

\chapternumberfont{\Large} 
\chaptertitlefont{\LARGE}

\def\beq{\begin{equation}}
\def\eeq{\end{equation}}
\def\bea{\begin{eqnarray*}}
\def\eea{\end{eqnarray*}}

\def\N0{ {\mathbb Z}_{+} }
\def\v0{ |d,r) }

\theoremstyle{plain}
\newtheorem{th1}{Theorem}[section]
\newtheorem{lem}[th1]{Lemma}

\theoremstyle{definition}

\theoremstyle{remark}

\begin{document}
\begin{center}
\begin{figure}[H]
\centering
\includegraphics{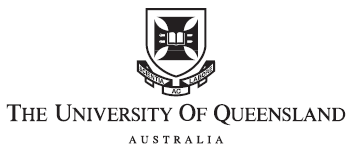}
\end{figure}
~~\\
{\LARGE \bf {Superintegrable systems, polynomial algebra structures and exact derivations of spectra} }
~~\\
~~\\
~~\\
\small {\bf MD FAZLUL HOQUE} \\
\small {B.Sc.(Honours) and M.Sc.(Thesis) in Mathematics} \\ 
 M.Phil. in Fuzzy Topology
 ~~\\
\vspace{80 mm}
\normalsize{\textit{A thesis submitted for the degree of Doctor of Philosophy at}} \\
\normalsize{\textit{The University of Queensland in 2017}}\\
~~\\
\normalsize{\textit{School of Mathematics and Physics}} \\

\end{center}
\thispagestyle{empty} 
\newpage
~~\\
\thispagestyle{empty} 

\newpage
\clearpage
\pagenumbering{roman} 

\begin{center}
{\LARGE\textbf{Abstract}}
\end{center}
Superintegrable systems are a class of physical systems which possess more conserved quantities than their degrees of freedom. The study of these systems has a long history and continues to attract significant international attention. This thesis investigates finite dimensional quantum superintegrable systems with scalar potentials as well as vector potentials with monopole type interactions. We introduce new families of $N$-dimensional superintegrable Kepler-Coulomb systems with non-central terms and double singular harmonic oscillators in the Euclidean space, and new families of superintegrable Kepler, MIC-harmonic oscillator and deformed Kepler systems interacting with Yang-Coulomb monopoles in the flat and curved Taub-NUT spaces. We show their multiseparability and obtain their Schr\"{o}dinger wave functions in different coordinate systems. We show that the wave functions are given by (exceptional) orthogonal polynomials and Painlev\'{e} transcendents (of hypergeometric type). We construct higher-order algebraically independent integrals of motion of the systems via the direct and constructive approaches. These integrals form (higher-rank) polynomial algebras with structure constants involving Casimir operators of certain Lie algebras. We obtain finite dimensional unitary representations of the polynomial algebras and present the algebraic derivations for degenerate energy spectra of these systems. Finally, we present a generalized superintegrable Kepler-Coulomb model from exceptional orthogonal polynomials and obtain its energy spectrum using both the separation of variable and the algebraic methods.

\newpage
\begin{center}
\Large{Declaration by author}
\end{center}

This thesis is composed of my original work, and contains no material previously published or written by another person except where due reference has been made in the text. I have clearly stated the contribution by others to jointly-authored works that I have included in my thesis.
~~\\~~\\
I have clearly stated the contribution of others to my thesis as a whole, including statistical assistance, survey design, data analysis, significant technical procedures, professional editorial advice, and any other original research work used or reported in my thesis. The content of my thesis is the result of work I have carried out since the commencement of my research higher degree candidature and does not include a substantial part of work that has been submitted to qualify for the award of any other degree or diploma in any university or other tertiary institution. I have clearly stated which parts of my thesis, if any, have been submitted to qualify for another award.
~~\\~~\\
I acknowledge that an electronic copy of my thesis must be lodged with the University Library and, subject to the policy and procedures of The University of Queensland, the thesis be made available for research and study in accordance with the Copyright Act 1968 unless a period of embargo has been approved by the Dean of the Graduate School
~~\\~~\\
I acknowledge that copyright of all material contained in my thesis resides with the copyright holder(s) of that material. Where appropriate I have obtained copyright permission from the copyright holder to reproduce material in this thesis.
%

\newpage 
\begin{center}
\Large{Publications during candidature}
\end{center}
{\bf Peer-reviewed publications}
\begin{enumerate}
\item
Hoque M F, Marquette I and Zhang Y-Z 2015 Quadratic algebra structure and spectrum of a new superintegrable system in $N$-dimension, \textit{Journal of Physics A: Mathematical and Theoretical} \textbf{48}, 185201.
\item
Hoque M F, Marquette I and Zhang Y-Z 2015 A new family of $N$ dimensional superintegrable double singular oscillators and quadratic algebra $Q(3)\oplus so(n)\oplus so(N-n)$, \textit{Journal of Physics A: Mathematical and Theoretical} \textbf{48}, 445207.
\item
Hoque M F, Marquette I and Zhang Y-Z 2016 Recurrence approach and higher rank cubic algebras for the $N$-dimensional superintegrable systems, \textit{Journal of Physics A: Mathematical and Theoretical} \textbf{49}, 125201.
\item
Hoque M F, Marquette I and Zhang Y-Z 2016 Quadratic algebra for superintegrable monopole system in a Taub-NUT space, \textit{Journal of Mathematical Physics} \textbf{57}, 092104.
\item
Hoque M F, Marquette I and Zhang Y-Z 2017 Quadratic algebra structure in the 5D Kepler system with non-central potentials and Yang-Coulomb monopole interaction, \textit{Annals of Physics} \textbf{380}, 121-134.
\end{enumerate}
{\bf Peer-reviewed conference publications}
\begin{enumerate}
\item[6.]
Hoque M F, Marquette I and Zhang Y-Z 2016 Family of $N$-dimensional superintegrable systems and quadratic algebra structures, \textit{Journal of Physics: Conference Series} \textbf{670}, 012024.
\end{enumerate}
{\bf Online e-print publications}
\begin{enumerate}
\item[7.]
Hoque M F, Marquette I and Zhang Y-Z 2016 Recurrence approach and higher order polynomial algebras for superintegrable monopole systems, \textit{arXiv:} \textbf{1605.06213}.
\item[8.]
Hoque M F, Marquette I, Post S and Zhang Y-Z 2017 Algebraic calculations for spectrum of superintegrable system from exceptional orthogonal polynomials, \textit{arXiv:} \textbf{1710.03589}.
\end{enumerate}

\newpage  
\begin{center}
\Large{Publications included in this thesis}
\end{center}
%
~~\\
1. Hoque M F, Marquette I and Zhang Y-Z 2015 Quadratic algebra structure and spectrum of a new superintegrable system in $N$-dimension, \textit{Journal of Physics A: Mathematical and Theoretical} \textbf{48}, 185201.
\\
-- Incorporated as Chapter \ref{ch3}.
\begin{table}[h]
	\centering
		\begin{tabular}{|l|c|c|c|}
                            \hline
			\textbf{Contributor} & \textbf{M. F. Hoque} &\textbf{I. Marquette} & \textbf{Y-Z. Zhang} \\ \hline\hline
			The Conceptualization of Key Ideas 	& 30\%  & 40\%  & 30\% \\
			The Technical Calculations 		           & 50\% & 25\% & 25\% \\
			The Drafting and Writing 			& 40\% & 30\% & 30\% \\
                            \hline
		\end{tabular}
\end{table}
\\
2. Hoque M F, Marquette I and Zhang Y-Z 2015 A new family of $N$ dimensional superintegrable double singular oscillators and quadratic algebra $Q(3)\oplus so(n)\oplus so(N-n)$, \textit{Journal of Physics A: Mathematical and Theoretical} \textbf{48}, 445207.
\\
-- Incorporated as Chapter \ref{ch4}.
\begin{table}[h]
	\centering
		\begin{tabular}{|l|c|c|c|}
                           \hline
			\textbf{Contributor} & \textbf{M. F. Hoque} &\textbf{I. Marquette} & \textbf{Y-Z. Zhang} \\ \hline\hline
			The Conceptualization of Key Ideas 	& 35\%  & 35\% & 30\% \\
			The Technical Calculations 			& 40\% & 30\% & 30\% \\
			The Drafting and Writing 			&  40\% & 30\% & 30\% \\
                           \hline
		\end{tabular}
\end{table}
\\
3. Hoque M F, Marquette I and Zhang Y-Z 2016 Family of $N$-dimensional superintegrable systems and quadratic algebra structures, \textit{Journal of Physics: Conference Series} \textbf{670}, 012024.
\\
-- Incorporated as Chapter \ref{ch5}.
\begin{table}[h]
	\centering
		\begin{tabular}{|l|c|c|c|}
                           \hline
			\textbf{Contributor} & \textbf{M. F. Hoque} &\textbf{I. Marquette} & \textbf{Y-Z. Zhang} \\ \hline\hline
			The Conceptualization of Key Ideas 	& 50\%  & 25\% & 25\% \\
			The Technical Calculations 			& 60\% & 20\% & 20\% \\
			The Drafting and Writing 			&  50\% & 25\% & 25\% \\
                           \hline
		\end{tabular}
\end{table}
\\
4. Hoque M F, Marquette I and Zhang Y-Z 2016 Recurrence approach and higher rank cubic algebras for the $N$-dimensional superintegrable systems, \textit{Journal of Physics A: Mathematical and Theoretical} \textbf{49}, 125201.
\\
-- Incorporated as Chapter \ref{ch6}.
\begin{table}[h]
	\centering
		\begin{tabular}{|l|c|c|c|}
                           \hline
			\textbf{Contributor} & \textbf{M. F. Hoque} &\textbf{I. Marquette} & \textbf{Y-Z. Zhang} \\ \hline\hline
			The Conceptualization of Key Ideas 	& 40\%  & 30\% & 30\% \\
			The Technical Calculations 			& 50\% & 25\% & 25\% \\
			The Drafting and Writing 			&  40\% & 30\% & 30\% \\
                           \hline
		\end{tabular}
\end{table}
\\
5. Hoque M F, Marquette I and Zhang Y-Z 2016 Quadratic algebra for superintegrable monopole system in a Taub-NUT space, \textit{J. Math. Phys.} \textbf{57}, 092104.
\\
-- Incorporated as Chapter \ref{ch7}.
\begin{table}[h]
	\centering
		\begin{tabular}{|l|c|c|c|}
                           \hline
			\textbf{Contributor} & \textbf{M. F. Hoque} &\textbf{I. Marquette} & \textbf{Y-Z. Zhang} \\ \hline\hline
			The Conceptualization of Key Ideas 	& 30\%  & 40\% & 40\% \\
			The Technical Calculations 			& 40\% & 30\% & 30\% \\
			The Drafting and Writing 			&  40\% & 30\% & 30\% \\
                           \hline
		\end{tabular}
\end{table}
\\
6. Hoque M F, Marquette I and Zhang Y-Z 2017 Quadratic algebra structure in the 5D Kepler system with non-central potentials and Yang-Coulomb monopole interaction, \textit{Annals of Physics} \textbf{380}, 121.
\\
-- Incorporated as Chapter \ref{ch8}.
\begin{table}[h]
	\centering
		\begin{tabular}{|l|c|c|c|}
                           \hline
			\textbf{Contributor} & \textbf{M. F. Hoque} &\textbf{I. Marquette} & \textbf{Y-Z. Zhang} \\ \hline\hline
			The Conceptualization of Key Ideas 	& 35\%  & 35\% & 30\% \\
			The Technical Calculations 			& 40\% & 30\% & 30\% \\
			The Drafting and Writing 			&  40\% & 30\% & 30\% \\
                           \hline
		\end{tabular}
\end{table}
\\
\\
\\
\\
7. Hoque M F, Marquette I and Zhang Y-Z 2016 Recurrence approach and higher order polynomial algebras for superintegrable monopole systems, \textit{arXiv:} \textbf{1605.06213}.
\\
-- Incorporated as Chapter \ref{ch9}.
\begin{table}[h]
	\centering
		\begin{tabular}{|l|c|c|c|}
                           \hline
			\textbf{Contributor} & \textbf{M. F. Hoque} &\textbf{I. Marquette} & \textbf{Y-Z. Zhang} \\ \hline\hline
			The Conceptualization of Key Ideas 	& 40\%  & 30\% & 30\% \\
			The Technical Calculations 			& 50\% & 25\% & 25\% \\
			The Drafting and Writing 			&  50\% & 25\% & 25\% \\
                           \hline
		\end{tabular}
\end{table}
\\
8. Hoque M F, Marquette I, Post S and Zhang Y-Z 2017 Algebraic calculations for spectrum of superintegrable system from exceptional orthogonal polynomials, \textit{arXiv:} \textbf{1710.03589}.
\\
-- Incorporated as Chapter \ref{ch10}.
\begin{table}[ht]
	\centering
{\footnotesize		
		\begin{tabular}{|l|c|c|c|c|}
                           \hline
			\textbf{Contributor} & \textbf{M. F. Hoque} &\textbf{I. Marquette} & \textbf{Y-Z. Zhang} & \textbf{S. Post}\\ \hline\hline
			The Conceptualization of Key Ideas 	& 55\%  & 15\% & 15\% & 15\% \\
			The Technical Calculations 			& 55\%  & 15\% & 15\% & 15\% \\
			The Drafting and Writing 			&  55\%  & 15\% & 15\% & 15\% \\
                           \hline
		\end{tabular}
		}
\end{table}
\newpage

\begin{center}
\Large{Contributions by others to the thesis}
\end{center}

My supervisors, Associate Professor and Reader Yao-Zhong Zhang and Dr. Ian Marquette, significantly contributed to the design of the project plan, directed me towards the relevant literature and helped to develop the techniques used in the thesis. They have also undertaken proofreading of the thesis.
~~\\
~~\\
\begin{center}
{\Large Statement of parts of the thesis submitted to qualify for the award of another degree}
\end{center}

None.

\newpage
\begin{center}
\Large{Acknowledgments}
\end{center}


I would like to start by thanking Almighty Allah, the Giver of Wisdom, Knowledge and Understanding, Whose divine help has been my source of sustenance.

I wish to express my deepest gratitude to my principal supervisor Associate Professor Yao-Zhong Zhang and associate supervisor Dr. Ian Marquette for all their professional guidance, constructive feedbacks and motivations throughout my graduate study at the University of Queensland. They introduced me to the interesting and vibrant field of superintegrable systems, symmetry algebras and their connection to mathematics and physics. They have dedicated their time, knowledge and skills to guide and help me in my study. This thesis would not be possible without their help.

Special thanks are to Dr. Sarah Post for her collaborations and kind hospitality in Hawaii during my visit in 2016. 

I would like to thank the Australian Government and the University of Queensland for the financial support via the International Postgraduate Research Scholarship (IPRS) and Australian Postgraduate Award (APA), without which this study may not have been a reality. Moreover, I thank the Graduate School of the University of Queensland for the Graduate School International Travel Award (GSITA).

I would like to thank Murray Kane for assisting me with so many non-mathematical aspects of my PhD study and David Agboola, Amir Moghaddam, Jason Werry, Inna Lukyanenko and Fahad Alsammari for helping me at various times at UQ. I would also like to acknowledge the kindness of my country mates and friends: Abdul Wahab, Narottam Saha, Rashed Abdulla, Sheuli Khanam, Kamruzzaman, Shimul Chaudhury, Jalal Khan, Runa, Mazhar, Hasan, Halim, Shafi, Nur and Nizhum Rahman.

I am extremely grateful to the authority of Pabna University of Science and Technology for giving me study leave to pursue my PhD study in Australia.  Also I thank my colleagues, friends and students in Pabna University of Science and Technology, specially Abu Sufian, Tahmina Sultana and Harun-Or-Roshid for their friendly support, great advice and encouragement throughout my study.

It is my great pleasure to thank my parents and brothers, sister and sisters-in-law for supporting me spiritually throughout my study and my life in general. It would be a great happy moment to inform my heavenly father Abul Hossain my achievements in this higher degree research study.

And finally I must thank my love for her inspiration, encouragement and beauty which give me spirit and energy to do hard works. Indeed, you have in my heart every moment, and I love you so much!

\newpage
\begin{center}
\Large{Keywords}
\end{center}
 

Superintegrable systems, quadratic algebras, polynomial algebras, Casimir operators, Lie algebras, deformed oscillator algebras, Schr\"{o}dinger equations, monopole interactions, special functions, (exceptional) orthogonal  polynomials  

 \newpage
 \begin{center}
{\Large {Australian and New Zealand Standard Research Classifications
(ANZSRC)}}
\end{center}
010501 Algebraic Structures in Mathematical Physics 40\% \\
010502 Integrable Systems (Classical and Quantum) 40\% \\
010503 Mathematical Aspects of Classical and Quantum Mechanics 20\% \\

 \begin{center}
{\Large {Fields of Research (FoR) Classification}}
\end{center}
FoR code: 0105 Mathematical Physics 100\% .

\newpage
\vspace{10 mm}
\begin{center}
\Large{{Dedication}}
\end{center}
\vspace{20 mm}
\begin{center}
\Large{To}
\end{center}
\vspace{10 mm}
\begin{center}
\large{My Mother\\ \textbf{Samina Khatun}}
\end{center}

\clearpage
\newpage
\pdfbookmark{\contentsname}{toc}
 \begin{tableofcontents}
 \end{tableofcontents}
 \chapter{Introduction}\label{ch1}
 \setcounter{page}{1}
\pagenumbering{arabic}


\section{Background and motivations}
Symmetries play a central role in a variety of branches in modern science. The general concepts of symmetry in physics are fundamentally important in formulating theories and models. Symmetry algebras such as Lie algebras lead to powerful methods for solving energy spectra and degeneracies of models in e.g., nuclear physics, quantum chemistry and particle physics. The harmonic oscillators, the hydrogen atom in quantum mechanics and the interacting boson model (IBM) with $su(6)$ Lie algebra symmetry in nuclear physics are the most famous applications of symmetries.  See \cite{lou2} for quantum models with $so(N+1)$ symmetry and \cite{jau1, lou1, hwa1} for those with $su(N)$ symmetry.

There are many classes of physical models that can be solved analytically. They are often called exactly solvable systems in the literature. An exciting subclass is the so-called superintegrable systems that allow the Schr\"{o}dinger eigenvalue problem to be solved exactly and algebraically \cite{eva4, fod1, mil1, tem1, tem2}. Superintegrable systems are systems that possess more conserved quantities than their degrees of freedom. The harmonic anisotropic oscillator and Kepler planetary orbit systems are famous examples of superintegrable systems. Thanks to the existence of more integrals of motion, superintegrable models provide deep quantitative insight into the physical systems under investigation. They also facilitate deeper conceptual understanding of modern physics, especially in quantum chemistry, atomic physics, nuclear physics, molecular physics and condensed matter physics. In fact, because they are exactly solvable physical systems, they offer an interesting field of research in multi-dimensional applications in physics and many domains of pure and applied mathematics. 

The modern theory of superintegrable systems allowing second-order integrals of motion in the 2D Euclidean space was inaugurated by Winternitz et. al. \cite{fri1, fri2, mak1} in the mid sixties. Wojciechowski popularly used the term "superintegrable" and applied superintegrability to the Calogero-Moser systems in 1983 \cite{woj1}. Earlier applications included systems with accidental degeneracy \cite{foc1, barg1} and dynamical symmetries \cite{fri1, fri2, mak1}. Calogero \cite{calo1, calo2, calo3, ruh1} and others (see \cite{gon1, pol1, fei1, hak1}) presented explicit solutions of some n-body problems which were really noble examples for the theory. The topic of superintegrability and the classifications of superintegrable systems have been extended in last fifteen years in the view point of symmetry algebras in mathematical physics \cite{kal2, kal3, kal4, kal6, kal9, mil1, pos2, abo1, mar2, mar3, mar12, mar11, mar15, mar16, ror1}. Classification of second order 2D and 3D superintegrable systems with their degenerate quadratic symmetry algebras is now more or less complete for conformal flat spaces. Recently, a lot of efforts have been devoted to relate distinct superintegrable systems and their nondegenerate and semidegenerate quadratic algebras by geometric contractions induced by B\^{o}cher contractions of the conformal Lie algebra $so(4,\mathbb{C})$ \cite{esc2, esc3, esc4}. However, classification in higher dimensions and with higher order integrals of motion is much more complicated.

Superintegrable systems also possess many attractive properties in the view point of mathematics, especially, their connections to special functions, (exceptional) orthogonal polynomials, Painlev\'{e} transcendents, hypergeometric Heun elliptic functions and higher-order analogs \cite{abo1, esc1, mar16, pos3}. Moreover, there has an interesting connection between superintegrable systems in 2D conformally flat spaces and the full Askey scheme of orthogonal polynomials of hypergeometric type \cite{kal10}. Such connections between superintegrable systems and special functions and orthogonal polynomials are intimately related to the conjecture that all superintegrable systems are exactly solvable \cite{tem1}. In addition, the In\"{o}n\"{u}-Winger type Lie algebra contractions and B\^{o}cher contractions have been applied to relate separable coordinate systems and the associated special functions \cite{ino1, izm1, izm2, esc2, esc3, esc4}.

Polynomial algebras are nonlinear generalizations of Lie algebras. They are often the symmetry algebras of superintegrable systems and play a basic role in algebraically deriving the energy spectra of these systems \cite{isa1} and explaining their degeneracies.  Algebraic derivations of the energy spectra for systems with Lie algebra symmetries are achieved using chains of second order Casimir operators to define appropriate quantum numbers \cite{lou2} and various embeddings in non-invariant algebras \cite{tru1}. For 2D superintegrable systems with quadratic algebra involving three generators as symmetry algebra a new algebraic approach was proposed in \cite{das2}. This method is based on the construction of the Casimir operators and the realization of the quadratic algebra in terms of the deformed oscillators.  The generalization of this approach to the $N$-dimensional superintegrable systems is a very attractive subject. Higher-dimensional superintegrable systems often possess higher rank polynomial algebras as symmetry algebras and accidental degeneracies. The structure of these algebras was less known. The accidental degeneracies come from the fact that the Schr\"{o}dinger equations for such systems are multiseparable in certain coordinate systems \cite{fri1, fri2, mak1}. For higher-dimensional superintegrable Hamiltonian systems, it is in general difficult to obtain their integrals of motion and the corresponding polynomial symmetry algebras, the Casimir operators and their realizations in terms of the deformed oscillator algebras via this direct approach in \cite{das2}.

The difficulties of this approach can be overcome by using the constructive approach. In classical and quantum mechanical systems, this constructive or recurrence approach is one of the powerful tools to derive integrals of motion and construct their higher-rank polynomial algebras. A lot of works have been devoted to construct integrals of motion and their corresponding symmetry algebras based on lower- (first and second) \cite{jau1, fri2, boy1, eva3, mar6} and higher-order ladder operators in various aspects (see e.g. \cite{kre1, adl1, jun1, dem1, mar3, rag1, mar9, mar13, mar14, mar5}). In fact, the constructive approach has a deep connection with special functions and (exceptional) orthogonal polynomials \cite{pos3, mar15, kal7, cal2, cal3}. It is interesting to generalize this approach to superintegrable systems in higher dimensions or with monopole interactions in both flat and curved spaces. Such a generalization has so far been only considered for a very limited number of cases.

Models in curved spaces are essential in modern physics. The Taub-NUT metrics discovered by Taub \cite{tau1} and Newman-Unti-Tamburino \cite{new1} with their plentiful symmetries provide an excellent background to investigate the classical and quantum conserved quantities in curved spaces. The Taub-NUT metric might give rise to the gravitational analog of the Yang-Mills instanton suggested by Hawking \cite{haw1}. This metric is the space part of the line element of the Kaluza-Klein monopole \cite{gro1, sor1} and is known to admit the Kepler-type symmetry and provide a non-trivial generalization of the oscillator and Kepler-Coulomb problems \cite{gib1, feh2, cor1, man1, ati1, gro2, cot1, mar8, iwa1, iwa2, iwa3, iwa4, bal2, bal3, lat1, kur1}.

Kepler problems involving magnetic monopoles, now known as MICZ-Kepler problems, were explored independently by McIntosh and Cisneros  \cite{mci1} and Zwanziger  \cite{zwa1}. The generalizations of the MICZ-Kepler problems \cite{mad6} lead to the intrinsic Smorodinsky-Winternitz systems \cite{fri1, eva2} with monopoles in 3D Euclidean space.  The extended Kepler systems with non-central potentials or  Yang's non-abelian $su(2)$ monopoles \cite{yan1} have been studied in both the flat Euclidean and curved Taub-NUT spaces. The majority of the works has so far been focused on systems in lower dimensions \cite{mil1}.

\section{Aim of the research}
This thesis will focus on finite dimensional classical and quantum superintegrable systems with scalar potentials as well as vector potentials with monopole type interactions. We introduce new families of superintegrable Hamiltonians in the $N$-dimensional Euclidean space. The main aims of this PhD thesis are to
\begin{itemize}
\item
study algebraic properties of the $N$-dimensional superintegrable systems and their connection to special functions and orthogonal polynomials.
\item
construct higher-rank finitely generated polynomial algebras and their Casimir operators related to the $N$-dimensional superintegrable systems using the direct approach relying on realizations of differential operators.
\item
develop new approaches to construct integrals of motion and the polynomial algebras satisfied by them. These approaches are based on the combination of the ladder, shift and intertwining operators, supercharge and various recurrence formulas of special functions and orthogonal polynomials (of hypergeometric type).
\item
construct finite-dimensional unitary representations of the higher-rank polynomial algebras and apply the results to derive the degenerate energy spectra of the corresponding superintegrable systems. 
\item
apply such approaches to monopole systems and systems from exceptional orthogonal polynomials (EOPs).
\end{itemize}

\section{Thesis structure}
After the introductory and preliminary chapters \ref{ch1} and \ref{ch2}, we present the thesis by the style of "thesis by publication". The thesis incorporates our peer-reviewed and published papers \cite{fh1, fh2, fh3, fh4, fh5, fh6, fh7, fh8} in such a way that each paper can be considered as a stand alone chapter. This thesis has thus been divided into eleven chapters according to the key mathematical themes of the publications \cite{fh1, fh2, fh3, fh4, fh5, fh6, fh7, fh8}.

Chapter \ref{ch2} gives the preliminary mathematical background and relevant algebraic tools for the study of superintegrable systems. Our main works start from  chapter \ref{ch3}. 

In chapter \ref{ch3}, we present a new superintegrable Hamiltonian system in $N$-dimensional Euclidean space. We show that its Schr\"{o}dinger wave function is multi-separable in hyperspherical and hyperparabolic coordinates and the wave function can be expressed in terms of special functions. We give an algebraic derivation of spectrum of the superintegrable system. We show how the $so(N+1)$ symmetry algebra of the $N$-dimensional Kepler-Coulomb system is deformed to a quadratic algebra with three generators and structure constants involving Casimir operator of $so(N-1)$ Lie algebra. We construct the quadratic algebra and the Casimir operator and derive the structure function of the deformed oscillator realization of the quadratic algebra which yields the energy spectrum. 

In chapter \ref{ch4}, we extend the symmetric double singular oscillators in 4D and 8D to arbitrary dimensions with any partition $(n, N-n)$ of the  coordinates. This provides a new family of quantum superintegrable systems. We show how the $su(N)$ symmetry algebra of the $N$-dimensional harmonic oscillator are broken to higher rank polynomial algebra of the form $Q(3)\oplus so(n)\oplus so(N-n)$. We construct integrals of the motion and the Casimir operator and obtain the realization of the polynomial algebra in terms of the deformed oscillator algebra. We obtain the finite dimensional unitary representations of the quadratic algebra and the degenerate energy spectrum of the superintegrable model. Moreover, we show that the model is multi-separable and obtain its wave function in $(n,N-n)$ double hyperspherical coordinates. 

In chapter \ref{ch5}, we present the common features of the quadratic algebra structures and algebraic derivations of families of $N$-dimensional superintegrable Hamiltonian models.

The applications of the direct approach in chapter \ref{ch3}, \ref{ch4} and \ref{ch5} show that it is quite involved to construct integrals of motion, their corresponding polynomial algebras, the Casimir operators and realizations in terms of the deformed oscillators. The difficulties of this direct approach can be overcome by the constructive approach in chapter \ref{ch6}. That is, in chapter \ref{ch6}, we apply the recurrence approach and coupling constant metamorphosis to construct higher order integrals of motion for the St\"{a}ckel equivalents of the $N$-dimensional superintegrable Kepler-Coulomb model with non-central terms and the double singular oscillator of type $(n,N-n)$. We present their higher rank cubic algebra $C(3)\oplus L_1\oplus L_2$ with structure constants involving Casimir operators of the certain Lie algebras $L_1$ and $L_2$. We realize this algebra in terms of the deformed oscillator and derive the degenerate energy spectra of the corresponding superintegrable systems.

In the next few chapters after chapter \ref{ch6}, we consider models with monopole interactions and from exceptional orthogonal polynomials. In chapter \ref{ch7}, we introduce a new superintegrable monopole system in the curved Taub-NUT space whose wave function is given by a product of Laguerre and Jacobi polynomials. By construction, algebraically independent integrals of motion of the model make it a superintegrable system with monopole interactions. We present the quadratic algebra, Casimir operator and algebraic derivation of energy spectrum of the monopole model.

In chapter \ref{ch8}, we construct integrals of motion for the 5D deformed Kepler system with $su(2)$ Yang-Coulomb monopole, which demonstrate superintegrability of the model. We present the higher-rank quadratic algebra formed by the integrals of motion and give an algebraic derivation of the energy spectrum of the model. Moreover, we show that the Schr\"{o}dinger wave equation of the model is multiseparable.

In the first part of chapter \ref{ch9}, we revisit the MIC-harmonic oscillator in the field of magnetic monopole in the flat space by means of the recurrence approach and present an algebraic derivation of the energy spectrum of the system. In second part, we introduce a new MIC-harmonic oscillator type Hartmann system with monopole interaction in the generalized Taub-NUT space. We construct its integrals of motion using recurrence formulas based on wave functions. We show that the integrals satisfy the polynomial algebra and apply this algebraic structure to derive the energy spectrum.

In chapter \ref{ch10}, we introduce a new three-parameter Kepler-Coulomb system and present the algebraic derivation of its spectrum via the recurrence approach based on the eigenvalue functions.

The final chapter \ref{conc} summaries the results of the thesis and gives comments and discussions on future research. 


\chapter{Superintegrable systems and polynomial algebras }\label{ch2}

In this chapter, we review the relevant theory of classical and quantum dynamical systems, focusing on their algebraic aspects. In sections \ref{CMS1} and \ref{QMS1}, we overview the Hamiltonian formalism in classical and quantum mechanics. In the section \ref{SIS1}, we give the definitions and basic properties of integrable and superintegrable systems. In section \ref{ABS1}, we describe the mathematical structures underlying superintegrable systems. We will give a brief review on Lie algebras and their nonlinear generalizations, polynomial algebras. Particular attention will be paid to quadratic algebras with three generators and their deformed oscillator realizations. More details on algebraic structures presented here can be found in the review paper \cite{mil1} and the book \cite{iac1}.

\section{Classical Mechanics}\label{CMS1}
In classical mechanics, the most important dynamical variable for a system with $N$-degrees of freedom is the Hamiltonian, i.e. the total energy, of the system which can be written as (in the unit of mass = 1)
\begin{eqnarray}
\mathcal{H}=\frac{1}{2}\sum_{j, k}g^{jk}(\textbf{x})p_j p_k+V(\textbf{x}),\label{CHm1}
\end{eqnarray}
where $g^{jk}$ is the metric tensor of the space and $x_j, p_k$ are the generalized coordinates of positions and momenta, respectively, in the $2N$-dimensional phase space. $\mathcal{H}$ is time independent for a conservative system with constant energy and it varies with time for a non-conservative system. All other dynamical variables $A(\textbf{x},\textbf{p},t)$, in general, are real-valued functions of the generalized coordinates $x_j$, $p_j$, $j=1,\dots, N$ and the time $t$.
The motion of the particle is determined by the Hamilton's canonical equations
\begin{eqnarray}
\dot{x}_j=\frac{\partial\mathcal{H}}{\partial p_j}, \quad \dot{p}_j=-\frac{\partial\mathcal{H}}{\partial x_j}. \label{HamE1}
\end{eqnarray} 
Let 
\begin{eqnarray}
\{A, B\}=\sum^n_{j=1}\left(\frac{\partial A}{\partial x_j}\frac{\partial B}{\partial p_j}-\frac{\partial A}{\partial p_j}\frac{\partial B}{\partial x_j}\right),\label{PB1}
\end{eqnarray}
denote the Poisson bracket of two dynamical variables $A$ and $B$, which satisfies the following properties:
\begin{enumerate}[(i)]
\item
Anti-symmetry: $\{A, B\}=-\{B, A\}$,
\item
Bilinearity: $\{A, aB+bC\}=a\{A, B\}+b\{A, C\}$,  $a, c$ are constants,
\item
Jacobi identity: $\{A, \{B,C\}\}+\{B, \{C,A\}\}+\{C, \{A,B\}\}=0$,  
\item
Leibniz rule: $\{A, BC\}=\{A, B\}C+ B\{A,C\}$.
\end{enumerate}
It is easily seen that the coordinates $(x_i,p_i)$ satisfy 
\begin{eqnarray}
\{p_j,p_k\}=0=\{x_j,x_k\}, \quad \{p_j,x_k\}=\delta_{jk},
\end{eqnarray}
where $\delta_{jk}$ is the usual Kronecker delta.

Then Hamiton's equations (\ref{HamE1}) can be rewritten in terms of the Poisson bracket as 
\begin{eqnarray}
\dot{x}_j=\{x_j,\mathcal{H}\}, \quad \dot{p}_j=\{p_j,\mathcal{H}\}.
\end{eqnarray}
The time evolution of dynamical variable $A(\textbf{x},\textbf{p})$ is determined by
\begin{eqnarray}
\frac{dA}{dt}=\{A,\mathcal{H}\}.\label{Cont1}
\end{eqnarray}
It follows that $A(\textbf{x},\textbf{p})$ is a constant if and only if $\{A,\mathcal{H}\}=0$. Such $A(\textbf{x},\textbf{p})$ are called constants of the motion of the system.

A system of $N$-degrees of freedom is integrable if it admits $N$ functionally independent constants of motion (including the Hamiltonian $\mathcal{H}$ of the system), $A_1=\mathcal{H}$, $A_2$,$\dots$, $A_N$, such that they are in involution:
\begin{eqnarray}
\{A_j, A_k\}=0, \quad 1\leq j,k\leq N.\label{int1}
\end{eqnarray} 
One of the most powerful methods for demonstrating integrability of a system is to explicitly exhibit a complete set of integrals by using the method of separation of variables. The exposition shows that one of the main assets of the Hamiltonian formalism is that it is well suited to utilizing symmetries of the system through the structures of the Poisson bracket.

\section{Quantum mechanics}\label{QMS1}

Recall that in classical mechanics, the dynamics of a system in $N$ dimensions is determined by its Hamiltonian $\mathcal{H}(\textbf{x},\textbf{p})$ (or the energy of the system) in the $2N$-dimensional phase space. Analogously in quantum mechanics, the dynamics of the system is determined by the self-adjoint Hamiltonian operator $H(\textbf{x},\textbf{p})$ in the Hilbert space, which also represents the energy of the system. Now $\textbf{x},\textbf{p}$ are self-adjoint operators with respect to suitable Hilbert space inner product, and satisfy the canonical commutation relations (i.e. the Heisenberg algebra),
\begin{eqnarray}
[x_i, x_j]=0=[p_i, p_j], \quad [x_i,p_j]=i\hbar\delta_{ij},\quad i, j=1,2,\dots,N.\label{CMR1}
\end{eqnarray}
In classical mechanics, as the state varies, the phase space coordinates $\textbf{x},\textbf{p}$ wander around in the phase space, according to Hamilton's equations. Analogously in quantum mechanics, quantum state vectors or simply quantum states wander around in the Hilbert space. In terms of Dirac's bra-ket notation, these states are denoted as $|\psi\rangle$. They vary in time according to  the time-dependent Schr\"{o}dinger equation
\begin{eqnarray}
i\hbar\frac{d}{dt}|\psi(t)\rangle=H|\psi(t)\rangle.
\end{eqnarray}
Analogous to classical mechanics, where in terms of the Poisson bracket, 
\begin{eqnarray}
\frac{d}{dt}A(\textbf{x},\textbf{p})=\{A,\mathcal{H}\},
\end{eqnarray}
the evolution of linear operators $A$ (possibly self-adjoint) in quantum mechanics is obtained from replacing
$\{A,\mathcal{H}\}$ by $\frac{1}{i\hbar}[A,H]$,
\begin{eqnarray}
\frac{d}{dt}A=\dot{A}=\frac{1}{i\hbar}[A,H].
\end{eqnarray}
Then the linear operator $A$ is a constant if and only if $[A,H]=0$. Such $A$ are called constants or integrals of motion. In particular, $H$ itself is a constant or integral of motion.

It is often convenient to use the wave functions $\psi(\textbf{x},t)$ to describe the quantum states of a quantum mechanical system in $N$ dimensions. For a given system, the set of all possible normalizable wave functions (at any given time) forms the (infinite-dimensional) Hilbert space. At any instant of time, wave functions $\psi(\textbf{x},t)$ are components of the quantum state vectors,
\begin{eqnarray}
|\psi(t)\rangle=\int d^N x \psi(\textbf{x},t)|\textbf{x}\rangle,
\end{eqnarray}
where $|\textbf{x}\rangle$ are eigenkets of the position operators $\textbf{x}$. That is, $\psi(\textbf{x},t)$ are continuous coefficients in the expansion of $|\psi(t)\rangle$ in the basis eigenkets $|\textbf{x}\rangle$. This is seen as follows. The $|\textbf{x}\rangle$ are the basis vectors in the Hilbert space which are orthonormal, $\int d^N x |\textbf{x}\rangle\langle\textbf{x}|=I$, so their inner product is
\begin{eqnarray}
\langle\textbf{x}|\textbf{x}\rangle=\delta(\textbf{x}-\textbf{x}').
\end{eqnarray}
Thus
\begin{eqnarray}
\psi(\textbf{x},t)=\int d^N\textbf{x}'\psi(\textbf{x}',t)\langle\textbf{x}|\textbf{x}'\rangle=\langle\textbf{x}|\psi(t)\rangle.
\end{eqnarray}
So in terms of the wave functions, the time dependent Schr\"{o}dinger equation is written as 
\begin{eqnarray}
i\hbar\frac{\partial}{\partial t}\psi(\textbf{x},t)=H\psi(\textbf{x},t).
\end{eqnarray}
Of all the integrals of the motion, the most interesting is $H$ itself, hence the eigenvalue problem for $H$ is especially interesting. Suppose $H$ has a complete orthogonal set of eigenvectors $\psi(\textbf{x})$ with corresponding eigenvalues $E$ so that
\begin{eqnarray}
H\psi(\textbf{x})=E\psi(\textbf{x}).\label{HE1}
\end{eqnarray}
Then the wave functions $\psi(\textbf{x},t)$ is given by 
\begin{eqnarray}
\psi(\textbf{x},t)=e^{-\frac{i}{\hbar}Et}\psi(\textbf{x}).
\end{eqnarray}
These $E$ are the possible energies of the system and they are real as $H$ is self-adjoint.
In the literature, Eq.(\ref{HE1}) is called the time-independent Schr\"{o}dinger equation and $\psi(\textbf{x})$ the time-independent wave functions. In the $\textbf{x}$-representation, $\textbf{p}\rightarrow -i\hbar\frac{\partial}{\partial\textbf{x}}$, $H$ in (\ref{HE1}) becomes a differential operator, which has the general form in the space with metric $g^{ij}$,
\begin{eqnarray}
H=-\frac{\hbar^2}{2}\sum_{i,j=1}^{N}g^{ij}(\textbf{x})\frac{\partial}{\partial x_i}\frac{\partial}{\partial x_j}+V(\textbf{x}).\label{QH1}
\end{eqnarray}

\section{Superintegrability}\label{SIS1}
Superintegrable systems are one important class of dynamical systems that can be solved exactly. In fact, the classical trajectories of superintegrable systems can be computed algebraically. 
In this section, we overview the solvability properties of classical Hamiltonian systems $\mathcal{H}=\frac{1}{2}\sum g^{jk}p_j p_k+V$ and their quantum equivalents $H=\Delta+V$, where $\Delta$ is the Laplace-Beltrami operator on the Riemannian manifold.  We will discuss integrability and superintegrability for both classical and quantum systems.

\subsection{Classical integrable and superintegrable systems}
In classical mechanics, an $N$-dimensional dynamical system with Hamiltonian
\begin{eqnarray}
\mathcal{H}=\frac{1}{2}\sum g^{ij}(\textbf{x})p_ip_j+V(\textbf{x}),
\end{eqnarray}
is integrable (Liouville integrable) if it has $N$ integrals of motion (including $\mathcal{H}$) $\mathcal{H}$, $X_a=f_a(\textbf{x},\textbf{p})$, $ a=2,\dots,N$, that are well-defined functionally independent functions in the phase space that are in involution
\begin{eqnarray}
\{\mathcal{H},X_a\}=0, \quad \{X_a,X_b\}=0, \quad a,b =2,\dots, N.
\end{eqnarray}
A classical Hamiltonian system is superintegrable if it is integrable and allows additional integrals of the motion $Y_b(\textbf{x},\textbf{p})$, 
\begin{eqnarray}
\{\mathcal{H}, Y_b\}=0, \quad  b=N+1,\dots, N+k,\quad k=1,\dots, N-1,
\end{eqnarray}
that are also well-defined functions in the phase space and the integrals $\{\mathcal{H},  X_2, \dots, X_{N},\\Y_{N+1}, \dots, Y_{N+k}\}$ are functionally independent. It is maximally superintegrable if the set contains $2N-1$ integrals and minimally superintegrable if it contains $N+1$ such integrals. The integrals of motion $Y_b$ are not required to be in evolution with $X_2,\dots, X_{N}$, nor with each other.

If $\mathcal{H}$ is the Hamiltonian in $N$-dimensions, then every constant of motion $\mathcal{S}$, polynomial or not, is a solution to the linear homogeneous first order partial differential equation $\{\mathcal{H},\mathcal{S}\}=0$ for $\mathcal{S}$ in $2N$ variables $\{x_i,p_i, i=1,2,\dots,N\}$. It is a well-known result that every solution of such differential equation can be expressed as a function $F(f_1,\dots, f_{2N-1})$ of $2N-1$ functionally independent solutions $\{f_1,f_2,\dots, f_{2N-1}\}$ \cite{cou1}. Thus there always exist the largest possible $2N-1$ independent functions, locally defined, in involution with the Hamiltonian $\mathcal{H}$. However, it is rare to find $2N-1$ such functions that are globally defined and polynomial in the momenta. Hence maximally superintegrable systems are very special.

The polynomial constants of motion for a system with Hamiltonian $\mathcal{H}$ generate a (polynomial) Poisson algebra. We have
\begin{lem}
Let $\mathcal{H}$ be a Hamiltonian with integrals of motion $\mathcal{L}$, $\mathcal{K}$. Then $\alpha\mathcal{L}+\beta\mathcal{K}$, $\mathcal{LK}$ and $\{\mathcal{L},\mathcal{K}\}$ are also integrals of motion.
\end{lem}

\subsection{Quantum integrable and superintegrable systems}
In quantum mechanics, the coordinates $x_j$ and momenta $p_j$ become hermitian operators in the Hilbert space, satisfying the canonical commutation relations, i.e. Heisenberg algebra. 

A quantum mechanical system in $N$-dimensions is integrable (of finite-order) if it allows $N$ integrals of motion, $L_j$, $j=1,\dots,N$ that satisfy the following conditions:
\begin{itemize}
\item
They are well-defined Hermitian operators in the enveloping algebra of the Heisenberg algebra $H_N$ or convergent series in the basis vectors $x_j, p_j$, $j=1,\dots,N$.
\item
They are algebraically independent in the sense that no Jordan polynomials formed entirely out of anti-commutators in $L_j$ vanish identically.
\item
The integrals $L_j$ commute pair-wise.
\end{itemize}
A quantum system is called superintegrable if it admits $N+k$, $k\geq 1$ algebraically independent finite-order partial differential operators $L_1=H,\dots,L_{N+k}$ such that $[H,L_j]=0$. In such case, the operators $L_j$ are called integrals of motion of the system. It is said to be maximally superintegrable if $k=N-1$ and minimally superintegrable if $k=1$.  The integrals of motion form a symmetry algebra $S_H$ of the quantum system, closed under scalar multiplication and commutation relations in analogy with classical case. That is,
\begin{lem}
Let $H$ be a Hamiltonian with integrals of motion $L$, $K$ and $\alpha, \beta$ be scalars. Then $\alpha L+\beta K$, $LK$ and $[L,K]$ are also integrals of motion.
\end{lem}

\section{Algebraic structures}\label{ABS1}

\subsection{Lie algebras}
The symmetry algebra of a quantum system is often a Lie algebra or polynomial algebra. So in this subsection, we recall some basic definitions for Lie algebras.

A Lie algebra is a vector space $\mathfrak{g}$ over a field $\mathbb{F}$ $(\mathbb{R}$ or $\mathbb{C})$ together with a bilinear mapping, known as the Lie bracket, $[\quad,\quad]: \mathfrak{g}\times \mathfrak{g}\longrightarrow \mathfrak{g}$, which satisfies the following axioms for all $X,Y,Z\in \mathfrak{g}$ and $a,b\in \mathbb{F}$,
\begin{enumerate}[(i)]
\item
Bilinearity: \hspace{1cm} $[aX+bY,Z]=a[X,Z]+b[Y,Z]$, \\ $\phantom{x}\hspace{17ex}[X,aY+bZ]=a[X,Y]+b[X,Z]$;
\item
Antisymmetric:\hspace{1cm} $[X,Y]=-[Y,X]$;  
\item
Jacobi identity:\hspace{1cm} $[X,[Y,Z]+[Y,[Z,X]]]+[Z,[X,Y]]=0$; 
\item
Commutativity:\hspace{1cm} $[X,X]=0$.  
\end{enumerate}
Let $\{X_i, i=1,2,..., d\}$ be the basis for a finite-dimensional Lie algebra $\mathfrak{g}$. Then the bracket operation is completely determined by 
\begin{eqnarray}
[X_i,X_j]=\sum_{k=1}^d C^k_{ij}X_k. 
\end{eqnarray}
The constants $C^k_{ij}$ are called structure constants of $\mathfrak{g}$.
A Lie algebra is real if $\mathbb{F}=\mathbb{R}$, while it is complex if $\mathbb{F}=\mathbb{C}$. Real Lie algebras have real structure constants, while the structure constants for complex Lie algebras can be real or complex. A subset $\mathfrak{g}'$ of a Lie algebra $\mathfrak{g}$ that is closed under the Lie bracket is called a Lie subalgebra. 
 
In the representation theory of Lie algebras, an important object is the universal enveloping algebra $U(\mathfrak{g})$ of the Lie algebra $\mathfrak{g}$. The universal enveloping algebra $U(\mathfrak{g})$ is an unital associative algebra constructed from the Lie algebra $\mathfrak{g}$ by taking the bracket to be the commutator $[X,Y]=XY-YX$. Thus for $\mathfrak{g}$ with basis $\{X_1,\dots,X_d\}$, $U(\mathfrak{g})$ is defined by the commutation relations,
\begin{eqnarray}
X_i X_j-X_j X_i=\sum_{k=1}^d C^k_{ij}X_k.
\end{eqnarray}
Casimir operators play an essential role in constructing representations of $\mathfrak{g}$.  A Casimir operator $K$ of $\mathfrak{g}$ is an operator which commutes with all elements of $\mathfrak{g}$,
\begin{eqnarray}
[K, X_i]=0 \quad \text{for all}\quad X_i\in \mathfrak{g}.
\end{eqnarray}
The Casimir operators can  be linear, quadratic or polynomial functions in $X_i$. In general, a Casimir operator of order $p$ in $\{X_\alpha\}$ has the form
\begin{eqnarray}
K_p=\sum_{\alpha_1,\alpha_2,\dots,\alpha_p}f^{\alpha_1,\alpha_2,\dots,\alpha_p}X_{\alpha_1}X_{\alpha_2}.\dots.X_{\alpha_p},
\end{eqnarray}
where $\alpha_1, \alpha_2,\dots, \alpha_p$ are constants.

\subsection{Polynomial algebras}
We now turn our attention to the most important ingredients, used in this thesis, polynomial algebras. A polynomial algebra $P(\mathfrak{g})$ is a non-linear (or polynomial) deformation of Lie algebra $\mathfrak{g}$.  Let $\{X_i, i=1,\dots, d\}$ be the generators of a polynomial algebra. Then the polynomial algebra $P(\mathfrak{g})$ is defined by the commutation relations,
\begin{eqnarray}
[X_i,X_k]=\sum C^q_{ik}X_q+\sum\sum C^{p,q}_{ik}X_p X_q+\sum\sum\sum C^{p,q,r}_{ik}X_p X_q X_r+\dots,
\end{eqnarray}
where $C^q_{ik}$,.., etc are structure constants which may involve the central elements of the Lie algebra $\mathfrak{g}$.

Quadratic algebras are the particular cases of polynomial algebras, which have important applications in many superintegrable systems. The general form of a quadratic algebra $Q(3)$ with three generators $\{A,B,C\}$ which often appears as the symmetry algebra of a superintegrable system has the following commutation relations \cite{das2, gra1}, 
\begin{eqnarray}
&&[A,B]=C,\label{AB}
\\&&
[A,C]=\alpha A^2+\gamma\{A,B\}+\delta A+\epsilon B+\zeta,\label{AC}
\\&&
[B,C]=a A^2+b B^2+c\{A,B\}+d A+e B+z,\label{BC}
\end{eqnarray}
where $A$, $B$ are the integrals of motion of the system, $\alpha, \gamma$, $a, b, c$ are constants, and $d, \delta, \epsilon, \zeta, z$ depend on the central elements $H$. 
The Jacobi identity for $A$, $B$ and $C$ induces the relation
\begin{eqnarray}
[A,[B,C]]=[B,[A,C]],
\end{eqnarray} 
which gives the relations $b=-\gamma$, $c=-\alpha$ and $e=-\delta$. Hence ((\ref{AB})-(\ref{BC})) take the form
\begin{eqnarray}
&&[A,B]=C,\label{AB1}
\\&&
[A,C]=\alpha A^2+\gamma\{A,B\}+\delta A+\epsilon B+\zeta,\label{AC1}
\\&&
[B,C]=a A^2-\gamma B^2-\alpha\{A,B\}+d A-\delta B+z,\label{BC1}
\end{eqnarray}
where   
\begin{eqnarray}
&&\delta=\delta(H)=\delta_0+\delta_1 H, \quad \epsilon=\epsilon(H)=\epsilon_0+\epsilon_1 H,\quad d=d(H)=d_0+d_1 H,\nonumber
\\&&
\zeta=\zeta(H)=\zeta_0+\zeta_1 H+\zeta_2 H^2, \quad z=z(H)=z_0+z_1 H+z_2H^2,
\end{eqnarray}
with $\delta_i$, $\epsilon_i$, $d_i$, $\zeta_i$, $z_i$ being constants.
The Casimir operator of the quadratic algebra $Q(3)$ is given by
\begin{eqnarray}
K&=&C^2-\alpha\{A^2,B\}-\gamma\{A,B^2\}+(\alpha\gamma-\delta)\{A,B\}+(\gamma^2-\epsilon)B^2+(\gamma\delta-2\zeta)B\nonumber\\&& +\frac{2a}{3}A^3+\left(d+\frac{a\gamma}{3}+\alpha^2\right)A^2+\left(\frac{a\epsilon}{3}+\alpha\delta+2z\right)A.\label{KK1}
\end{eqnarray}

\subsection{Deformed oscillator algebras}
The realizations of the polynomial algebras in terms of the deformed oscillator algebras are essential in deriving the energy spectra and degeneracies of quantum superintegrable systems.
In this subsection let review the results about the deformed oscillator algebra obtained in \cite{das2, das1}
\begin{eqnarray}
[\aleph,b^{\dagger}]=b^{\dagger},\quad [\aleph,b]=-b,\quad bb^{\dagger}=\Phi (\aleph+1),\quad b^{\dagger} b=\Phi(\aleph),\label{DOA1}
\end{eqnarray}
where $\aleph $ is the number operator and $\Phi(x)$ is well behaved real function satisfying
\begin{eqnarray}
 \Phi(0)=0,\quad \Phi(x)>0,\quad \text{for all}\quad x>0.
\end{eqnarray}
 The realization of the quadratic algebra $Q(3)$ in terms of (\ref{DOA1}) is given by
\begin{eqnarray}
 A=A(\aleph),\quad B=b(\aleph)+b^{\dagger} \rho(\aleph)+\rho (\aleph)b, \label{AOB1}
\end{eqnarray} 
 where $A(x)$, $b(x)$ and $\rho(x)$ are functions. Defining
 \begin{eqnarray}
 \Delta A(\aleph)= A(\aleph+1)-A(\aleph),
\end{eqnarray}
  then it can be shown that the following relations hold:
 \begin{eqnarray}
 &&[A(\aleph),b^\dagger]=b^\dagger \Delta A(\aleph), \quad \{A(\aleph),b^\dagger\}=b^\dagger (\Delta A(\aleph+1)+A(\aleph)), \nonumber\\&& [A(\aleph),b]=-\Delta A(\aleph)b,  \quad \{A(\aleph),b\}=(\Delta A(\aleph+1)+A(\aleph))b.
 \end{eqnarray}
By means of (\ref{AOB1}), one can obtain the realization of the generator $C$ (\ref{AB1})
\begin{eqnarray}
C=[A,B]=b^\dagger \Delta A(\aleph)\rho(\aleph)-\rho(\aleph)\Delta A(\aleph)b.
\end{eqnarray}
It thus follows from (\ref{AC1}) that \cite{das2}
\begin{eqnarray}
[A,C]&=&b^\dagger (\Delta A(\aleph))^2\rho(\aleph)+\rho(\aleph)(\Delta A(\aleph))^2 b\nonumber\\&=& b^\dagger(\gamma\Delta A(\aleph+1)+A(\aleph)+\epsilon)\rho(\aleph)+\rho(\aleph)(\gamma\Delta A(\aleph+1)+A(\aleph)+\epsilon)b \nonumber\\&&+\alpha A(\aleph)^2+2\gamma A(\aleph)b(\aleph)+\delta A(\aleph)+\epsilon b(\aleph)+\zeta.
\end{eqnarray}
This provides two constraints for $A(\aleph)$ and $b(\aleph)$:
\begin{eqnarray}
&&\Delta A(\aleph))^2=\gamma\Delta A(\aleph+1)+A(\aleph)+\epsilon,\label{DA1}
\\&&\alpha A(\aleph)^2+2\gamma A(\aleph)b(\aleph)+\delta A(\aleph)+\epsilon b(\aleph)+\zeta=0.\label{DA2}
\end{eqnarray}
Solving (\ref{DA1}) and (\ref{DA2}) yields two distinct cases \cite{das2}
\begin{eqnarray}
    A(\aleph)&=&\left\{
                \begin{array}{ll}
                  \sqrt{\epsilon}(\aleph+u),\quad \text{$\gamma=0$},\\
                  \frac{\gamma}{2}\left((\aleph+u)^2-\frac{\epsilon}{\gamma^2}-\frac{1}{4}\right),\quad \text{$\gamma\neq 0$},
                \end{array}
              \right.
 \\ b(\aleph)&=&-\frac{\alpha\{(\aleph+u)^2-1/4\}}{4} +\frac{\alpha\epsilon-\delta\gamma}{2\gamma^2}\nonumber\\&& -\frac{\alpha\epsilon^2-2\delta\epsilon\gamma+4\gamma^2\zeta}{4\gamma^4}\frac{1}{(\aleph+u)^2-\frac{1}{4}},\quad \text{$\gamma\neq 0$},
\\ \rho(\aleph)&=&\frac{1}{3.2^{12}.\gamma^8(\aleph+u)(1+\aleph+u)\{1+2(\aleph+u)\}^2}, \quad \text{$\gamma\neq 0$},    
\end{eqnarray}
where $u$ is a parameter. Moreover, using (\ref{DOA1}) and (\ref{AOB1}), one can show that (\ref{BC1}) becomes
\begin{eqnarray}
&&2\Phi(\aleph+1)\{\Delta A(\aleph)+\frac{\gamma}{2}\}\rho(\aleph)-2\Phi(\aleph)\{\Delta A(\aleph-1)-\frac{\gamma}{2}\}\rho(\aleph-1)
\nonumber\\&&=a A(\aleph)^2-\gamma b(\aleph)^2-2\gamma A(\aleph)b(\aleph)+d A(\aleph)-\delta(\aleph)+z,\label{PHi1}
\end{eqnarray}
and the Casimir (\ref{KK1}) takes the form,
\begin{eqnarray}
K&=&\Phi(\aleph+1)\{\gamma^2-\epsilon-2\gamma A(\aleph)-\Delta A(\aleph)^2\}\rho(\aleph)+\Phi(\aleph)\{\gamma^2-\epsilon-2\gamma A(\aleph)\nonumber\\&&-\Delta A(\aleph-1)^2\}\rho(\aleph-1)-2\alpha A(\aleph)^2b(\aleph)+\{\gamma^2-\epsilon-2\gamma A(\aleph)\}b(\aleph)^2\nonumber\\&&+2(\alpha\gamma-\delta)A(\aleph)b(\aleph)+(\gamma\delta-2\zeta)b(\aleph)+\frac{2}{3}aA(\aleph)^3+(d+\frac{1}{3}a\gamma+\alpha^2)A(\aleph)^2\nonumber\\&& +(\frac{1}{3}a\epsilon+\alpha\delta+2z)A(\aleph).\label{PHi2}
\end{eqnarray}
(\ref{PHi1}) and (\ref{PHi2}) are linear functions of $\Phi(\aleph)$ and $\Phi(\aleph+1)$. Hence the function $\Phi(\aleph)$ can be determined by using $\rho(\aleph)$, and is given by \cite{das2}
\begin{eqnarray}
\Phi(\aleph)&=&-3072\gamma^6 K\{2(\aleph+u)-1\}^2-48\gamma^6(\alpha^2\epsilon-\alpha\delta\gamma+a\epsilon\gamma-d\gamma^2)\{2(\aleph+u)-3\}\nonumber\\&& \times\{2(\aleph+u)-1\}^4\{2(\aleph+u)+1\}+\gamma^8(3\alpha^2+4a\gamma)\{2(\aleph+u)-3\}^2\nonumber\\&& \times\{2(\aleph+u)-1\}^4\{2(\aleph+u)+1\}^2+768(\alpha\epsilon^2-2\delta\epsilon\gamma+4\gamma^2\zeta)^2\nonumber\\&&+32\gamma^4\{2(\aleph+u)-1\}^2\{12(\aleph+u)^2-12(\aleph+u)-1\}(3\alpha^2\epsilon^2-6\alpha\delta\epsilon\gamma+2a\epsilon^2\gamma \nonumber\\&& +2\delta^2\gamma^2-4d\epsilon\gamma^2+8\gamma^3 z+4\alpha\gamma^2\zeta)-256\gamma^2\{2(\aleph+u)-1\}^2(3\alpha^2\epsilon^3-9\alpha\delta\epsilon^2\gamma\nonumber\\&&+a\epsilon^3\gamma+6\delta^2\epsilon\gamma^2-3d\epsilon^2\gamma^2+2\delta^2\gamma^4+2d\epsilon\gamma^4+12\epsilon\gamma^3 z-4\gamma^5 z+12\alpha\epsilon\gamma^2\zeta\nonumber\\&&-12\delta\gamma^3\zeta+4\alpha\gamma^4\zeta).
\end{eqnarray}

\subsection{Finite-dimensional representations of polynomial algebras}
We now study the finite-dimensional representations of $Q(3)$ by means of its realization in terms of the deformed oscillator algebra \cite{das1, das2}. 

Let $|n;E\rangle$ be the common eigenstate of the number operator $\aleph$ and the Hamiltonian $H$, 
\begin{eqnarray}
&&\aleph|n;E\rangle=n|n;E\rangle,
\nonumber\\&&
H|n;E\rangle=E|n;E\rangle, \quad n=0,1,2,\dots.
\end{eqnarray}
Then the action of the operators $b$ and $b^\dagger$ on $|n;E\rangle$ are given by
\begin{eqnarray}
&&b^\dagger|n;E\rangle=\sqrt{\Phi(n+1;E,u)}|n+1;E\rangle,
\nonumber\\&&
b|n;E\rangle=\sqrt{\Phi(n;E,u)}|n-1;E\rangle,
\end{eqnarray}
where $u$ is certain representation-dependent parameter.
For the above representation to be unitary and finite-dimensional, we impose the constraints on the structure function
\begin{eqnarray}
&&\Phi(p+1;E,u)=0,\quad \Phi(0;E,u)=0,\nonumber\\&& \Phi(n;E,u)>0,\quad n=1,2,\dots,p,\label{CSF1}
\end{eqnarray}
where $p$ is a positive integer. The constraints (\ref{CSF1}) give rise to $p+1$-dimensional unitary representations of $Q(3)$. Their solutions provide the energy eigenvalues $E$ of the associated model.


\chapter{Superintegrable Kepler-Coulomb system }\label{ch3}

{\bf \large{Acknowledgement}}
\\This chapter is based on the work that was published in  Ref. \cite{fh1}. I have incorporated text of that paper \cite{fh1}. In this chapter we introduce a new superintegrable Kepler-Coulomb system with non-central terms in $N$-dimensional Euclidean space. We show this system is multiseparable and allows separation of variables in hyperspherical and hyperparabolic coordinates. We present the wave function in terms of special functions. We give an algebraic derivation of spectrum of the superintegrable system. We show how the $so(N+1)$ symmetry algebra of the $N$-dimensional Kepler-Coulomb system is deformed to a quadratic algebra with only 3 generators and structure constants involving a Casimir operator of $so(N-1)$ Lie algebra. We construct the quadratic algebra and the Casimir operator. We show this algebra can be realized in terms of deformed oscillator and obtain the structure function which yields the energy spectrum.

\section{Introduction}
Superintegrable systems form a fundamental part of mathematical theories and modern physics such as quantum chemistry and nuclear physics. They possess many properties in particular analytic and algebraic solvability. Moreover, they have connections to special functions, (exceptional) orthogonal polynomials and Painlev\'{e} transcendents. Though it has much deeper historical roots, the modern theory of superintegrability was only started 45 years ago \cite{fri2}. A systematic classification of maximally superintegrable systems is now complete for 2 and 3 dimensional Hamiltonians on conformally flat spaces. The classification in higher dimensions and with higher order integrals of motion is much more complicated. In lower dimensions much work has been done for systems involving spins, magnetic fields and monopoles \cite{mil1}. We refer the reader to this review paper for an extended list of references, description of the properties, definitions of superintegrability and symmetry algebra in classical and quantum mechanics. One important property of such systems is that they possess non-abelian symmetry algebras generated by integrals of motion. They can be embedded in non-invariance algebras involving non-commuting operators. These symmetry algebras are in general finitely generated polynomial algebras and only exceptionally finite dimensional Lie algebras. The most known examples whose symmetry algebras are Lie algebras generated by integrals of motion are $N$-dimensional hydrogen atom and harmonic oscillator. See \cite{foc1, barg1, sud1, ban1, lou2, ras1} for systems with $so(N+1)$ symmetry and \cite{jau1, bak1, lou1, bar1, hwa1} for those with $su(N)$ symmetry.

Quadratic algebras have been used to provide algebraic derivation of the energy spectrum of superintegrable systems such as the Hartmann system that models the Benzene molecule \cite{gra1}. A systematic approach for 2D superintegrable systems with quadratic algebra involving three generators was proposed in \cite{das2}. This method is based on the construction of the Casimir operators and the realization of the quadratic algebra as deformed oscillator. It has recently been generalized to 2D superintegrable systems with cubic, quartic and more generally polynomial algebras \cite{isa1}. In some cases degeneracy patterns for the energy level are non-trivial and one needs to consider a union of finite dimensional unitary representations to obtain the correct total degeneracies. It has been pointed out how the method can be adapted to study 3D, 4D, 5D and 8D superintegrable systems \cite{mar7, mar10, mar13, tan1}. However, the generalization of this approach to $N$-dimensional superintegrable systems is an unexplored subject. Higher-dimensional superintegrable systems often lead to higher rank polynomial algebras and the structure of these algebras is unknown. 

The purpose of this chapter is to show how we can provide an algebraic derivation of the complete energy spectrum and the total number degeneracies of the $N$-dimensional superintegrable Kepler-Coulomb system with non-central terms. It is based on the quadratic algebra symmetry of the system with structure constant involving Casimir operator of $so(N-1)$ Lie algebra. To our knowledge these type of calculations for higher dimensions or even arbitrary dimensions have not been explored before. This is a first step in the study of the polynomial algebra approach to general $N$-dimensional systems. 

The structure of the chapter is as follows: In Section 2, we present a new superintegrable Hamiltonian system in $N$-dimensional Euclidean space and show that its Schrodinger wave function is multi-separable in hyperspherical and hyperparabolic coordinates. We present the wave function in terms of special functions and obtain its energy spectrum. In Section 3, we give an algebraic derivation of the energy spectrum of the system. We construct the quadratic symmetry algebra and its Casimir operators. We investigate the realization of the quadratic algebra in terms of deformed oscillator algebra of Daskaloyannis \cite{das2} and obtain the structure functions which yield the energy spectrum. Finally, in Section 4, we present some discussions with a few remarks on the physical and mathematical relevance of these algebras.

\section{New quantum superintegrable system and separation of variables}
Let us consider the following $N$-dimensional superintegrable Kepler-Coulomb system with non central terms 
\begin{equation}
H=\frac{1}{2}p^{2}-\frac{c_{0}}{r}+\frac{c_{1}}{r(r+x_{N})}+\frac{c_{2}}{r(r-x_{N})}, \label{ahamil}
\end{equation}
where $ \vec{r}=(x_{1},x_{2},...,x_{N})$, $\vec{p}=(p_{1},p_{2},...,p_{N})$, $r^{2}=\sum_{i=1}^{N}x_{i}^{2}$, $p_{i}=-i \hbar \partial_{i}$ and $c_0$, $c_1$, $c_2$ are positive real constants. This system is a generalization of the 3D system that appears in the classification of quadratically superintegrable systems on three dimensional Euclidean space \cite{eva1,kib1}. The 3D system has been considered using the method of separation of variables and various results obtained. This new $N$-dimensional superintegrable Hamiltonian includes as particular 3D case the Hartmann potential which has applications in quantum chemistry \cite{har1, har2}. Moreover, the classical analog of the Hartmann system possesses closed trajectories and thus periodic motion \cite{win1}. Such 3D models with non central potentials have also found applications in context of pseudo-spin symmetries in relativistic quantum mechanics and regard of the Klein Gordon and Dirac equations \cite{che1,guo1,zha1}. The $N$-dimensional model introduced here could have wider applicability and other aspects as quasi-exactly solvable deformations could be investigated. Moreover, it has also been shown that the 3D and 5D particular cases are still superintegrable with appropriate vector potentials, that are in fact respectively an Abelian monopole and nonAbelian $su(2)$ monopole \cite{pie1, mad5, tru1}. In addition, it has also been demonstrated that these 3D and 5D models with vector potentials admit dual that are also superintegrable. These properties make the $N$-dimensional generalisation to be highly interesting. In this section we apply separation of variables to (\ref{ahamil}).

\subsection{Hyperspherical Coordinates}
The $N$-dimensional hyperspherical coordinates are given by 
\begin{eqnarray}
&x_{1}&=r \sin(\Phi_{N-1})\sin(\Phi_{N-2})\cdots \sin(\Phi_{1}),
\nonumber\\& x_{2}&=r \sin(\Phi_{N-1})\sin(\Phi_{N-2})\cdots \cos(\Phi_{1}),
\nonumber\\&...&
\nonumber\\&...&
\nonumber\\&  x_{N-1}&=r\sin(\Phi_{N-1})\cos(\Phi_{N-2}),
\nonumber\\&  x_{N}&=r\cos(\Phi_{N-1}),\label{akfpf1}
\end{eqnarray}
where the $N$ $x_{i}$'s are Cartesian coordinates in the hyperspherical coordinates, $\{\Phi_{1},\dots, \Phi_{N-1}\}$ are the hyperspherical angles and $r$ is the hyperradius. The Schrodinger equation $H\psi=E\psi$ in $N$-dimensional hyperspherical coordinates can be expressed as
\begin{eqnarray}
&&\left[\frac{\partial^2}{\partial r^2}+\frac{N-1}{r}\frac{\partial}{\partial r}-\frac{1}{r^2}\Lambda^2(N)+\frac{2c'_{0}}{r}-\frac{2c'_{1}}{r^2(1+\cos\Phi_{N-1})} \right.\nonumber\\&&\qquad\left.-\frac{2c'_{2}}{r^2(1-\cos\Phi_{N-1})}+2E'\right]\psi(r,\Omega)=0, \label{aschro}
\end{eqnarray}
where $c'_{0}=\frac{c_{0}}{\hbar{^2}}$, $c'_{1}=\frac{c_{1}}{\hbar{^2}}$, $c'_{2}=\frac{c_{2}}{\hbar{^2}}$ and $E'=\frac{E}{\hbar^{2}}$; and $\Lambda^2(N)$ is the grand angular momentum operator which satisfies the recursive formula
\begin{eqnarray}
-\Lambda^2(N)= \frac{\partial^2}{\partial\Phi^2_{N-1}}+(N-2)\cot(\Phi_{N-1})\frac{\partial}{\partial\Phi_{N-1}}-\frac{\Lambda^2(N-1)}{\sin^2(\Phi_{N-1})},
\end{eqnarray}
valid for all $N$ and $\Lambda^2(1)=0.$ The separation of the radial and angular parts of (\ref{aschro})
\begin{equation}
\psi(r,\Omega)=R(r)y(\Omega_{N-1})
\end{equation}
 gives rise to
\begin{eqnarray}
&& \frac{d^2R}{dr^2} +\frac{N-1}{r}\frac{dR}{dr}+\left(\frac{2c'_{0}}{r}+2E'-\frac{A}{r^2}\right)R=0,\label{arad1} 
\\&&\left[\Lambda^2(N)+\frac{2c'_{1}}{1+\cos(\Phi_{N-1})}+\frac{2c'_{2}}{1-\cos(\Phi_{N-1})}-A\right]y(\Omega_{N-1})=0,\label{aan1}
\end{eqnarray}
where $A$ is the separation constant  and $N>1$. Again we may separate the variables of (\ref{aan1}) \cite{sae1,ave1}
\begin{eqnarray}
y(\Omega_{N-1})=\Theta(\Phi_{N-1})y(\Omega_{N-2}), 
\end{eqnarray}
we obtain
\begin{eqnarray}
&&\left[\frac{\partial^2}{\partial\Phi_{N-1}^2}+(N-2)\cot(\Phi_{N-1})\frac{\partial}{\partial\Phi_{N-1}}-\frac{2c'_{1}}{1+\cos\Phi_{N-1}}-\frac{2c'_{2}}{1-\cos\Phi_{N-1}}\right. \nonumber\\&&\qquad\left. +A-\frac{I_{N-2}(I_{N-2}+N-3)}{\sin^2\Phi_{N-1}}\right]\Theta(\Phi_{N-1})=0\label{aan2}
\end{eqnarray}
and
\begin{eqnarray}
\left[\Lambda^2(N-1)-I_{N-2}(I_{N-2}+N-3)\right]y(\Omega_{N-2})=0, \qquad (N>2),\label{aan3}
\end{eqnarray}
where $I_{N-2}(I_{N-2}+N-3)$ is the separation constant and $I_{N-2}\in \mathbb{Z}$.  
The solution of (\ref{aan3}) is obtained recursively in $N$.
 
We now turn to (\ref{aan2}), which can be converted, by setting $z=\cos(\Phi_{N-1})$ and $\Theta(z)=(1+z)^{a}(1-z)^{b} f(z)$, to
\begin{eqnarray}
&&(1-z^2)f''(z)+\{2a-2b-(2a+2b+N-1)z\}f'(z)\nonumber \\&&\quad\quad+\{A-(a+b)(a+b+N-2)\}f(z)=0,\label{aan4}
\end{eqnarray}
where $2a=\delta_{1}+I_{N-2}$, \quad  $2b=\delta_{2}+I_{N-2}$ and
\begin{eqnarray}
\delta_{i}=\{\sqrt{(I_{N-2}+\frac{N-3}{2})^2+4c'_{i}}-\frac{N-3}{2}\}-I_{N-2},\quad i=1, 2. \label{aprova2}
 \end{eqnarray}
Comparing (\ref{aan4}) with the Jacobi differential equation  
\begin{equation}
(1-x^{2})y''+\{\beta-\alpha-(\alpha+\beta+2)x\}y'+\lambda(\lambda+\alpha+\beta+1)y=0,
\end{equation}
we obtain the  separation constant 
\begin{equation}
A=(l+\frac{\delta_{1}+\delta_{2}}{2})(l+\frac{\delta_{1}+\delta_{2}}{2}+N-2),\label{acon1} 
\end{equation}
where $l=\lambda+I_{N-2}$. 
Hence the solutions of (\ref{aan4}) are given in terms of the Jacobi polynomials as
\begin{eqnarray}
&\Theta(\Phi_{N-1})&=\Theta_{l I_{N-2}}(\Phi_{N-1; \delta_{1}, \delta_{2}})
\nonumber\\&&=F_{l I_{N-2}}(\delta_{1}, \delta_{2})(1+\cos(\Phi_{N-1}))^{\frac{(\delta_{1}+I_{N-2})}{2}}(1-\cos(\Phi_{N-1}))^{\frac{(\delta_{2}+I_{N-2})}{2}}\nonumber\\&&\qquad\times P^{(\delta_{2}+I_{N-2}, \delta_{1}+I_{N-2})}_{l-I_{N-2}}(\cos(\Phi_{N-1})),\label{ajp1}
\end{eqnarray}
where $P^{(\alpha, \beta)}_{\lambda}$ denotes a Jacobi polynomial and $l\in \mathbb{N}$. The normalization constant $F_{l I_{N-2}}(\delta_{1},\delta_{2})$ in (\ref{ajp1}) is given by 
\begin{eqnarray}  
&F_{l I_{N-2}}&(\delta_{1},\delta_{2})=\frac{(-1)^{(I_{N-2}-|I_{N-2}|)/2}}{2^{|I_{N-2}|}}\nonumber\\&&\times\sqrt{\frac{(2l+\delta_{1}+\delta_{2}+N-2)(l-|I_{N-2}|)!\Gamma(l+I_{N-2}+\delta_{1}+\delta_{2}+N-2)}{2^{\delta_{1}+\delta_{2}+N-1}\pi\Gamma(l+\delta_{1}+N-2)\Gamma(l+\delta_{2}+N-2)}}.
\end{eqnarray}
Let us now turn to the radial equation. Using (\ref{acon1}), we have
\begin{eqnarray}
\frac{d^2R}{dr^2}+\frac{N-1}{r}\frac{dR}{dr}+\left[\frac{2c'_{0}}{r}+2E'-\frac{1}{r^2}(l+\frac{\delta_{1}+\delta_{2}}{2})(l+\frac{\delta_{1}+\delta_{2}}{2}+N-2)\right]R=0.\label{aan5}
\end{eqnarray}
(\ref{aan5}) can be converted, by setting  
 $z=\varepsilon r$, $R(z)=z^{l+\frac{\delta_{1}+\delta_{2}}{2}} e^{-\frac{z}{2}}f(z)$ and $E'=\frac{-\varepsilon^2}{8}$, to
\begin{eqnarray}
z\frac{d^2f(z)}{dz^2}+\{(2l+\delta_{1}+\delta_{2}+N-1)-z\}\frac{df(z)}{dz}-\left(l+\frac{\delta_{1}+\delta_{2}}{2}+\frac{N-1}{2}-\frac{2c'_{0}}{\varepsilon}\right)f(z)=0.\label{aan6}
\end{eqnarray}
Set  
\begin{eqnarray}
n=\frac{2c'_{0}}{\varepsilon}-\frac{\delta_{1}+\delta_{2}}{2}-\frac{N-3}{2}.\label{aan7}
\end{eqnarray}
Then (\ref{aan6}) can be written as
\begin{equation}
z\frac{d^2f(z)}{dz^2}+\{(2l+\delta_{1}+\delta_{2}+N-1)-z\}\frac{df(z)}{dz}-(-n+l+1)f(z)=0.\label{aan8}
\end{equation}
This is the confluent hypergeometric equation. Hence we can write the solution of (\ref{aan5}) in terms of the confluent hypergeometric function as  
\begin{eqnarray}
&R(r)&\equiv R_{nl}(r;\delta_{1}, \delta_{2})=F_{nl}(\delta_{1},\delta_{2})(\varepsilon r)^{l+\frac{\delta_{1}+\delta_{2}}{2}} e^{\frac{-\varepsilon r}{2}}\nonumber\\&&
\quad \times {}_1 F_1(-n+l+1, 2l+\delta_{1}+\delta_{2}+N-1; \varepsilon r).\label{aan9}
\end{eqnarray}
The normalization constant $F_{nl}(\delta_{1},\delta_{2})$ from the above relation is given by   
\begin{eqnarray}
&F_{nl}(\delta_{1},\delta_{2})&=\frac{2(-c'_{0})^{3/2}}{(n+\frac{\delta_{1}+\delta_{2}}{2})^2}\frac{1}{\Gamma(2l+\delta_{1}+\delta_{2}+N-1)}\nonumber\\&&\qquad\times\sqrt{\frac{\Gamma(n+l+\delta_{1}+\delta_{2}+N-2))}{(n-l-1)!}}.
\end{eqnarray}
In order to have a discrete spectrum the parameter $n$ needs to be positive integer. From (\ref{aan7}) 
\begin{equation}
\varepsilon=\frac{2c_{0}}{\hbar^{2}(n+\frac{\delta_{1}+\delta_{2}}{2}+\frac{N-3}{2})}\label{en1}
\end{equation} and hence the energy $E=\frac{-\varepsilon^2 \hbar^2}{8}$ is given by
\begin{equation}
E\equiv E_{n}=-\frac{c^{2}_{0}}{2\hbar^{2}(n+\frac{\delta_{1}+\delta_{2}}{2}+\frac{N-3}{2})^{2}},\qquad n=1,2,3,....\label{aen1}
\end{equation}
Here $n$ is the principal quantum number.  
\subsection{Hyperparabolic Coordinates}
The $N$-dimensional hyperparabolic coordinates are given by
\begin{eqnarray}
&x_{1}&=\sqrt{\xi \eta} \sin(\Phi_{N-2})\cdots \sin( \Phi_{1}),
\nonumber\\& x_{2}&=\sqrt{\xi \eta} \sin(\Phi_{N-2})\cdots \cos( \Phi_{1}),
\nonumber\\&...&
\nonumber\\&...&
\nonumber\\&x_{N-1}&=\sqrt{\xi \eta} \cos(\Phi_{N-2}), 
\nonumber\\& x_{N}&=\frac{1}{2}(\xi -\eta), 
\nonumber\\& r&= \frac{\xi+\eta}{2},\label{akfpf2}
\end{eqnarray}
where the $N$ $x_{i}$'s are Cartesian coordinates in the hyperparabolic coordinates, $\{\Phi_{1},\dots, \Phi_{N-1}\}$ are the hyperparabolic angles and the parabolic coordinates $\xi$, $\eta$ range from $0$ to $\infty$.  The Schrodinger equation $H\psi=E\psi$ in the hyperparabolic coordinates can be written as 
\begin{eqnarray}
&&\left[-\frac{2}{\xi+\eta}\left\{\Delta(\xi)+\Delta(\eta)-\frac{\xi+\eta}{4\xi\eta}\Lambda^2(\Omega_{N-1})\right\}-\frac{2c'_{0}}{\xi+\eta}+\frac{2c'_{1}}{\xi(\xi+\eta)}\right.\nonumber\qquad\\&&\left.+\frac{2c'_{2}}{\eta(\xi+\eta)}\right]U(\xi, \eta, \Omega_{N-1})=E'U(\xi, \eta, \Omega_{N-1}),\label{afk1}
\end{eqnarray} 
where $\Lambda^2(N)$ is the grand angular momentum operator defined in the previous subsection and 
\begin{eqnarray*}
&&\Delta(\xi) = \xi^{-\frac{N-3}{2}}\frac{\partial}{\partial \xi}  \xi^{\frac{N-1}{2}}\frac{\partial}{\partial \xi},\quad  \Delta(\eta) = \eta^{-\frac{N-3}{2}}\frac{\partial}{\partial \eta}  \eta^{\frac{N-1}{2}}\frac{\partial}{\partial \eta},\\&& c'_{0}=\frac{c_{0}}{\hbar{^2}}, \quad c'_{1}=\frac{c_{1}}{\hbar{^2}},\quad c'_{2}=\frac{c_{2}}{\hbar{^2}}\quad \text{and} \quad E'=\frac{E}{\hbar^{2}}.
\end{eqnarray*}
The equation can be separated in radial and angular parts by setting 
\begin{eqnarray}
U(\xi, \eta, \Omega_{N-1})=U_{1}(\xi, \eta)y(\Omega_{N-1}),
\end{eqnarray}
we obtain two equations 
\begin{eqnarray}
&&\left[\Delta(\xi)+\Delta(\eta)-\frac{c'_{1}}{\xi}-\frac{c'_{2}}{\eta}+\frac{E'}{2}\xi+\frac{E'}{2}\eta+c'_{0}-\frac{1}{4\xi}I_{N-2}(I_{N-2}+N-3)\right.\nonumber\\&&\left.-\frac{1}{4\eta}I_{N-2}(I_{N-2}+N-3)\right]U_{1}(\xi,\eta)=0\label{afk2}
\end{eqnarray}
and
\begin{equation}
\Lambda^2(\Omega_{N-1})y(\Omega_{N-1})=I_{N-2}(I_{N-2}+N-3)y(\Omega_{N-1}),\label{afk3}
\end{equation}
where $I_{N-2}(I_{N-2}+N-3)$ being the general form of the separation constant. The solution of (\ref{afk3}) is well-known. By looking for solution of (\ref{afk2}) of the form 
\begin{eqnarray}
U_{1}(\xi,\eta)=f_{1}(\xi)f_{2}(\eta),
\end{eqnarray}
we get two coupled equations
\begin{eqnarray}
&&\left[\Delta(\xi)-\frac{c'_{1}}{\xi}+\frac{E'}{2}\xi-\frac{1}{4\xi}I_{N-2}(I_{N-2}+N-3)+v_{1}\right]f_{1}(\xi)=0,\label{afk4}
\\&&
\left[\Delta(\eta)-\frac{c'_{2}}{\eta}+\frac{E'}{2}\eta-\frac{1}{4\eta}I_{N-2}(I_{N-2}+N-3)+v_{2}\right]f_{2}(\eta)=0,\label{afk5}
\end{eqnarray}
where $v_{2}=-v_{1}-c'_{0}$ and $v_{1}$ is the separating constant. Putting $z_{1}=\varepsilon \xi $ in (\ref{afk4}), $z_{2}=\varepsilon \eta $ in (\ref{afk5}), and $E'=-\varepsilon^2$, these two equations become
\begin{eqnarray}
z_{i}\frac{d^2f_{i}}{dz^2_{i}}+\left(\delta_{1}+I_{N-2}+\frac{N-1}{2}-z_{i}\right)\frac{df_{i}}{dz_{i}}-\left(\frac{\delta_{1}+I_{N-2}+\frac{N-1}{2}}{2}-\frac{1}{\varepsilon}v_{i}\right)f_{i}=0,\label{afk6}
\end{eqnarray}
where 
\begin{eqnarray*}
 \delta_{i}=\left\{\sqrt{(I_{N-2}+\frac{N-3}{2})^2+4c'_{i}}-\frac{N-3}{2}\right\}-I_{N-2},\quad i=1, 2.
\end{eqnarray*}
Let us now denote 
\begin{eqnarray}
n_{i}=-\frac{1}{2}\left(\delta_{i}+I_{N-2}+\frac{N-1}{2}\right)+\frac{1}{\varepsilon}v_{i}, \quad i=1, 2.\label{afk8}
\end{eqnarray}
Then (\ref{afk6}) can be identified with the Laguerre differential equation. Thus we have the normalized wave function
 \begin{eqnarray} 
U(\xi, \eta, \Omega_{N-1})&=&U_{n_{1}n_{2}I_{N-2}}(\xi, \eta, \Omega_{N-1}; \delta_{1},\delta_{2})\nonumber\\&=&\frac{\hbar \varepsilon^2}{\sqrt{-8c_{0}}}f_{n_{1}I_{N-2}}(\xi;\delta_{1})f_{n_{2}I_{N-2}}(\eta;\delta_{2})\frac{e^{iI_{N-2}\Omega_{N-1}}}{\sqrt{2\pi}}, 
\end{eqnarray}
where
\begin{eqnarray*}
 &f_{n_{i} I_{N-2}}(t_{i};\delta_{i})& \equiv f_{i}(t_{i})=\frac{1}{\Gamma(I_{N-2}+\delta_{i}+\frac{N-1}{2})}\sqrt{\frac{\Gamma(n_{i}+I_{N-2}+\delta_{i}+\frac{N-1}{2})}{ n_{i}!}} 
 \nonumber\\&&\times(\frac{\varepsilon}{2}t_{i})^{(I_{N-2}+\delta_{i})/2} e^{-\varepsilon t_{i}/4}\times {}_1F_{1}(-n_{i}, I_{N-2}+\delta_{i}+\frac{N-1}{2}; \frac{\varepsilon}{2}t_{i}),\label{afk7}   
\end{eqnarray*}
 $i=1, 2$ and $ t_{1}\equiv\xi , t_{2}\equiv\eta $. We look for the discrete spectrum and thus $n_{1}$ and $n_{2}$ are both positive integers.

An expression for the energy of the system in terms of $n_{1}$ and $n_{2}$  can be found by using $E=-\hbar^2 \varepsilon^2$ in (\ref{afk8}) to be 
\begin{equation}
E\equiv E_{(n_{1},n_{2})}=\frac{-c^{2}_{0}}{\hbar^{2}\{n_{1}+n_{2}+\frac{1}{2}(\delta_{1}+\delta_{2}+2I_{N-2}+N-1)\}^{2}}.\label{aen2}
\end{equation}
We can relate the quantum numbers in (\ref{aen1}) and (\ref{aen2}) by the following relation
\begin{eqnarray}
n_{1}+n_{2}+I_{N-2}=n-1,
\end{eqnarray}
where $n_{1}, n_{2}=0, 1, 2,....$
\section{Algebraic derivation of the energy spectrum}
In this section we present an algebraic derivation of the energy spectrum for the non-central Kepler-Coulomb system in $N$-dimension. For this purpose we recall some facts about the central Kepler-Coulomb system in the next subsection.

\subsection{Kepler-Coulomb System}
The Hamiltonian of the (central) Kepler-Coulomb system in $N$-dimensional Euclidean space is given by
\begin{equation}
H=\frac{1}{2}p^{2}-\frac{c_{0}}{r},\label{aham1}
\end{equation}
where $ \vec{r}=(x_{1},x_{2},...,x_{N})$, $\vec{p}=(p_{1},p_{2},...,p_{N})$, $r^{2}=\sum_{i=1}^{N}x_{i}^{2}$ and $p_{i}=-i \hbar \partial_{i}.$
This system has integrals of motion given by the Runge-Lenz vector
\begin{eqnarray}
&M_{j}&=\frac{1}{2} \sum_{i=1}^{N} ( L_{ji}p_{i}-p_{i}L_{ij} )- \frac{c_{0}x_{j}}{r}\\&&=-x_{j}(\frac{1}{2}p^{2}+H)+\sum_{i=1}^{N} x_{i}p_{i}p_{j}-\frac{N-1}{2}i\hbar p_{j}- \frac{c_{0}x_{j}}{r}
\end{eqnarray}\label{arg1}
and the angular momentum 
\begin{equation}
L_{ij}=x_{i}p_{j}-x_{j}p_{i}
\end{equation}
for $i,j=1,2,...,N$.
They commute with the Hamiltonian (\ref{aham1}),  
\begin{eqnarray*}
[ L_{ij},H]=[ M_{i},H]=0.
\end{eqnarray*}  
The Runge-Lenz vector and angular momentum components generate a Lie algebra isomorphic to $so(N+1)$ for bound states and $so(N,1)$ for scattering state,
\begin{eqnarray*}
 &[L_{ij},L_{kl}]&=i ( \delta_{ik}L_{jl}+ \delta_{jl}L_{ik}-\delta_{il}L_{jk}-\delta_{jk}L_{il})\hbar,  
 \\&[M_{i},M_{j}]&=-2i\hbar H L_{ij}, \quad  [M_{k},L_{ij}]=i\hbar (\delta_{ik}M_{j}-\delta_{jk}M_{i}).
\end{eqnarray*}  
An algebraic derivation of the energy spectrum was obtained using a chain of second order Casimir operators (i.e., subalgebra chain $so(N+1) \supset so(N) \supset ... \supset so(2)$ ) to define appropriate quantum numbers \cite{foc1,sud1,ban1,lou2,ras1,bar1}. Another derivation consists in using higher order Casimir operators. This has been performed for the five dimensional hydrogen atom for which the $so(6)$ Casimir operators of order two, three and four, and the related eigenvalues were used to calculate the energy spectrum \cite{tru1}. The calculation was involved and to our knowledge no such calculation for higher dimensions or even arbitrary dimensions have been done. Let us also remark that the symmetry algebra is not the only algebraic structure of interest and one can use various embeddings of the symmetry algebra into a larger one called non-invariance algebra to perform algebraic derivation in particular for $so(4,2)$, $so(7, 4)$ and $sp(8, R)$ \cite{bar3,kib2,san1}. 

\subsection{Quadratic Poisson algebra in the non-central Kepler-Coulomb system}
We now consider the non-central Kepler-Coulomb system with Hamiltonian given by (\ref{ahamil}).
This system is superintegrable. The system has the following second order integrals of motion 
\begin{eqnarray}
&A&=\sum_{i<j}^{N}L_{ij}^{2}+ \frac{2 r c_{1}}{r+x_{N}}+\frac{2 r c_{2}}{r-x_{N}},\label{akf5}
\\&B&=-M_{N}+\frac{c_{1}(r-x_{N})}{r(r+x_{N})}-\frac{c_{2}(r+x_{N})}{r(r-x_{N})}.\label{akf6}
\end{eqnarray}
This can be checked by proving  
\begin{eqnarray*}
\{H,A\}_{p}=0=\{H,B\}_{p},
\end{eqnarray*}
where $\{\quad,\quad\}_{p}$ is the Poisson bracket defined as $\{X,Y\}_{p}=\sum^{n}_{j=1}(\frac{\partial X}{\partial p_{j}}\frac{\partial Y}{\partial q_{j}}-\frac{\partial X}{\partial q_{j}}\frac{\partial Y}{\partial p_{j}})$. We still have first order integrals of motion 
\begin{eqnarray*}
L_{i j}=x_{i}p_{j}-x_{j}p_{i} \quad for \quad i, j=1,\dots, N-1
\end{eqnarray*}
as $\{H, L_{ij}\}_{p}=0.$ The system is minimally superintegrable as it allows $N+1$ integrals of motion and in 3D the potential reduces to the Hartmann system which described axially symmetric systems (i.e. ring-shaped molecules)\cite{har1, har2} which is also known to be minimally superintegrable. The integrals are $H$, $A$, $B$ and $N-2$ components of the angular momentum $L_{ij}$. Define
\begin{eqnarray}
J^{2}=\sum_{i<j}^{N-1}L_{ij}^{2}.
\end{eqnarray}
$J^2$ is the Casimir operator of the $so(N-1)$ Lie algebra and is also a central element of the Poisson algebra. Other Casimir operators are associated to this $so(N-1)$ Lie algebra. 
After a long computation, we can show that the integrals of motion generate the quadratic Poisson algebra,
\begin{eqnarray}
&\{A,B\}_{p}&=C,\label{akf1}
\\&\{A,C\}_{p}&=-4 A B + 4 (c_{1}-c_{2})c_{0},\label{akf2}
\\&\{B,C\}_{p}&=2B^{2}-8 H A + 4 J^{2}H+8(c_{1}+c_{2})H-2 c_{0}^{2},\label{akf3}
\end{eqnarray}
where
\begin{eqnarray}
&C&=-\sum_{i,j}^{N}2x_{i}x_{j}p_{i}p_{j}p_{N}+\sum_{i}^{N}[2r^2 p^2_{i}p_{N}+\frac{2c_{0}}{r}x_{i}x_{N}p_{i}-\frac{2c_{1}}{r}x_{i}p_{i}\nonumber\\&&\qquad+\frac{2c_{2}}{r}x_{i}p_{i}] -2c_{0}r p_{N}+\frac{4c_{1}r p_{N}}{r+x_{N}}+ \frac{4c_{2}r p_{N}}{r-x_{N}}.\label{akf7}
\end{eqnarray}
The first order integrals of motion generate a $so(N-1)$ Lie algebra
\begin{eqnarray*}
&\{L_{ij},L_{kl}\}_{p}&= \delta_{ik}L_{jl}+ \delta_{jl}L_{ik}-\delta_{il}L_{jk}-\delta_{jk}L_{il}, 
\end{eqnarray*}
for $i, j, k, l=1, ..., N-1.$ Moreover, $\{A, L_{ij}\}_{p}=0=\{B, L_{ij}\}_{p}$. So the full symmetry algebra is a direct sum of the quadratic algebra and $so(N-1)$ Lie algebra. Thus $so(N+1)$ symmetry algebra in the central Kepler-Coulomb system is deformed to the quadratic algebra with defined by (\ref{akf1})-(\ref{akf3}). Its Casimir operator is given by
\begin{eqnarray}
&&K=C^{2}+4 A B^{2}-8 (c_{1}-c_{2})c_{0}B - 8 H A^{2} + [16 (c_{1}+c_{2})H+8 J^{2}-4 c_{0}^{2}]A.\label{akf4}
\end{eqnarray}
Using the realization for $A$, $B$, $C$ (i.e. (\ref{akf5}), (\ref{akf6}) and (\ref{akf7})), we can show that the Casimir operator (\ref{akf4}) becomes in terms of the central elements $H$ and $J^2$  
\begin{eqnarray}
K=8(c_{1}-c_{2})^2 H-8(c_{1}+c_{2})c^2_{0}-4c^2_{0}J^2.
\end{eqnarray}
The study of the Poisson algebra and its Casimir operator is important as they will correspond to the lowest order terms in $\hbar$ the quadratic algebra and Casimir operator of the corresponding quantum system.

\subsection{Quadratic algebra in the quantum non-central Kepler-Coulomb system}
We now consider the Hamiltonian of the quantum non-central Kepler-Coulomb system 
\begin{equation}
H=\frac{1}{2}p^{2}-\frac{c_{0}}{r}+\frac{c_{1}}{r(r+x_{N})}+\frac{c_{2}}{r(r-x_{N})}.
\end{equation}
Similar to the classical case, the integrals of motion are 
\begin{eqnarray}
&A&=\sum_{i<j}^{N}L_{ij}^{2}+ \frac{2 r c_{1}}{r+x_{N}}+\frac{2 r c_{2}}{r-x_{N}},\label{apr1}
\\&B&=-M_{N}+\frac{c_{1}(r-x_{N})}{r(r+x_{N})}-\frac{c_{2}(r+x_{N})}{r(r-x_{N})},\label{apr2}
\\&J^{2}&=\sum_{i<j}^{N-1}L_{ij}^{2}.
\end{eqnarray}
We still have a set of first order integrals of motion
\begin{eqnarray*}
L_{i j}=x_{i}p_{j}-x_{j}p_{i} \quad for \quad i, j=1, ... , N-1.
\end{eqnarray*}
The integral of motion $A$ is associated with the separation of variables in hyperspherical coordinates and the integral of motion $B$ is associated with the separation of variables in hyperparabolic coordinates. There is no other coordinates systems in which the Schrodinger equation, related to this model admits separation of variables, as this is connected to the existence of second order integrals of motion. We can easily verify the commutation relations
\begin{eqnarray*}
[H,A]=[H,B]=[H,J^{2}]=[A,J^{2}]=[B,J^{2}]=[H,L_{ij}]=[L_{ij},J^{2}]=0. 
\end{eqnarray*}
For later convenience we present a diagram representation of the above commutation relations 
\begin{equation}
\begin{xy}
(0,0)*+{A}="a"; (20,0)*+{J^{2}}="f"; (40,0)*+{B}="b"; (20,20)*+{H}="h";  (60,0)*+{J^{2}}="j"; (100,0)*+{L_{ij}}="l"; (80,20)*+{H}="g"; 
"a";"h"**\dir{--}; 
"h";"b"**\dir{--};
"h";"f"**\dir{--};
"f";"b"**\dir{--};
"f";"a"**\dir{--};
"j";"l"**\dir{--}; 
"l";"g"**\dir{--};
"g";"j"**\dir{--};
\end{xy}
\end{equation}
The left figure shows that $J^2$ is a central element. The right figure illustrates $J^2$ is the Casimir operator of $so(N-1)$ Lie algebra realized by angular momentum $L_{ij}$, $i, j=1, 2,..., N-1$. 
\\After a long direct but involving analytical computations and the use of various commutation identities and relations, we can show that the integrals of motion close to the quadratic algebra $Q(3)$,
\begin{eqnarray}
&[A,B]&=C,\label{aprova5}
\\&[A,C]&=2 \hbar^{2} \{A,B\}+(N-1)(N-3) \hbar^{4} B -4 (c_{1}-c_{2}) \hbar^{2} c_{0},\label{aprova6}
\\&[B,C]&=-2 \hbar^{2} B^{2} +8 \hbar^{2} H A -4 \hbar^{2} J^{2} H + (N-1)^{2} \hbar^{4} H \nonumber\\&&\quad-8 \hbar^{2}(c_{1}+c_{2})H + 2 \hbar^{2} c_{0}^{2},\label{aprova1}
\end{eqnarray}
where \begin{eqnarray}
&C&=-\sum_{i,j}^{N}2i\hbar x_{i}x_{j}p_{i}p_{j}p_{N}+\sum_{i}^{N}[2 i \hbar r^2 p^2_{i}p_{N}+2\hbar^2 x_{N}p^2_{i}-2N\hbar^2 x_{i} p_{i}p_{N}\nonumber\\&&\quad+ \frac{2i\hbar c_{0}}{r} x_{i}x_{N}p_{i}-\frac{2i\hbar c_{1}}{r}x_{i}p_{i}+\frac{2i\hbar c_{2}}{r}x_{i}p_{i}]+\frac{i\hbar^3}{2}(N-1)^2 p_{N}\nonumber\\&&\quad-2i\hbar c_{0}r p_{N} +\frac{4r c_{1}}{r+x_{N}}i\hbar p_{N}+\frac{4r c_{2}}{r-x_{N}}i\hbar p_{N}+\frac{(N-1)c_{0}}{r}\hbar^2 x_{N}\nonumber\\&&\quad -\frac{(N+1)r+(N-3)x_{N}}{r(r+x_{N})}c_{1}\hbar^2+\frac{(N+1)r-(N-3)x_{N}}{r(r-x_{N})}c_{2}\hbar^2.\label{apr3}
\end{eqnarray}
This quadratic algebra is the quantization of the Poisson algebra in the previous subsection. It can be shown that the Casimir operator is 
\begin{eqnarray}
&K&=C^{2}-2 \hbar^{2} \{A,B^{2}\}+ [ 4 \hbar^{4}-(N-1)(N-3) \hbar^{4}] B^{2} + 8 (c_{1}-c_{2})\hbar^{2}c_{0} B\nonumber\\&&+ 8 \hbar^{2} H A^{2} +2[ -4 \hbar^{2} J^{2} H + (N-1)^{2} \hbar^{4} H -8 \hbar^{2} (c_{1}+c_{2})H^{2} +2 \hbar^{2} c_{0}^{2}]A.\label{aprova4} 
\end{eqnarray}
By means of the explicit expressions of $A$, $B$, $C$ (i.e. (\ref{apr1}), (\ref{apr2}) and (\ref{apr3})), we can show that the Casimir operator (\ref{aprova4}) becomes in terms of the central elements $H$ and $J^2$,   
\begin{eqnarray}
&K&=2(N-3)(N-1)\hbar^{4} H J^{2} -8 \hbar^{2}(c_{1}-c_{2})^{2} H + 4  (N-3) (N-1)(c_{1}+c_{2}) \hbar^{4} H \nonumber\\&&\quad 
 -\hbar^{6}(N-3)(N-1)^{2} H +4 \hbar^{2} c_{0}^{2} J^{2} +8 \hbar^{2} (c_{1}+c_{2})c_{0}^{2} -2 (N-3) \hbar^{4} c_{0}^{2}.\label{aprova3} 
\end{eqnarray}
The first order integrals also generate a $so(N-1)$ Lie algebra as in the classical case
\begin{eqnarray}
[L_{ij},L_{kl}]=i ( \delta_{ik}L_{jl}+ \delta_{jl}L_{ik}-\delta_{il}L_{jk}-\delta_{jk}L_{il})\hbar,
\end{eqnarray}
for $i, j, k, l=1, .., N-1.$ Furthermore, $[A,L_{ij}]=0=[B, L_{ij}]$. So the full symmetry algebra is the direct sum of $Q(3)$ and $so(N-1)$ (i.e. $Q(3)\oplus so(N-1)$). A chain of second and higher order Casimir operators are associated with this $so(N-1)$ component. However for the purpose of the algebraic derivation of the spectrum we rely mainly on the quadratic algebra and its Casimir operator.

The quadratic algebra (\ref{aprova5})-(\ref{aprova1}) can be realized in terms of the generator of the deformed oscillator algebra $\{\aleph, b^{\dagger}, b\}$ \cite{das1} which satisfies
\begin{eqnarray}
[\aleph,b^{\dagger}]=b^{\dagger},\quad [\aleph,b]=-b,\quad bb^{\dagger}=\Phi (\aleph+1),\quad b^{\dagger} b=\Phi(\aleph),
\end{eqnarray}
where $\aleph $ is the number operator and $\Phi(x)$ is real function such that $\Phi(0)=0$ and $\Phi(x)>0$ for all $x>0$. The realization of the quadratic algebra $Q(3)$ is of the form $A=A(\aleph)$, $B=b(\aleph)+b^{\dagger} \rho(\aleph)+\rho (\aleph)b$, where $A(x)$, $b(x)$ and $\rho(x)$ are functions. Similarly as the case of quadratic algebra for 2D superintegrable systems \cite{das2}, we obtain
\begin{eqnarray}
&&A(\aleph)=\hbar^2\{(\aleph+u)^2-\frac{(N-1)(N-3)}{4}-\frac{1}{4}\},
\\&&
b(\aleph)=\frac{-1}{4\hbar^4\{(\aleph+u)^2-\frac{1}{4}\}},
\\&&
\rho(\aleph)=\frac{1}{3.2^{20}.\hbar^{16}(\aleph+u)(1+\aleph+u)\{1+2(\aleph+u)\}^2},
\end{eqnarray}
where $u$ is a constant determined from the constraints on the structure function.
We now construct the structure function $\Phi (x)$ by using the Casimir operator (\ref{aprova4}) and the quadratic algebra ((\ref{aprova5}), (\ref{aprova6}),(\ref{aprova1})) as
\begin{eqnarray}
&\Phi (x, u, H)&=3145728 c^2_{0} (c_{1} - c_{2})^2 h^{12} - 196608 h^{12} [8 c^2_{0} (c_{1} + c_{2}) h^2 \nonumber\\&&\quad- 8 (c_{1} - c_{2})^2 h^2 H + 4 c^2_{0} h^2 J^2 - 2 c^2_{0} h^4 (N-3) + 4 (c_{1} + c_{2}) h^4 H \nonumber\\&&\quad (N-3) (N-1) + 2 h^4 H J^2 (N-3) (N-1) - h^6 H (N-3)\nonumber\\&&\quad (N-1)^2] \{-1 + 2 (x+u)\}^2 - 1024 h^4 [-128 h^{10} \{2 c^2_{0} h^2\nonumber\\&&\quad - 8 (c_{1} + c_{2}) h^2 H - 4 h^2 H J^2 + h^4 H (N-1)^2\} + 256 h^{14} H \nonumber\\&&\quad (N-3) (N-1) + 96 h^{10} \{2 c^2_{0} h^2 - 8 (c_{1} + c_{2}) h^2 H\nonumber\\&&\quad  - 4 h^2 H J^2 + h^4 H (N-1)^2\} (N-3) (N-1) - 96 h^{14} H  \nonumber\\&&\quad (N-3)^2 (N-1)^2] \{-1 + 2 (x+u)\}^2  + 98304 h^{18} H  \nonumber\\&&\quad\{-3 + 2 (x+u)\} \{-1 + 2 (x+u)\}^4 \{1 + 2 (x+u)\} \nonumber\\&&\quad + 512 h^8 [64 h^6 \{2 c^2_{0} h^2  - 8 (c_{1} + c_{2}) h^2 H - 4 h^2 H J^2 \nonumber\\&&\quad+ h^4 H (N-1)^2\} - 128 h^{10} H (N-3)(N-1)] \{-1 \nonumber\\&&\quad  + 2 (x+u)\}^2\{-1 - 12 (x+u) + 12 (x+u)^2\}.\label{afk9}
\end{eqnarray}
Here we have also used the expression (\ref{aprova3}) for the Casimir.

A set of quantum numbers can be defined in same way as \cite{ras1} with a subalgebra chain for $so(N-1)$ Lie algebra and the related Casimir operators. Thus the eigenvalue of $J^2$ is $\hbar^2 I_{N-2}(I_{N-2}+N-3)$. Also as we act with the structure function $\Phi(x)$ on Fock basis $|n, E>$ with $\aleph|n, E>=n|n, E>$, $H$ in $\Phi(x, u, H)$ can be replaced by $E$. 

To obtain unitary representations we should impose the following three constraints on the structure function :
\begin{equation}
\Phi(p+1, u, E)=0,\quad \Phi(0,u,E)=0,\quad \Phi(x)>0,\quad \forall x>0,\label{apro2}
\end{equation}
where $p$ is a positive integer. These constraints ensure the representations are unitary and finite $(p+1)$-dimensional. The solution of these constrains gives us the energy $E$ and the arbitrary constant $u$. 

From (\ref{afk9}) and eigenvalues of $J^2$ and $H$, the structure function takes the following factorized form that will greatly simplifies the analysis of finite dimensional unitary representations:
\begin{eqnarray}
&\Phi(x)&=6291456 E\hbar^{18}[x+u-\frac{1-m_{1}-m_{2}}{2}][x+u-\frac{1-m_{1}+m_{2}}{2}] \nonumber\\&&\quad \times[x+u-\frac{1+m_{1}-m_{2}}{2}][x+u-\frac{1+m_{1}+m_{2}}{2}]\nonumber\\&&\quad \times[x+u- (\frac{1}{2}-\frac{c_{0}}{\hbar\sqrt{-2E}}  )] [x+u-( \frac{1}{2}+\frac{c_{0}}{\hbar\sqrt{-2E}}   )] 
\end{eqnarray}
with $\hbar^{2}m_{1,2}^{2} =16 c_{1,2}+(4(I_{N-2}(I_{N-2}+N-3)) +(N-3)^{2})\hbar^{2}.$
 
From the condition (\ref{apro2}), we obtain all possible structure functions and energy spectra, for $\epsilon_{1}=\pm 1$,  $\epsilon_{2}=\pm 1$.
\\Set-1:
\begin{eqnarray}
u=\frac{1}{2}+\frac{c_{0}}{\hbar\sqrt{-2E}},  \qquad E=\frac{-2 c^2_{0}}{h^2 (2 + 2 p +\epsilon_{1} m_{1} +\epsilon_{2} m_{2} )^2}
\end{eqnarray}
and
\begin{eqnarray}
&\Phi(x)&=\frac{786432 c^2_{0} h^{16} x[2 + 2 p +x+ \epsilon_{1} m_{1} +\epsilon_{2} m_{2}] }{(2 + 2 p + \epsilon_{1} m_{1} +\epsilon_{2} m_{2} )^2}  \nonumber\\&&\quad \times [2 + 2p - 2x+(1 + \epsilon_{1})m_{1}+ (1 + \epsilon_{2}) m_{2} ]\nonumber\\&&\quad \times[2x- 2 - 2p +(1 -\epsilon_{1}) m_{1} - (1 +\epsilon_{2})m_{2}  ]  \nonumber\\&&\quad\times [2x-2p-2+(1-\epsilon_{1})m_{1}+(1-\epsilon_{2})m_{2}]\nonumber\\&&\quad \times[2x-2p-2-(1+\epsilon_{1})m_{1}+(1-\epsilon_{2})m_{2}]. 
 \end{eqnarray}
\\Set-2:
\begin{eqnarray}
u=\frac{1}{2}-\frac{c_{0}}{\hbar\sqrt{-2E}},\qquad E=\frac{-2 c^2_{0}}{h^2 (2 + 2 p +\epsilon_{1} m_{1} +\epsilon_{2} m_{2} )^2}
\end{eqnarray}
and
\begin{eqnarray}
&\Phi(x)&=\frac{786432 c^2_{0} h^{16} x [2+2p-x+\epsilon_{1}m_{1}+\epsilon_{2}m_{2}]}{(2 + 2 p + \epsilon_{1} m_{1} +\epsilon_{2} m_{2} )^2}  \nonumber\\&&\quad \times[ 2 + 2p+2x+(1 + \epsilon_{1})m_{1}+ (1 + \epsilon_{2}) m_{2}]\nonumber\\&&\quad \times[2 + 2 p + 2x-(1-\epsilon_{1}) m_{1} -(1-\epsilon_{2})m_{2}]
\nonumber\\&&\quad \times[2+2p+2x+(1+\epsilon_{1})m_{1}-(1-\epsilon_{2})m_{2}]\nonumber\\&&\quad \times[2+2p+2x-(1-\epsilon_{1})m_{1}+(1+\epsilon_{2})m_{2}]. 
 \end{eqnarray}
\\Set-3:
\begin{eqnarray}
u=\frac{1}{2}(1+\epsilon_{1}m_{1}+\epsilon_{2}m_{2}), \qquad  E=\frac{-2 c^2_{0}}{h^2 (2 + 2 p +\epsilon_{1} m_{1} +\epsilon_{2} m_{2} )^2}
\end{eqnarray}
and
\begin{eqnarray}
&\Phi(x)&=\frac{786432 c^2_{0} h^{16}[1+p-x]}{(2 + 2 p + \epsilon_{1} m_{1} +\epsilon_{2} m_{2} )^2}[1+p+x+\epsilon_{1}m_{1}+\epsilon_{2}m_{2}]  \nonumber\\&&\quad \times[ 2 + 2p+(1 +\epsilon_{1})m_{1}- (1 - \epsilon_{2}) m_{2}][2 + 2 p+(1+\epsilon_{1}) m_{1} +(1+\epsilon_{2})m_{2}] \nonumber\\&&\quad \times[2+2p-(1-\epsilon_{1})m_{1}+(1+\epsilon_{2})m_{2}][2x-(1-\epsilon_{1})m_{1}-(1-\epsilon_{2})m_{2}].
\end{eqnarray}
The structure functions are positive for the constraints $\varepsilon_{1}=1$, $\varepsilon_{2}=1$ and $m_{1}, m_{2}>0$.
Using formula (\ref{aprova2}), we can write $m_{1}$ and $m_{2}$ in terms of $\delta_{1}$, $\delta_{2}$ and $I_{N-2}$ as $m_{1}=\frac{1}{2}(3-2I_{N-2}-N-2\delta_{1})$, $ m_{2}=\frac{1}{2}(3-2I_{N-2}-N-2\delta_{2})$. Making the identification $p=n_{1}+n_{2}$, the energy spectrum becomes (\ref{aen2}).

\section{Conclusion}
One of the main results of this chapter is the construction of the quadratic algebra for the $N$-dimensional non-central Kepler-Coulomb system. We obtain the Casimir operators and derive the structure function of the deformed oscillator realization of the quadratic algebra. The finite dimensional unitary representations of the algebra yield the energy spectrum. We compare our results with those obtained from separation of variables. 
 
Algebra structures appearing in $N$-dimensional superintegrable systems are an unexplored area. More complicated polynomial algebra structures are expected in general and it is non-trivial to generalise the present approach to these cases. Let us mention the possible generalizations to monopole interaction and their dual based on \cite{mar7,mar10}. Moreover, the classification of certain families of superintegrable systems  with quadratic integrals of motion in $N$-dimensional curved spaces have been done and their quadratic algebra structures should be studied \cite{bal1}. In recent a paper \cite{rig1} a superintegrable system with spin has been obtained. An algebraic derivation of the spectrum would be of interest.

Let us point out that 2D superintegrable systems and their quadratic algebras have been related to the full Askey scheme of orthogonal polynomials via a contraction process. This illustrates a deep conection between superintegrable systems, orthogonal polynomials and quadratic algebras \cite{kal10}. The relations between the quadratic algebras of superintegrable systems involving Dunkl operators and special functions have been studied in a series of papers \cite{gen2, gen3}. It would be interesting to generalize the results to $N$-dimensional superintegrable systems.


\chapter{Superintegrable harmonic oscillators}\label{ch4}

{\bf \large{Acknowledgement}}
\\This chapter is based on the work that was published in  Ref. \cite{fh2}. I have incorporated text of that paper \cite{fh2}. In this chapter, we introduce a new family of $N$-dimensional quantum superintegrable model consisting of double singular oscillators of type $(n,N-n)$. The special cases $(2,2)$ and $(4,4)$ were previously identified as the duals of 3- and 5-dimensional deformed Kepler-Coulomb systems with $u(1)$ and $su(2)$ monopoles, respectively. The models are multiseparable and their wave functions are obtained in $(n,N-n)$ double-hyperspherical coordinates. We obtain the integrals of motion and construct the finitely generated polynomial algebra that is the direct sum of a quadratic algebra $Q(3)$ involving three generators, $so(n)$, $so(N-n)$ (i.e. $Q(3)\oplus so(n) \oplus so(N-n)$ ). The structure constants of the quadratic algebra themselves involve the Casimir operators of the two Lie algebras $so(n)$ and $so(N-n)$. Moreover, we obtain the finite dimensional unitary representations (unirreps) of the quadratic algebra and present an algebraic derivation of the degenerate energy spectrum of the superintegrable model.

\section{Introduction}

The isotropic and anisotropic harmonic oscillators \cite{jau1, bak1, mos1, lou1, bud1, bar1, fra1, hwa1} are among the most well known maximally superintegrable systems with applications in various areas of physics. However, the aspect of symmetry algebras generated by well-defined integrals of motion is a complicated issue in quantum mechanics as recognized in early work by Jauch and Hill \cite{jau1}. In the case of the isotropic harmonic oscillator in $N$-dimensional Euclidean space one can apply finite dimensional unitary representations (unirreps) and Gel'fand invariants and their eigenvalues of the Lie algebra $su(n)$ to provide an algebraic derivation  of the spectrum and the degeneracies \cite{hwa1}. The 3D case of isotropic harmonic oscillator was discussed using what is now called the Fradkin tensor and connected to $su(3)$ generators \cite{fra1}. There are however various embeddings of the symmetry algebra, some with operators that do not commute with the Hamiltonian \cite{lou1}, in terms of $su(n+1)$, $su(n,1)$ and $sp(n)$. The anisotropic case has been for a long time the subject of research due to its various applications in nuclear physics. An analysis based on well-defined integrals of motion and polynomial algebras of arbitrary order is known only in the 2D case \cite{bon1}.\par
%
%
It is also known for a long time that the harmonic oscillator in 2-dimensional Euclidean space is related to the 2D Kepler-Coulomb system  via the so-called Levi-Civita or regularisation transformation \cite{lev1}. This transformation however can only be extented to certain specific dimensions. The 3- and the 5-dimensional Kepler-Coulomb systems are related to the harmonic oscillators in 4- and 8-dimensional Euclidean space respectively via the Kustaanheimo-Stiefel and the Hurwitz transformations \cite{kus1,hur1} and these results can be extended to curved spaces \cite{kal1}. The Kustaanheimo-Stiefel transformations are connected to the so-called St\"{a}ckel transformations and were used to classify superintegrable systems in conformally flat spaces \cite{kal2,kal3,mil1}. These are specific Levi-Civita, Kustaanheimo-Stiefel and Hurwitz transformations. The above-mentioned connections between various models can be reinterpreted in terms of monopole interactions: the 4D harmonic oscillator has a duality relation with 3-dimensional Kepler system with an Abelian $u(1)$ monopole ( also refered to as MICZ-Kepler systems) \cite{mci1,zwa1,jac1,bar2,bac1,dho1,feh1} and the 8D harmonic oscillator is in fact dual to 5D Kepler-Coulomb system with a non-Abelian $su(2)$ monopole (Yang -Coulomb monopole) \cite{mad3, mad2, mad4, ner1, mad5, ple2, ple1, ple3, tru1}. \par
%
%
It has been discovered that some of these duality relations can be extended to deformed MICZ-Kepler systems and 4D singular oscillators that are sums of two 2D singular oscillators. It was proven that the superintegrability and multiseparability properties are preserved \cite{mad6, mad9, ran1, sal1, tre1, pet1}. The dual of the 4D singular oscillator has interesting properties and its classical analog has period motion \cite{tre1}. It has been recognized this is also the case for the dual of the 8D singular oscillator \cite{mar10}. Moreover, it has been shown that a quadratic algebra structure exists for these two models with monopole interactions and their duals \cite{mar10, mar7}. \par
%
%
In this chapter, we introduce a new family of $N$-dimensional superintegrable singular oscillators with arbitrary partition ($n$, $N-n$) of the coordinates. This model is a generalization of the 4D and 8D systems \cite{pet1, mar10, mar7} obtained from monopole systems via the Hurwitz transformations. However, even in 4 and 8 dimensions only the symmetric cases $(2,2)$ and $(4,4)$ were studied \cite{mad9, mar7}. We show that quadratic algebra structures exist for all members of the family and can be used to obtain the energy spectrum. Another main objective of this chapter is to extend the analysis presented in \cite{fh1} to more complicated algebraic structures. \par
%
%
The chapter is organized as follow. In Section 2, we present the integrals of motion of the new family of superintegrable systems. In Section 3, we obtain the quadratic algebra and present the Poisson analog and their Casimir operator. We also highlight the structure of higher rank quadratic algebra and the decomposition. In Section 4, we generalize the realizations as deformed oscillator algebra, construct the Fock space and obtain the finite-dimensional unitary representations (unirreps). We also using this analysis provide an algebraic derivation of the energy spectrum. In Section 5, we use the method of separation of variables in double hyperspherical coordinates and compare with the results obtain algebraically. In the closing section 6, we present some discussion with few remarks on the physical and mathematical relevance of these algebras. 

\section{New family of superintegrable system}
Let us consider a family of $N$-dimensional superintegrable Hamiltonians involving singular terms with any two partitions of the coordinates $(n, N-n)$
\begin{eqnarray}
H=\frac{p^2}{2}+\frac{\omega^2 r^2}{2}+\frac{c_1}{x^2_1+...+x^2_n}+\frac{c_2}{x^2_{n+1}+...+x^2_N},\label{bhamil}
\end{eqnarray}
where $ \vec{r}=(x_{1},x_{2},...,x_{N})$, $\vec{p}=(p_{1},p_{2},...,p_{N})$, $r^{2}=\sum_{i=1}^{N}x_{i}^{2}$, $p_{i}=-i \hbar \partial_{i}$ and $c_1$, $c_2$ are positive real constants. This family of systems represents the sum of two singular oscillators of dimensions $n$ and $N-n$. It is a generalization of the 4D and 8D systems obtained via the Hurwitz transformation for specific case (2,2) in 4D \cite{pet1,mar7} and (4,4) in 8D \cite{mar10}. In fact model (\ref{bhamil}) not only contains cases (2,2) and (4,4), but also cases (1,3) and (1,7), (2,6), (3,5) for 4D and 8D respectively. An algebraic derivation of the energy spectrum has previously been obtained only for the two symmetric cases (2,2) and (4,4). 

The system (\ref{bhamil}) has the following integrals of motion
\begin{eqnarray}
&&A=-\frac{h^2}{4}\left\{\sum^N_{i,j=1}x^2_i\partial^2_{x_j}-\sum^N_{i,j=1}x_i x_j \partial_{x_i}\partial_{x_j}-(N-1)\sum^N_{i=1}x_i \partial_{x_i}\right\}\nonumber\\&&\qquad +\frac{1}{2}\sum^N_{i=1}x^2_i\left\{\frac{c_1}{x^2_1+...+x^2_n}+\frac{c_2}{x^2_{n+1}+...+x^2_N}\right\},\label{bkpA}
\\&&
B=\frac{1}{2}\left\{\sum^n_{i=1}p^2_i-\sum^N_{i=n+1}p^2_i\right\} + \frac{\omega^2}{2}\left\{\sum^n_{i=1}x^2_i-\sum^N_{i=n+1}x^2_i\right\}\nonumber\\&&\qquad +\frac{c_1}{x^2_1+...+x^2_n}-\frac{c_2}{x^2_{n+1}+...+x^2_N},\label{bkpB}
\end{eqnarray}
which can be verified to fulfill the commutation relation
\begin{eqnarray*}
[H,A]=0=[H,B].
\end{eqnarray*}
The first order integrals of motion are given by 
\begin{eqnarray}
&&J_{ij}=x_i p_j-x_j p_i,\qquad i, j= 1,2,....,n,
\\&&K_{ij}=x_i p_j-x_j p_i,\qquad i, j= n+1,....,N.
\end{eqnarray}
The integrals of motion $A$ and $B$ are associated with the separation of variables in double hyper-Eulerian and double hyperspherical coordinates respectively.
Let
\begin{eqnarray}
J_{(2)}=\sum_{i<j}J^2_{ij}, \qquad  K_{(2)}=\sum_{i<j}K^2_{ij}\label{bkpJ}.
\end{eqnarray}
$J_{(2)}$ and $K_{(2)}$ represent the second order Casimir operators and fulfill the commutation relations
\begin{eqnarray}
&&[H,J_{(2)}]=0=[H,K_{(2)}],\quad [A,J_{(2)}] =0=[B,J_{(2)}]\nonumber\\&&[B,K_{(2)}]=0=[A,K_{(2)}], \quad [J_{(2)},K_{(2)}]=0.
\end{eqnarray}
These commutation relations can be conveniently represented by the following diagrams
\begin{equation}
\begin{xy}
(10,0)*+{J_{(2)}}="f"; (50,0)*+{K_{(2)}}="k"; (0,30)*+{A}="a"; (60,30)*+{B}="b"; (30,60)*+{H}="h";  
"f";"k"**\dir{--}; 
"h";"b"**\dir{--};
"h";"f"**\dir{--};
"h";"a"**\dir{--};
"h";"k"**\dir{--};
"h";"f"**\dir{--};
"a";"f"**\dir{--}; 
"a";"k"**\dir{--};
"b";"f"**\dir{--};
"b";"k"**\dir{--};
\end{xy}
\end{equation}

\begin{equation}
\begin{xy}
(0,0)*+{K_{(2)}}="m"; (40,0)*+{K_{ij}}="n"; (20,20)*+{H}="r";   (60,0)*+{J_{(2)}}="j"; (100,0)*+{J_{ij}}="l"; (80,20)*+{H}="g"; 
"j";"l"**\dir{--}; 
"l";"g"**\dir{--};
"g";"j"**\dir{--};
"m";"n"**\dir{--}; 
"n";"r"**\dir{--};
"r";"m"**\dir{--};
\end{xy}
\end{equation}
The first diagram shows that $J_{(2)}$ and $K_{(2)}$ are central elements. The second and third diagrams show that $J_{(2)}$ and $K_{(2)}$ are the Casimir operators of $so(n)$ and $so(N-n)$ Lie algebras realized by angular momentum $J_{ij}$, $ i, j= 1,2,....,n$ and $K_{ij}$, $i, j= n+1,....,N$ respectively. The models are minimally superintegrable and the total number of algebraically independant integrals is $N+1$. The construction of the minimally superintegrable systems is interesting, as research has so far mainly focused on maximally superintegrable systems. In the next section, we construct the quadratic algebra, its Casimir operator and finite dimensional unitary representations which give the energy spectrum of the superintegrable systems.

\section{Quadratic algebra structure}

We present in this section quadratic algebra structure $Q(3)$ of the superintegrable Hamiltonian systems (\ref{bhamil}). We show how the $su(N)$ symmetry algebra of isotropic harmonic oscillator is broken to $Q(3)\oplus so(n)\oplus so(N-n).$ 

\subsection{The isotropic harmonic oscillator and $su(N)$}
In this subsection we review some facts about isotropic harmonic oscillator and $su(N)$. The Hamiltonian (\ref{bhamil}) in the limit $c_{1}=0$ and $c_{2}=0$ reduces to the isotropic harmonic oscillator in $N$-dimension is given by 
\begin{eqnarray}
H=\frac{p^2}{2}+\frac{\omega^2 r^2}{2}.\label{bIohamil}
\end{eqnarray}
In this limiting case, we define the following operators \cite{bud1} by 
\begin{eqnarray}
&&a_i=\frac{1}{\sqrt{2}}(p_i+i\omega x_i),\quad a^+_i=\frac{1}{\sqrt{2}}(p_i-i\omega x_i),
\end{eqnarray}
which satisfy the commutation relation $[a_i,a^+_j]=-\omega\hbar\delta_{ij}$. Furthermore, we can define the operators 
\begin{eqnarray}
F^j_i=\frac{1}{\sqrt{2\omega\hbar}}\{a_i, a^+_j\}
\end{eqnarray}
are hermitian and satisfy the commutation relations
\begin{eqnarray}
[F^i_j, F^k_l]=\delta_{il} F^k_j-\delta_{jk} F^i_l.
\end{eqnarray}
Hence $su(N)$ is the symmetry algebra of the isotropic harmonic oscillator.
We present some key briefly key elements related to an algebraic derivation of the energy spectrum. The $su(N)$ symmetry algebra can be embedded in the non-compact algebra $sp(N)$, define the operators
\begin{eqnarray}
F^{ij}_0=\frac{-1}{\sqrt{2\omega\hbar}}\{a^+_i, a^+_j\}, \quad F_{ij}^0=\frac{1}{\sqrt{2\omega\hbar}}\{a_i, a_j\}
\end{eqnarray}
and satisfying the commutators
\begin{eqnarray}
[F^i_j, F^{lk}_0]=\delta_{jk} F^{il}_0+\delta_{jl} F^{ik}_0,
\quad
[F^i_j, F_{lk}^0]=-\delta_{ik} F_{jl}^0-\delta_{il} F_{jk}^0.
\end{eqnarray}
One can construct the Casimir operator of $sp(N)$ in the form
\begin{eqnarray}
Q_2=B_2+\frac{H^2}{2}+\frac{1}{\sqrt{2\omega\hbar}}\{F_{ij}^0,F^{ij}_0\},
\end{eqnarray}
where $B_2$ is the Casimir operator of $su(N)$.
The eigenvalues of $Q_2$ and $B_2$ in the representations of $sp(N)$ and $su(N)$ are as 
\begin{eqnarray}
Q_2=-\frac{N}{2}(N+\frac{1}{2}), \quad B_2=-\frac{l(l+N)(N-1)}{N},
\end{eqnarray}
where all representations belonging to the $H$-eigenstates are obtained for $l=1,2,\dots ,$
and the energy spectrum
\begin{eqnarray}
E=\omega \hbar (l+\frac{N}{2}),\quad l=0,1,2,\dots
\end{eqnarray}

\subsection{Quadratic Poisson algebra }
We study in this subsection the Poisson algebra of the classical version of the superintegrable system (\ref{bhamil}) and its Casimir operators. The system has the second order integrals of motion, 
\begin{eqnarray}
A&=&\frac{1}{4}\left\{\sum^N_{i,j=1}x^2_ip^2_j-\sum^N_{i,j=1}x_i x_j p_i p_j\right\}\nonumber\\&& +\frac{1}{2}\sum^N_{i=1}x^2_i\left\{\frac{c_1}{x^2_1+...+x^2_n}+\frac{c_2}{x^2_{n+1}+...+x^2_N}\right\},
\end{eqnarray}
$B$ and $J_{(2)}$, $K_{(2)}$ are given by (\ref{bkpB}) and (\ref{bkpJ}) respectively. 
It can be verified they satisfy 
\begin{eqnarray}
\{H,A\}_{p}=0=\{H,B\}_p,\quad \{H,J_{(2)}\}_p=0=\{H,K_{(2)}\}_p,
\end{eqnarray}
where $\{,\}_{p}$ is the usual Poisson bracket. Also we can check these following Poisson brackets
\begin{eqnarray*}
\{A,J_{(2)}\}_p=\{A,K_{(2)}\}_p=\{B,J_{(2)}\}_p=0=\{B,K_{(2)}\}_p=\{J_{(2)},K_{(2)}\}_p.
\end{eqnarray*}
The above commutation relations show that $J_{(2)}$ and $K_{(2)}$ are second order Casimir operators and central elements. We now construct a new integral of motion from $A$ and $B$ as 
\begin{eqnarray}
\{A,B\}_p=C.
\end{eqnarray} 
The integral of motion $C$ is cubic function of momenta. After a direct but involving computation relying on properties of the Poisson bracket and identities, we can show that the integrals of motion generate the quadratic Poisson algebra $QP(3)$, 
\begin{eqnarray}
&&\{A,B\}_p=C,
\\&&\{A,C\}_p= -4AB+ J_{(2)}H- K_{(2)}H+2(c_1-c_2)H,
\\&&\{B,C\}_p=2 B^2-2 H^2+16\omega^2 A-4\omega^2 J_{(2)}-4\omega^2 K_{(2)}-8\omega^2 (c_1+c_2).
\end{eqnarray}
The Casimir operator of this quadratic Poisson algebra can be shown to be cubic and explicitly given by 
\begin{eqnarray}
K&=&C^2+4AB^2-2[J_{(2)}H-K_{(2)}H+2(c_1-c_2)H]B+16\omega^2 A^2\nonumber\\&&-2[8\omega^2 (c_1+c_2)+4\omega^2 J_{(2)}+4\omega^2 K_{(2)}+2H^2]A.
\end{eqnarray}
By means of explicit expressions as functions of the coordinates and the momenta for the generators $A$, $B$, $C$ and the central elements, the Casimir operator becomes
\begin{eqnarray*}
K_1&=&-2 J_{(2)}H^2-2 K_{(2)}H^2-4(c_1+c_2)H^2-\omega^2 J^2_{(2)}-\omega^2 K^2_{(2)}+2\omega^2 J_{(2)} K_{(2)}\\&&-4\omega^2(c_1-c_2)J_{(2)}+4\omega^2(c_1-c_2)K_{(2)}-4\omega^2(c_1-c_2)^2.
\end{eqnarray*}
The quadratic Poisson algebra and the Casimir operator are useful in deriving the quadratic algebra and Casimir operator: the lowest order terms of $\hbar$ in quantum case coincide with Poisson analog. The first order integrals of motion generate a Poisson algebra isomorphic to $so(n)$ Lie algebra
\begin{eqnarray*}
&\{J_{ij},J_{kl}\}_{p}&= \delta_{ik}J_{jl}+ \delta_{jl}J_{ik}-\delta_{il}J_{jk}-\delta_{jk}J_{il}, 
\end{eqnarray*}
for $i, j, k, l=1, ..., n$ and $so(N-n)$ Lie algebra
\begin{eqnarray}
&\{K_{ij},K_{kl}\}_{p}&= \delta_{ik}K_{jl}+ \delta_{jl}K_{ik}-\delta_{il}K_{jk}-\delta_{jk}K_{il},
\end{eqnarray}
for $i, j, k, l=n+1, ..., N-n$.
So the full symmetry algebra is a direct sum of the quadratic Poisson algebra $QP(3)$, $so(n)$ and $so(N-n)$ Lie algebras.

\subsection{Quadratic algebra }
We now construct integral of motion $C$ of the quantum system from (\ref{bkpA}) and (\ref{bkpB}) via commutator
\begin{eqnarray}
[A,B]=C, \label{bkpC}
\end{eqnarray}
where $C$ represents a new integral of motion and is a cubic function of momenta. The cubicness of $C$ explains the impossibility of expressing $C$ as a polynomial function of other integrals of motion, which are quadratic function of momenta. After an involving but direct computations that is based on properties of commutators and various identities, we obtain the following quadratic algebra $Q(3)$ of the integrals of motion $H$, $A$ and $B$ 
\begin{eqnarray}
&&[A,C]= 2\hbar^2 \{A,B\}-\hbar^2 J_{(2)}H+\hbar^2 K_{(2)}H-\frac{\hbar^2}{4} \{8c_1-8c_2\nonumber\\&&\qquad\qquad -(N-4)(N-2n)\hbar^2\}H+\frac{\hbar^4}{4}N(N-4)B,\label{bkpAC}
\\&&
[B,C]=-2\hbar^2 B^2+2\hbar^2 H^2-16\hbar^2\omega^2A+4\hbar^2\omega^2 J_{(2)}+4\hbar^2\omega^2 K_{(2)}\nonumber\\&&\qquad\qquad +8\hbar^2\omega^2\{c_1+c_2-\frac{\hbar^2}{4}n(N-n)\}.\label{bkpBC}
\end{eqnarray}
It can be demonstrated, the Casimir operator is a cubic expression of the generators and is explicitly given in terms of the generators ($A,B$ and $C$) as
\begin{eqnarray}
K&=&C^2-2\hbar^2\{A,B^2\}+\frac{\hbar^4}{4}\{16-N(N-4)\}B^2+2\hbar^2\left[J_{(2)}H-K_{(2)}H\right.\nonumber\\&&\left.+\frac{1}{4}\{8c_1-8c_2-(N-4)(N-2n)\hbar^2\}H\right]B-16\hbar^2\omega^2 A^2\nonumber\\&&+2\hbar^2\left[8\omega^2\{c_1+c_2-\frac{\hbar^2}{4}n(N-n)\}+4\omega^2 J_{(2)}+4\omega^2 K_{(2)}+2H^2\right]A.\label{bkpK}
\end{eqnarray}
Using the realization of the integrals of motion $A$, $B$, $C$ and the central element as differential operators, we can represent the Casimir operator (\ref{bkpK}) only in terms of the central elements $H$, $J_{(2)}$ and $K_{(2)}$, 
\begin{eqnarray}
K&=&2\hbar^2 J_{(2)}H^2+2\hbar^2 K_{(2)}H^2+\frac{\hbar^2}{4}\left[16c_1+16c_2-\{4(N-4)\right.\nonumber\\&&\left.-(N-2n)^2\}\hbar^2\right]H^2+\hbar^2\omega^2 J^2_{(2)}+\hbar^2\omega^2 K^2_{(2)}-2\hbar^2\omega^2 J_{(2)} K_{(2)}\nonumber\\&&+4\hbar^2\omega^2\{c_1-c_2-\frac{1}{4}(N-4)(N-n)\hbar^2\}J_{(2)}-4\hbar^2\omega^2\{c_1-c_2\nonumber\\&&+\frac{1}{4}n(N-4)\hbar^2\}K_{(2)}+4\hbar^2\omega^2\left[(c_1-c_2)^2-\frac{1}{2}(N-n)(N-4)\hbar^2 c_1\right.\nonumber\\&&\left.-\frac{1}{2}n(N-4)\hbar^2 c_2+\frac{1}{4}n(N-n)(N-4)\hbar^4\right].\label{bkpK1}
\end{eqnarray}
This is key step in the application of the deformed oscillator algebra approach which relies on both form of the Casimir operators. The first order integrals of motion, that are simply components of angular momentum, generate an algebra isomorphic to the $so(n)$ Lie algebra
\begin{eqnarray*}
&[J_{ij},J_{kl}]&= i(\delta_{ik}J_{jl}+ \delta_{jl}J_{ik}-\delta_{il}J_{jk}-\delta_{jk}J_{il})\hbar, 
\end{eqnarray*}
for $i, j, k, l=1, ..., n$ and  $so(N-n)$ Lie algebra
\begin{eqnarray*}
&[K_{ij},K_{kl}]&=i( \delta_{ik}K_{jl}+ \delta_{jl}K_{ik}-\delta_{il}K_{jk}-\delta_{jk}K_{il})\hbar,
\end{eqnarray*}
for $i, j, k, l=n+1, ..., N-n$. So the full symmetry algebra is a direct sum of the quadratic algebra $Q(3)$, $so(n)$ and $so(N-n)$ Lie algebras. Thus the $su(N)$ Lie algebra generated by the integrals of motion of the $N$-dimensional isotropic harmonic oscillators is deformed into higher rank quadratic algebra $Q(3) \oplus so(n) \oplus so(N-n)$ for the family of superintegrable systems (\ref{bhamil}). The structure constants of the quadratic algebra involve three central elements, the Hamiltonian and the two Casimir operators of the Lie algebras approached in the decomposition. Quadratic algebras involving three generators have been obtained and studied by various authors \cite{gra1, gen1, gen2, das2}. In fact, the one involved in the decomposition would be related to case $QR(3)$, called quadratic Racah algebra.

\section{Energy spectrum and unirreps}

We now consider the realizations of the quadratic algebra ((\ref{bkpC})-(\ref{bkpBC})) in terms of deformed oscillator algebra \cite{das2, das1} $\{\aleph, b^{\dagger}, b\}$ defined by
\begin{eqnarray}
[\aleph,b^{\dagger}]=b^{\dagger},\quad [\aleph,b]=-b,\quad bb^{\dagger}=\Phi (\aleph+1),\quad b^{\dagger} b=\Phi(\aleph),
\end{eqnarray}
where $\aleph $ is the number operator and the function $\Phi(x)$ is well behaved real function satisfying the boundary condition
\begin{eqnarray}
\Phi(0)=0, \quad \Phi(x)>0, \quad \forall x>0.\label{bkpbc}
\end{eqnarray}
The $\Phi(x)$ is the so-called structure function. The realization of $Q(3)$ is of the form $A=A(\aleph)$, $B=b(\aleph)+b^{\dagger} \rho(\aleph)+\rho (\aleph)b$, where $A(x)$, $b(x)$ and $\rho(x)$ are functions. Similar to the quadratic algebra for 2D superintegrable systems \cite{das2}, we have
\begin{eqnarray}
&&A(\aleph)=\hbar^2\{(\aleph+u)^2-\frac{(N-2)^2}{16}\},
\\&&
b(\aleph)=\frac{8c_1-8c_2+4J_{(2)}-4K_{(2)}+(4N-8n+2nN-N^2)\hbar^2}{16\hbar^2\{(\aleph+u)^2-\frac{1}{4}\}},
\\&&
\rho(\aleph)=\frac{1}{3.2^{20}.\hbar^{16}(\aleph+u)(1+\aleph+u)\{1+2(\aleph+u)\}^2},
\end{eqnarray}
where $u$ is a constant ( in fact a representation dependent constant ) to be determined from the constraints on the structure function. We construct the structure function $\Phi(x)$ by using the realization, the quadratic algebra ((\ref{bkpC}), (\ref{bkpAC}), (\ref{bkpBC})) and the two forms of the Casimir operator (\ref{bkpK}) and (\ref{bkpK1})
\begin{eqnarray}
\Phi(x;u,H)&=&12288 \hbar^{12} \left[64 c_1^2 + 64 c_2^2 - 48 \hbar^4 - 32 \hbar^2 J_{(2)} + 16 J_{(2)}^2 -32\hbar^2 K_{(2)}\nonumber \right.\\&&\left.- 32 J_{(2)} K_{(2)} + 16 K_{(2)}^2 - 64 \hbar^2 J_{(2)} n + 64 \hbar^2 K_{(2)} n + 48 \hbar^4 n^2 \nonumber \right.\\&&\left.+ 32 \hbar^4 N + 32 \hbar^2 J_{(2)} N- 32 \hbar^2 K_{(2)} N - 48 \hbar^4 n N + 16 \hbar^2 J_{(2)} n N\nonumber \right.\\&&\left. - 16 \hbar^2 K_{(2)} n N - 32 \hbar^4 n^2 N + 8 \hbar^4 N^2 - 8 \hbar^2 J_{(2)} N^2 + 8 \hbar^2 K_2 N^2 \nonumber\right.\\&&\left.+ 32 \hbar^4 n N^2 + 4 \hbar^4 n^2 N^2 - 8 \hbar^4 N^3 - 4 \hbar^4 n N^3 + \hbar^4 N^4 \nonumber\right.\\&&\left.-16 c_2 [4 (J_{(2)} - K_{(2)}) + \hbar^2 \{(N-4)(2n-N)+ 4 (1 - 2(x+u))^2\}]\nonumber \right.\\&&\left.- 16 c_1 [8 c_2 - 4 J_{(2)} + 4 K_{(2)} + \hbar^2\{(N-4)(N-2n) \nonumber \right.\\&&\left.+ 4 (1 - 2(x+u))^2\}]+ 32 \hbar^2 [4 (J_{(2)} + K_{(2)}) + \hbar^2 \{2 n^2 + (N-2)^2 \nonumber \right.\\&&\left.- 2 n N\}] (x+u) - 32 \hbar^2 [4 (J_{(2)}+ K_{(2)}) + \hbar^2 \{2 (n^2-2)\nonumber \right.\\&&\left.- 2 (n+2) N + N^2\}] (x+u)^2 - 512 \hbar^4 (x+u)^3 + 256 \hbar^4 (x+u)^4\right]\nonumber \\&&\times [H^2 - h^2 \{1 - 2 (x+u)\}^2 \omega^2].\label{bkpST}
\end{eqnarray}
We will show how the finite dimensional unirreps can be obtained using an appropriate Fock space. In 2D, we need $|n, E>$ such that $\aleph|n, E>=n|n, E>$. However, in our case one needs to define quantum numbers associated to certain subalgebra chain. We use two subalgebra chains, $so(n) \supset so(n-1) \supset ... \supset so(2)$  and $so(N-n) \supset so(N-n-1) \supset ... \supset so(2)$ related chains of quadratic Casimir operators \cite{ras1} $J_{2}^{(\alpha)}$ and $K_{2}^{(\alpha)}$ can be written as
\begin{eqnarray}
&&J_{2}^{(\alpha)}=\sum^{\alpha}J_{ij},\quad \alpha=2,...,n,
\\&&
K_{2}^{(\alpha)}=\sum^{\alpha}K_{ij},\quad \alpha=n+2,...,N.
\end{eqnarray}
In fact, $|n, E>$ means $|n,E,l_{N},...,l_{n+2},l_{n},...,l_{2}>$. Then the eigenvalues of $J_{(2)}$ and $K_{(2)}$ are $\hbar^2 l_{n}(l_{n}+n-2)$ and $\hbar^2 l_{N-n}(l_{N-n}+N-n-2)$ respectively. The constraint (\ref{bkpbc}) can be imposed on Fock type representation of the deformed oscillator algebra, $|n, E>$ with $\aleph|n, E>=n|n, E>$,  and $H$ in $\Phi(x, u, H)$ replaced by $E$.  Hence by using the eigenvalues of  $J_{(2)}$, $K_{(2)}$ and $H$, the structure function becomes the following factorized form:
\begin{eqnarray}
\Phi(x)&=&-12582912 \hbar^{18}\omega^2 [x+u-\frac{1}{4}(2+m_1+m_2)] [x+u-\frac{1}{4}(2-m_1+m_2)]\nonumber\\&&\times[x+u-\frac{1}{4}(2+m_1-m_2)] [x+u-\frac{1}{4}(2-m_1-m_2)]\nonumber\\&&\times(x+u -\frac{-H +\hbar \omega}{2\hbar \omega})(x - \frac{H + \hbar \omega}{2 \hbar \omega}),\label{bpro3}
\end{eqnarray}
where
$\hbar^2 m_1^2 = 8 c_1  + 4 J_{(2)} +\hbar^2 (n-2)^2$ and 
$\hbar^2 m_2^2 = 8 c_2  + 4 K_{(2)} + \hbar^2(N-n-2)^2$.
For unirreps to be finite dimensional, we impose the following constrains on the structure function:
\begin{equation}
\Phi(p+1; u, E)=0;\quad \Phi(0;u,E)=0;\quad \Phi(x)>0,\quad \forall x>0,\label{bpro2}
\end{equation}
where $p$ is a positive integer and $u$ is arbitrary constant. We then obtain finite $(p+1)$-dimensional unirreps. The solution of the constraints (\ref{bpro2}) gives the energy $E$ and constant $u$. Thus we obtain the following possible structure functions and energy spectra, for $\epsilon_1=\pm 1$, $\epsilon_2= \pm 1$, $\eta =24576 \hbar^{18}\omega^2$.
\\Set-1:
\begin{eqnarray*}
u = \frac{-E + \hbar \omega}{2\hbar \omega},\qquad E = 2\hbar\omega(p+1+\frac{\epsilon_1 m_1 +\epsilon_2  m_2 }{4} ),
\end{eqnarray*}
\begin{eqnarray*} 
\Phi(x) &=& \eta x \hbar^{18}\omega^2 [4+4p-4x-(1-\epsilon_1) m_1 +(1+\epsilon_2)m_2]\\&&\times[4+4p-4x+(1+\epsilon_1) m_1 -(1-\epsilon_2)m_2][4+4p-4x\nonumber\\&&-(1-\epsilon_1) m_1 -(1-\epsilon_2)m_2][4+4p-4x+(1+\epsilon_1) m_1 \nonumber\\&&+(1+\epsilon_2)m_2][4+4p-2x+ \epsilon_1 m_1 +\epsilon_2 m_2].
\end{eqnarray*}
\\Set-2:
\begin{eqnarray*}
u = \frac{E + \hbar \omega}{2\hbar \omega},\qquad E = 2\hbar\omega(p+1+\frac{\epsilon_1 m_1 +\epsilon_2  m_2 }{4} ),
\end{eqnarray*}
\begin{eqnarray*} 
\Phi(x) &=& - x [4+4p+4x-(1-\epsilon_1) m_1 +(1+\epsilon_2)m_2]\\&&\times[4+4p+4x+(1+\epsilon_1) m_1 -(1-\epsilon_2)m_2][4+4p+4x\nonumber\\&&-(1-\epsilon_1) m_1 -(1-\epsilon_2)m_2][4+4p+4x+(1+\epsilon_1) m_1 \nonumber\\&&+(1+\epsilon_2)m_2][4+4p+2x+ \epsilon_1 m_1 +\epsilon_2 m_2].
\end{eqnarray*}
\\Set-3:
\begin{eqnarray*}
u = \frac{1}{4}(2+\epsilon_1 m_1+\epsilon_2 m_2),\qquad E = 2\hbar\omega(p+1+\frac{\epsilon_1 m_1 +\epsilon_2  m_2 }{4} ),
\end{eqnarray*}
\begin{eqnarray*} 
\Phi(x)& =& \eta (p+1-x)[4x-(1-\epsilon_1) m_1 -(1-\epsilon_2)m_2]\\&&\times[4x+(1+\epsilon_1) m_1 -(1-\epsilon_2)m_2][4x-(1-\epsilon_1) m_1 +(1+\epsilon_2)m_2]\\&&\times[4x+(1+\epsilon_1) m_1 +(1+\epsilon_2)m_2][2+2p+2x+ \epsilon_1 m_1 +\epsilon_2 m_2].
\end{eqnarray*}
The structure functions $\Phi(x)>0$ for $\varepsilon_{1}=1$, $\varepsilon_{2}=1$ and $m_{1}, m_{2}>0$. In the limit $c_1=0$ and $c_2=0$, this results coincide with $N$-dimensional harmonic oscillator with the following relation among the quantum numbers $l=2p+l_n+l_{N-n}$ and the algebraic derivation using the $su(N)$ and $sp(N)$ Lie algebra and their Casimir operators and eigenvalues. Let us mention that the value of $\eta$ do not play a role, only the sign needs to be taken into account for the constraint to obtain the finite- dimensional unirreps.

\section{Separation of variables }  
Each member of the family of superintegrable Hamiltonian systems (\ref{bhamil}) is multiseparable and allows for the Schordinger equation the separation of variables in double hyper Eulerian and double hyperspherical coordinates. We can rewrite (\ref{bhamil}) into the sum of two singular oscillators of dimensions $n$ and $N-n$ as 
\begin{eqnarray}
H=H_1+H_2,
\end{eqnarray}
where
 \begin{eqnarray}
 &&H_1=\frac{1}{2}(p^2_1+...+p^2_n)+\frac{\omega^2 r^2_1}{2}+\frac{c_1}{r^2_1},
 \\&&
 H_2=\frac{1}{2}(p^2_{n+1}+...+p^2_N)+\frac{\omega^2 r^2_2}{2}+\frac{c_2}{r^2_2}
 \end{eqnarray}
 and the position vectors $r^2_1=x^2_1+...+x^2_n, \quad r^2_2=x^2_{n+1}+...+x^2_N.$
The Schrodinger equation $H\psi(r,\Omega)=E\psi(r, \Omega)$ can also be written as 
\begin{eqnarray}
 H_1\psi(r_1,\Omega)=E_1\psi(r_1, \Omega), \qquad H_2\psi(r_2,\Omega)=E_2\psi(r_2, \Omega),
\end{eqnarray} 
where $E_1$ and $E_2$ are the eigenvalues of $H_1$ and $H_2$ respectively.
The $\mathcal{N}$-dimensional hyperspherical coordinates are given by 
\begin{eqnarray}
&x_{1}&=r \sin(\Phi_{\mathcal{N}-1})\sin(\Phi_{\mathcal{N}-2})\cdots \sin(\Phi_{1}),
\nonumber\\& x_{2}&=r \sin(\Phi_{\mathcal{N}-1})\sin(\Phi_{\mathcal{N}-2})\cdots \cos(\Phi_{1}),
\nonumber\\&...&
\nonumber\\&...&
\nonumber\\&  x_{\mathcal{N}-1}&=r \sin(\Phi_{\mathcal{N}-1})\cos(\Phi_{\mathcal{N}-2}),
\nonumber\\&  x_{\mathcal{N}}&=r \cos(\Phi_{\mathcal{N}-1}),
\end{eqnarray}
where the $\mathcal{N}$ $x_{i}$'s are Cartesian coordinates in the hyperspherical coordinates, $\{\Phi_{1},\dots, \Phi_{\mathcal{N}-1}\}$ are the hyperspherical angles and $r$ is the hyperradius. We can also introduce a double type of hyperspherical coordinates system by considering two copies and setting $\mathcal{N}=n$ and $\mathcal{N}=N-n$ respectively for the $H_{1}$ component and the $H_{2}$ component. The Schrodinger equation of $H_1$ in $n$-dimensional hyperspherical coordinates 
\begin{eqnarray}
\left[\frac{\partial^2}{\partial r^2_1}+\frac{n-1}{r_1}\frac{\partial}{\partial r_1}-\frac{1}{r^2_1}\Lambda^2(n)-\omega'^2 r^2_1-\frac{2c'_1}{r^2_1}+2E'_1\right]\psi(r_1,\Omega)=0,\label{bkpH1}
\end{eqnarray}
where $c'_1=\frac{c_1}{\hbar^2}, \quad\omega'=\frac{\omega}{\hbar}, \quad E'_1=\frac{E_1}{\hbar^2}$ and the grand angular momentum operator $\Lambda^2(n)$ which satisfies the recursive formula
\begin{eqnarray}
-\Lambda^2(n)=\frac{\partial^2}{\partial\Phi^2_{n-1}}-(n-2) \cot(\Phi_{n-1})\frac{\partial}{\partial\Phi_{n-1}}-\frac{\Lambda^2(n-1)}{\sin^2(\Phi_{n-1})}, 
\end{eqnarray}
valid for $n>0$ and $\Lambda^2(1)=0$. The separation of radial and angular parts  of (\ref{bkpH1}) can be perfomed by setting $\psi(r_1,\Omega)=R_1(r_1)y(\Omega_{n-1})$. Thus, we obtain
\begin{eqnarray}
\frac{\partial^2 R_1(r_1)}{\partial r^2_{1}}+\frac{n-1}{r_1} \frac{\partial R_1(r_1)} {\partial r_1}+\{2E'_1-\omega'^2 r^2_1-\frac{2c'_1}{r^2_1}-\frac{l_{n}(l_{n}+n-2)}{r^2_1}\}R_1(r_1)=0,\label{bKF1}
\end{eqnarray}
\begin{eqnarray}
\Lambda^2(n)y(\Omega_{n-1})=l_{n}(l_{n}+n-2)y(\Omega_{n-1}),
\end{eqnarray}
where $l_{n}(l_{n}+n-2)$ being the general form of the separation constant.
Now the radial equation (\ref{bKF1}) can be converted, by setting $u=a r^2_1$, $R_1(u)=u^\alpha f(u)$ and $f(u)=e^{\beta u}f_1(u)$, to
\begin{eqnarray}
u f''_1(u)+\{2(\delta_1+\frac{1}{2}l_{n}+\frac{n}{4})-u\}f'_1(u)+\{\frac{E'_1}{2\omega'}-(\delta_1+\frac{1}{2}l_{n}+\frac{n}{4})\}f_1(u)=0,\label{bkp17}
\end{eqnarray}
where
\begin{eqnarray}
\delta_1&=&\left\{\sqrt{(\frac{1}{2}l_{n}+\frac{n-2}{4})^2+\frac{1}{2}c'_1}-\frac{n-2}{4}\right\}-\frac{1}{2}l_{n}.
\end{eqnarray}
Set 
\begin{eqnarray}
N_1=\frac{E'_1}{2\omega'}-\left(\delta_1+\frac{1}{2}l_{n}+\frac{n}{4}\right).\label{bkpN1}
\end{eqnarray}
Then the equation (\ref{bkp17}) represents the confluent hypergeometric equation and it gives the solution in terms of special function  
\begin{eqnarray}
\psi_{N_1 l_n}(u)&=&\sqrt{\frac{2\Gamma\{N_1+2(\delta_1+\frac{1}{2}l_{n}+\frac{n}{4})\}}{N_1!}}\frac{a e^{-\frac{u}{2}} u^{(\delta_1+\frac{1}{2}l_{n})/2}}{\sqrt{2(\delta_1+\frac{1}{2}l_{n}+\frac{n}{4})}}\nonumber\\&& \times{}_1F_1\left(-N_1; 2\{\delta_1+\frac{1}{2}l_{n}+\frac{n}{4}\}; u\right)\label{bkpWf}
\end{eqnarray}
In order the wavefunctions to be square integrable, the parameter $N_1$ needs to be positive. Hence we obtain the discrete energy $E_1$ of $H_1$ from (\ref{bkpN1}) as
\begin{eqnarray}
E_1=2\hbar \omega\left(N_1+\frac{\alpha_1}{2}+\frac{1}{2}\right),\label{bEKP1}
\end{eqnarray}
where $\alpha_1=2\delta_1+l_{n}+\frac{n-2}{2}$. The wave equation of $H_2$ in (N-n)-dimensional hyperspherical coordinates provides similar solution $\psi_{N_2 l_{N-n}}$, replacing $r_1$, $n$, $l_n$, $c'_1$, $\delta_1$  by $r_2$, $N-n$, $l_{N-n}$, $c'_2$, $\delta_2$ respectively in (\ref{bkpWf}) and the energy $E_2$, replacing $N_1$, $\alpha_1$ by $N_2$, $\alpha_2$ respectively in (\ref{bEKP1}). Hence the energy spectrum of the $N$-dimensional double singular oscillators
\begin{eqnarray}
E=2\hbar \omega\left(p+1+\frac{\alpha_1+\alpha_2}{2}\right),
\end{eqnarray}
where $p=N_1+N_2$ which coincides with the energy expression obtained algebraically. In fact, this multiseparability properties is also shared by its classical analog as the Hamilton-Jacobi can also be separated in the double type coordinates systems.

\section{Conclusion}

In this chapter, we have extended the symmetric double singular oscillators in 4D and 8D  to arbitrary dimensions with any partition $(n, N-n)$ of the coordinates. This provide a new family of minimally superintegrable systems. As main results, we construct and obtain its realization as deformed oscillator algebra. We also construct the finite dimensional unitary representations which enable the algebraic derivation of the energy spectrum. This is compared with the results from the separation of variables method. Moreover, the new family may also include duals of deformed higher dimensional Kepler-Coulomb systems involving nonAbelian monopoles \cite{le2, men1}. 

Our systems are generalisations of one of the four 2D Smorodinsky-Winternitz models \cite{fri2}. Further generalisations are possible. For example,
\begin{eqnarray}
&H&=\frac{p^2}{2}+\frac{\omega^2 r^2}{2}+\frac{c_1}{x^2_1+\dots +x^2_{n_1}}\nonumber\\&&+\frac{c_2}{x^2_{{n_1}+1}+\dots +x^2_{n_2}}+\dots +\frac{c_\lambda}{x^2_{n_{\lambda-1}+1}+\dots +x^2_{n_\lambda}} 
\end{eqnarray}
generalizes the $N$-dimensional Smorodinsky-Winternitz model \cite{eva2} to the one with any partition ($n_1, n_2,\dots,n_\lambda$) such that $n_1+n_2+\dots +n_\lambda=N$. This model would also be superintegrable but with a more complicated quadratic algebra and embedded structure seen in \cite{das3, tan1}.


\chapter{Family of $N$-dimensional superintegrable systems}\label{ch5}

{\bf \large{Acknowledgement}}\\
This chapter is based on the talk which I have presented in highly significant mathematical physics conference \textit{'The XXIII International Conference on Integrable Systems and Quantum Symmetries (ISQS-23), Prague, Czech Republic'}, that was published in Ref.\cite{fh7}. I have also presented partially of this talk in \textit{'The Australian and New Zealand Association of Mathematical Physics (ANZAMP)'} conference at Newcastle, Australia. I overview the models (\ref{ahamil}) and (\ref{bhamil}), and represent their common algebraic features.

\section{Introduction}
Algebraic methods are powerful tools in modern physics. Well known examples include the $N$-dimensional hydrogen atom and harmonic oscillator which were studied using the $so(N+1)$ \cite{barg1, lou2} and $su(N)$ \cite{lou1, hwa1} Lie algebras respectively. In particular the spectrum of the 5D hydrogen atom have been calculated using its $so(6)$ Lie algebra and the corresponding Casimir operators of order two, three and four \cite{tru1}. Superintegrable models are an important class of quantum systems which can be solved using algebraic approaches. An important property of such systems is the existence of non-Abelian symmetry algebras generated by integrals of motion. These symmetry algebras can be embedded in certain non-invariance algebras involving non-commuting operators. Such symmetry algebras are in general finitely generated polynomial algebras and only exceptionally finite dimensional Lie algebras.

Quadratic algebras have been used to obtain energy spectrum of superintegrable systems \cite{gra1}. The structure of a class of quadratic algebras with only three generators was studied and applied to 2D superintegrable systems in \cite{das2}. Many researchers have studied quadratic and polynomial symmetry algebras of superintegrable systems and their representation theory ( see e.g. \cite{kal4, kal5, tan1, mil1, isa1, gen1, gen2, mar14}).  

We review the results of our recent two papers \cite{fh1, fh2} and present the algebraic derivation of the complete energy spectrum of the $N$-dimensional superintegrable Kepler-Coulomb system with non-central terms and superintegrable double singular oscillators.

\section{The main results}
Let us consider the $N$-dimensional superintegrable Kepler-Coulomb system with non-central terms and superintegrable Hamiltonian with double singular oscillators of type $(n, N-n)$ introduced in \cite{fh1, fh2} 
\begin{eqnarray}
&&H_{KC}=\frac{1}{2}p^{2}-\frac{c_{0}}{r}+c_1\chi_1(r,x_N)+c_2\chi_2(r,x_N), \label{hamil1}
\\&&
H_{dso}=\frac{p^2}{2}+\frac{\omega^2 r^2}{2}+c_1\phi_1(x_1,\dots,x_n)+c_2\phi_2(x_{n+1},\dots,x_{N}),\label{hamil2}
\end{eqnarray}
where $ \vec{r}=(x_{1},x_{2},...,x_{N})$, $\vec{p}=(p_{1},p_{2},...,p_{N})$, $r^{2}=\sum_{i=1}^{N}x_{i}^{2}$, $p_{i}=-i \hbar \partial_{i}$, $\chi_1(r,x_N)=\frac{1}{r(r+x_N)}$, $\chi_2(r,x_N)=\frac{1}{r(r-x_N)}$, $\phi_1(x_1,\dots,x_n)=\frac{1}{x_1^2+\dots+x^2_n}$, $\phi_2(x_{n+1},\dots,x_{N})=\frac{1}{x^2_{n+1}+\dots+x^2_{N-n}}$ and $c_0$, $c_1$, $c_2$ are positive real constants. The system (\ref{hamil1}) is a generalization of the 3D superintegrable system in $E_3$ \cite{kib1}. It includes as a particular 3D case the Hartmann potential which has applications in quantum chemistry and its classical analog possesses closed trajectories and periodic motion \cite{tru1}. The model (\ref{hamil2}) is the generalization of the 4D and 8D systems obtained via the Hurwitz transformation for which only the symmetry cases (2,2) and (4,4) were studied \cite{hur1, mar7, mar10}.

The model (\ref{hamil1}) is multiseparable and allows separation of variables in hyperspherical and hyperparabolic coordinates. The wave function is 
\begin{equation}
\psi(r,\Omega)=R(r)\Theta(\Phi_{N-1})y(\Omega_{N-2})
\end{equation}
in terms of special functions
\begin{eqnarray}
&&R(r)\propto(\varepsilon r)^{l+\frac{\delta_{1}+\delta_{2}}{2}} e^{\frac{-\varepsilon r}{2}}{}_1 F_1(-n+l+1, 2l+\delta_{1}+\delta_{2}+N-1; \varepsilon r),
\\&&
\Theta(z)\propto(1+z)^{\frac{(\delta_{1}+I_{N-2})}{2}}(1-z)^{\frac{(\delta_{2}+I_{N-2})}{2}} P^{(\delta_{2}+I_{N-2}, \delta_{1}+I_{N-2})}_{l-I_{N-2}}(z),
\\&&
\Lambda^2(N-1)y(\Omega_{N-2})=I_{N-2}(I_{N-2}+N-3)y(\Omega_{N-2}),\label{fk3}
\end{eqnarray}
where $z=\cos(\Phi_{N-1})$, $I_{N-2}(I_{N-2}+N-3)$ being the general form of the separation constant,
$P^{(\alpha, \beta)}_{\lambda}$ denotes a Jacobi polynomial and $l\in{\rm I\!N}$, $\delta_{i}=\{\sqrt{(I_{N-2}+\frac{N-3}{2})^2+4c'_{i}}-\frac{N-3}{2}\}-I_{N-2},c'_i=\frac{c_i}{\hbar^2}, i=1, 2$. The energy spectrum of the system (\ref{hamil1}) is \cite{fh1}
\begin{eqnarray}
E_{KC}\equiv E_n=\frac{-c^{2}_{0}}{2\hbar^{2}\left( n+\frac{\delta_{1}+\delta_{2}}{2}+\frac{ N-3}{2}\right)^{2}},\quad n=1,2,3,\dots\label{en2}
\end{eqnarray}
The system (\ref{hamil2}) is also multiseparable and allows separation of variables in double hyper Eulerian and double hyperspherical coordinates. We can split (\ref{hamil2}) into the sum of two singular oscillators of dimensions $n$ and $N-n$ as $H_{dso}=H_1+H_2$, where 
\begin{eqnarray}
 &&H_1=\frac{1}{2}(p^2_1+...+p^2_n)+\frac{\omega^2 r^2_1}{2}+c_1\phi_1(x_1,\dots,x_n),
 \\&&
 H_2=\frac{1}{2}(p^2_{n+1}+...+p^2_N)+\frac{\omega^2 r^2_2}{2}+c_2\phi_2(x_{n+1},\dots,x_{N}).
 \end{eqnarray}
 The wave function $\psi(r_1,\Omega)=R_1(r_1)y(\Omega_{n-1})$ of $H_1$ in terms of special functions is given by
 \begin{eqnarray}
\psi_{n_1 l_n}(u)\propto\frac{a e^{-\frac{u}{2}} u^{(\delta_1+\frac{1}{2}l_{n})/2}}{\sqrt{2(\delta_1+\frac{1}{2}l_{n}+\frac{n}{4})}} \times{}_1F_1\left(-n_1; 2\{\delta_1+\frac{1}{2}l_{n}+\frac{n}{4}\}; u\right),\label{kpWf}
\end{eqnarray}
where 
$\delta_1=\left\{\sqrt{(\frac{1}{2}l_{n}+\frac{n-2}{4})^2+\frac{1}{2}c'_1}-\frac{n-2}{4}\right\}-\frac{1}{2}l_{n}$, $n_1=\frac{E'_1}{2\omega'}-\left(\delta_1+\frac{1}{2}l_{n}+\frac{n}{4}\right)$, $c'_1=\frac{c_1}{\hbar^2}$, $\omega'=\frac{\omega}{\hbar}$ and $E'_1=\frac{E_1}{\hbar^2}$. The wave equation of $H_2$ in $(N-n)$-dimensional hyperspherical coordinates has similar solution. The energy spectrum of the system (\ref{hamil2}) is \cite{fh2} 
\begin{eqnarray}
E_{dso}=2\hbar \omega\left(p+1+\frac{\alpha_1+\alpha_2}{2}\right)\label{Edso},
\end{eqnarray}
where $\alpha_1=2\delta_1+l_{n}+\frac{n-2}{2}$, $\alpha_2=2\delta_2+l_{N-n}+\frac{N-n-2}{2}$ and $p=n_1+n_2$.

The integrals of motion of (\ref{hamil1}) are given by
\begin{eqnarray}
&A&=\sum_{i<j}^{N}L_{ij}^{2}+ 2r^2[c_1\chi_1(r,x_N)+c_2\chi_2(r,x_N)],\label{kf5}
\\&B&=-M_{N}+c_{1}(r-x_{N})\chi_1(r,x_N)-c_{2}(r+x_{N})\chi_2(r,x_N),\label{kf6}
\\&J^{2}&=\sum_{i<j}^{N-1}L_{ij}^{2},\quad L_{i j}=x_{i}p_{j}-x_{j}p_{i}, \quad  i, j=1, ... , N-1,
\end{eqnarray}
and for the model (\ref{hamil2}) they are
\begin{eqnarray}
&A&=-\frac{h^2}{4}\left\{\sum^N_{i,j=1}x^2_i\partial^2_{x_j}-\sum^N_{i,j=1}x_i x_j \partial_{x_i}\partial_{x_j}-(N-1)\sum^N_{i=1}x_i \partial_{x_i}\right\}\nonumber\\&& +\frac{c_1\phi_1(x_1,\dots,x_n)+c_2\phi_2(x_{n+1},\dots,x_{N})}{2}\left[\frac{1}{\phi_1(x_1,\dots,x_n)}+\frac{1}{\phi_2(x_{n+1},\dots,x_{N})}\right],\label{kpA}
\\
&B&=\frac{1}{2}\left\{\sum^n_{i=1}p^2_i-\sum^N_{i=n+1}p^2_i\right\} + \frac{\omega^2}{2}\left\{\frac{1}{\phi_1(x_1,\dots,x_n)}-\frac{1}{\phi_2(x_{n+1},\dots,x_{N})}\right\}\nonumber\\&&\quad+c_1\phi_1(x_1,\dots,x_n)-c_2\phi_2(x_{n+1},\dots,x_{N}),\label{kpB}
\\
&J_{(2)}&=\sum_{i<j}J^2_{ij},\quad J_{ij}=x_i p_j-x_j p_i, i, j= 1,2,....,n, 
\\
&K_{(2)}&=\sum_{i<j}K^2_{ij},\quad K_{ij}=x_i p_j-x_j p_i, i, j= n+1,....,N,\label{kpJ}
\end{eqnarray}
where the Runge-Lenz vector
$M_{j}=\frac{1}{2} \sum_{i=1}^{N} ( L_{ji}p_{i}-p_{i}L_{ij} )- \frac{c_{0}x_{j}}{r}$.
The integrals of motion \{$A,B,C$\} close to the general form of the quadratic algebra $Q(3)$ \cite{das2},
\begin{eqnarray}
&[A,B]&=C,\label{prova5}
\\&[A,C]&=\alpha A^2+\gamma \{A,B\}+\delta A+\epsilon B+\zeta,\label{prova6}
\\&[B,C]&=aA^2-\gamma B^{2}-\alpha\{A,B\} +d A -\delta B + z\label{prova1}
\end{eqnarray}
and the Casimir operator
\begin{eqnarray}
&K&=C^{2}-\alpha\{A^2,B\}-\gamma \{A,B^{2}\}+(\alpha \gamma-\delta)\{A,B\}+ (\gamma^2-\epsilon) B^{2} \nonumber\\&&+ (\gamma\delta-2\zeta) B+\frac{2a}{3}A^3 +(d+\frac{a\gamma}{3}+\alpha^2) A^{2} +(\frac{a\epsilon}{3}+\alpha\delta+2z)A.\label{prova4} 
\end{eqnarray}
Here coefficients $\alpha, \gamma,\delta, \epsilon, \zeta, a, d, z$ are given in the following table for models $H_{KC}$ and $H_{dso}$ \cite{fh1, fh2}.
\begin{table}[h] 
\begin{center}
    \begin{tabular}{ | l | p{6cm} | p{6cm} |}
    \hline
    &${\bf H_{KC}}$ & ${\bf H_{dso}}$ 
    \\ \hline
     $\alpha$ & 0 & 0 
    \\ \hline
   $\gamma$ & $2 \hbar^{2}$ & $2 \hbar^{2}$ 
    \\ \hline
     $\delta$ & 0 &  0
     \\ \hline
     $\epsilon$ &$(N-1)(N-3) \hbar^{4}$& $\frac{\hbar^4}{4}N(N-4)$
         \\ \hline
     $\zeta$ & $-4 (c_{1}-c_{2}) \hbar^{2} c_{0}$ &  $-\hbar^2 J_{(2)}H+\hbar^2 K_{(2)}H -\frac{\hbar^2}{4}\{8c_1-8c_2-(N-4)(N-2n)\hbar^2\}H$
         \\ \hline
     $a$ & 0 &  0
         \\ \hline
     $d$ & $8 \hbar^{2} H$ & $-16\hbar^2\omega^2$
         \\ \hline
     $z$ & $-4 \hbar^{2} J^{2} H + (N-1)^{2} \hbar^{4} H-8 \hbar^{2}(c_{1}+c_{2})H + 2 \hbar^{2} c_{0}^{2} $ &  $2\hbar^2 H^2+4\hbar^2\omega^2 J_{(2)}+4\hbar^2\omega^2 K_{(2)}+8\hbar^2\omega^2\{c_1+c_2-\frac{\hbar^2}{4}n(N-n)\}$
    \\
    \hline
    \end{tabular}
    \caption{\label{tabone} Coefficients in $Q(3)$ and $K$.}
\end{center}
\end{table}
\\
By means of the explicit expressions of $A, B, C$, we can write the Casimir operator in terms only of central elements as
\begin{eqnarray}
&K_{KC}&=2(N-3)(N-1)\hbar^{4} H J^{2} -8 \hbar^{2}(c_{1}-c_{2})^{2} H + 4  (N-3) (N-1) (c_{1}+c_{2}) \hbar^{4} H 
 \nonumber\\&&\quad-\hbar^{6}(N-3)(N-1)^{2} H +4 \hbar^{2} c_{0}^{2} J^{2} +8 \hbar^{2} (c_{1}+c_{2})c_{0}^{2} -2 (N-3) \hbar^{4} c_{0}^{2},\label{prova3} 
\end{eqnarray}
\begin{eqnarray}
&K_{dso}&=2\hbar^2 J_{(2)}H^2+2\hbar^2 K_{(2)}H^2+\frac{\hbar^2}{4}\left[16c_1+16c_2-\{4(N-4)-(N-2n)^2\}\hbar^2\right]H^2\nonumber\\&&+\hbar^2\omega^2 J^2_{(2)}+\hbar^2\omega^2 K^2_{(2)}-2\hbar^2\omega^2 J_{(2)} K_{(2)}+4\hbar^2\omega^2\{c_1-c_2-\frac{1}{4}(N-4)(N-n)\hbar^2\}J_{(2)}\nonumber\\&&-4\hbar^2\omega^2\{c_1-c_2+\frac{1}{4}n(N-4)\hbar^2\}K_{(2)}+4\hbar^2\omega^2\left[(c_1-c_2)^2-\frac{1}{2}(N-n)(N-4)\hbar^2 c_1\right.\nonumber\\&&\left.-\frac{1}{2}n(N-4)\hbar^2 c_2+\frac{1}{4}n(N-n)(N-4)\hbar^4\right].\label{kpK1}
\end{eqnarray}
The first order integrals $L_{ij}$, $i, j, k, l=1, .., N-1$ of $H_{KC}$ generate $so(N-1)$ Lie algebra 
\begin{eqnarray}
[L_{ij},L_{kl}]=i ( \delta_{ik}L_{jl}+ \delta_{jl}L_{ik}-\delta_{il}L_{jk}-\delta_{jk}L_{il})\hbar
\end{eqnarray}
and
$J_{ij}=x_i p_j-x_j p_i$, $i, j= 1,2,....,n$ and 
$K_{ij}=x_i p_j-x_j p_i$, $i, j= n+1,....,N$ of $H_{dso}$ generate $so(n)$ and $so(N-n)$ Lie algebras
\begin{eqnarray}
&[J_{ij},J_{kl}]&= i(\delta_{ik}J_{jl}+ \delta_{jl}J_{ik}-\delta_{il}J_{jk}-\delta_{jk}J_{il})\hbar, 
\\&[K_{ij},K_{kl}]&=i( \delta_{ik}K_{jl}+ \delta_{jl}K_{ik}-\delta_{il}K_{jk}-\delta_{jk}K_{il})\hbar.
\end{eqnarray}
Thus the full symmetry algebra is $Q(3)\oplus so(N-1)$ for $H_{KC}$ and $Q(3)\oplus so(n)\oplus so(N-n)$ for $H_{dso}$. Chains of second order Casimir operators associated with $so(N-1)$, $so(n)$ and $so(N-n)$ components may be used to label quantum states for  $H_{KC}$ and $H_{dso}$ respectively. 

These quadratic algebras can be realized in terms of the deformed oscillator algebra \cite{das1, das2} \begin{eqnarray}
[\aleph,b^{\dagger}]=b^{\dagger},\quad [\aleph,b]=-b,\quad bb^{\dagger}=\Phi (\aleph+1),\quad b^{\dagger} b=\Phi(\aleph),
\end{eqnarray}
where $\aleph$ is the number operator and $\Phi(x)$ is a well behaved real function satisfying $\Phi(x)>0$ for all $x>0$. Then the structure function has the form \cite{fh1, fh2} 
\begin{eqnarray}
\Phi(x; u,E)=\nu_0\prod^6_{i=1} [x+u-\nu_i],\label{St}
\end{eqnarray}
where $\nu_i$ are given in the following table.
\\
\begin{table}[h] 
\begin{center}
    \begin{tabular}{ | l | l | l |}
    \hline
    &${\bf H_{KC}}$ & ${\bf H_{dso}}$ 
    \\ \hline
     $\nu_0$ & 6291456$E\hbar^{18}$ & -12582912$\hbar^{18}\omega^2$
    \\ \hline
   $\nu_1$ & $\frac{1}{2}(1+m_{1}+m_{2})$ & $\frac{1}{4}(2+m'_{1}+m'_{2})$ 
    \\ \hline
     $\nu_2$ & $\frac{1}{2}(1+m_{1}-m_{2})$ &  $\frac{1}{4}(2+m'_{1}-m'_{2})$
     \\ \hline
     $\nu_3$ &$\frac{1}{2}(1-m_{1}+m_{2})$& $\frac{1}{4}(2-m'_{1}+m'_{2})$
         \\ \hline
     $\nu_4$ &$\frac{1}{2}(1-m_{1}-m_{2})$ &  $\frac{1}{4}(2-m'_{1}-m'_{2})$
         \\ \hline
     $\nu_5$ & $\frac{1}{2}+\frac{c_0}{\hbar\sqrt{-2E}}$ &  $\frac{\hbar\omega+H}{2\hbar\omega}$
         \\ \hline
     $\nu_6$ & $\frac{1}{2}-\frac{c_0}{\hbar\sqrt{-2E}}$ & $\frac{\hbar\omega-H}{2\hbar\omega}$
            \\
    \hline
    \end{tabular}
 \caption{\label{tabone} Coefficients $\nu_i$ in (\ref{St}). }
\end{center}
\end{table}
\\
In the above table, $\hbar^2 m_{1,2}^{2} =16 c_{1,2}+\{4J^2 +(N-3)^{2}\}\hbar^2$, $\hbar^2 m'^2_1 =8 c_1+4J_{(2)} +(n-2)^2\hbar^2$ and $\hbar^2 m'^2_2 =8 c_2+4K_{(2)} +(N-n-2)^2\hbar^2$.

For a finite-dimensional unitary representation, we have the following constraints
\begin{equation}
\Phi(p+1; u, E)=0;\quad \Phi(0;u,E)=0;\quad \Phi(x)>0,\quad 0<x<p+1\label{pro2}
\end{equation}
on the structure function (\ref{St}). The solutions of these constraints provide the energy spectra $E_{KC,dso}$ and the arbitrary constant $u_{KC, dso}$ ($\epsilon_1=\pm 1, \epsilon_2=\pm 1$) as 
\begin{eqnarray}
&&E_{KC}=\frac{-2 c^2_{0}}{h^2 (2 + 2 p +\epsilon_{1} m_{1} +\epsilon_{2} m_{2} )^2},\quad u_{KC}=\frac{1}{2}+\frac{c_{0}}{\hbar\sqrt{-2E}},
\\
Case\quad 1:&& E_{dso}=2\hbar\omega(p+1+\frac{\epsilon_{1} m'_{1} +\epsilon_{2} m'_{2}}{4}),\qquad u_{dso}=\frac{ E+\hbar\omega}{2\hbar\omega},
\\
Case\quad 2:&& E_{dso}=2\hbar\omega(p+1+\frac{\epsilon_{1} m'_{1} +\epsilon_{2} m'_{2}}{4}),\qquad u_{dso}=\frac{- E+\hbar\omega}{2\hbar\omega},
\\
Case\quad 3:&& E_{dso}=2\hbar\omega(p+1+\frac{\epsilon_{1} m'_{1} +\epsilon_{2} m'_{2}}{4}),\qquad u_{dso}=\frac{1}{4}(2+\epsilon_1 m_1+\epsilon_2 m_2).
\end{eqnarray}
We have four possible structure functions for $H_{KC}$ \cite{fh1} and twenty four for $H_{dso}$ \cite{fh2}. 
Making the identification $n_1+n_2+I_{N-2}=n-1$, $p=n_{1}+n_{2}$, $m_{i}=\frac{1}{2}(3-2I_{N-2}-N-2\delta_{i}), m'_i=2\alpha_i, i=1, 2$, the energy spectra $E_{KC}$ and $E_{dso}$ coincide (\ref{en2}) and (\ref{Edso}) respectively. 

\section{Conclusion}
We have presented some of the results in \cite{fh1, fh2} and shown how the $su(N)$ and $so(N+1)$ symmetry algebras of the $N$-dimensional Kepler-Coulomb (i.e., $c_1=c_2=0$) and the $N$-dimensional harmonic oscillator (i.e., $c_1=c_2=0$) are broken to higher rank polynomial algebra of the form $Q(3)\oplus L_1\oplus L_2\oplus\dots$ for non-zero $c_1$ and $c_2$, where $L_1, L_2,\dots$, are certain Lie algebras. $Q(3)$ is a quadratic algebra involving Casimir operators of the Lie algebras in its structure constants.  We have also presented the realizations of these quadratic algebras in terms of deformed oscillator algebras and obtained the finite dimensional unitary representations which yield the energy spectra of these superintegrable systems. In addition, we have compared them with the physical spectra obtained from the separation of variables.

The systems (\ref{hamil1}) and (\ref{hamil2}) could be studied using new approaches such as the recurrence method related to special functions and orthogonal polynomials and approaches combining ladder, shift, intertwining and supercharges operators \cite{mar7, kal7}. The generalizations of these systems to include monopole interactions and their duals \cite{mar10, mad6} could be also investigated.


\chapter{Constructive approach on superintegrable systems}\label{ch6}

{\bf \large{Acknowledgement}}
\\This chapter is based on the work that was published in  Ref. \cite{fh3}. I have incorporated text of that paper \cite{fh3}. In this chapter, we apply the recurrence approach and coupling constant metamorphosis to construct higher order integrals of motion for the St\"{a}ckel equivalents of the $N$-dimensional superintegrable Kepler-Coulomb model with non-central terms and double singular oscillators of type ($n, N-n$). We show how the integrals of motion generate higher rank cubic algebra $C(3)\oplus L_1\oplus L_2$ with structure constants involving Casimir operators of the Lie algebras $L_1$ and $L_2$.  The realizations of the cubic algebras in terms of deformed oscillators enable us to construct finite dimensional unitary representations and derive the degenerate energy spectra of the corresponding superintegrable systems.

\section{Introduction}
A systematic approach for obtaining algebraic derivations of spectra of two dimensional superintegrable systems with quadratic and cubic algebras involving three generators was introduced in \cite{das2, mar1, mar2}. This algebraic method is based on Casimir operators and realizations in terms of deformed oscillator algebras \cite{das1} for obtaining finite dimensional unitary representation (unirreps) \cite{mar1, mar12}. This approach was extended to classes of higher order polynomial algebras with three generators of arbitrary order \cite{isa1}. However, superintegrable systems in higher dimensional spaces are usually associated with symmetry algebras taking the form of higher rank polynomial algebras which have typically a quite complicated embedded structure. It was discovered recently that there are systems which are related to classes of quadratic algebras that display a decomposition as direct sums of Lie algebras and polynomial algebras of three generators only \cite{fh1,fh2}. The structure constants contain Casimir operators of certain Lie algebras \cite{fh1,fh2}. This specific structure was exploited to obtain  algebraic derivations of the spectra. However, in general it is quite complicated to apply this direct approach to obtain the corresponding polynomial algebras, Casimir operators and their realizations in terms of deformed oscillators. In fact, it is not even guaranteed that the integrals close to some polynomial algebra.

The difficulties of this direct approach can be overcome using a constructive approach and in particular using ladder operators in order to build integrals of motion for models allowing separation of variables in Cartesian coordinates. Such ideas have been used by several authors in the case of first or second order ladder operators \cite{jau1, fri2, boy1, eva2, eva3, mar6}. It facilitates the construction of the corresponding polynomial algebras. One has to distinguish the case of constructing integrals using higher order (i.e., greater than 2) ladder operators as such operators themselves need to be generated using various methods. one of them consists in exploiting supersymmetric quantum mechanics (SUSYQM) \cite{jun1} and using combinations of ladders and supercharges \cite{dem1, mar3, rag1, mar13} or combining only supercharges \cite{mar6, kre1, adl1, mar9, mar14}.

In recent years, Kalnins et al\cite{kal7} introduced a recurrence approach and applied it to models separable in polar coordinates. They also pointed out the close relation between such approach and the study of special functions and orthogonal polynomials.  Calzada et al\cite{cal2, cal3} introduced an operator version of the recurrence relations. In this scheme an intermediate set of non-polynomial integrals of motion was obtained. From these formal algebraic relations, and by decomposing integrals into polynomial and non-polynomial parts, a final well defined set of integrals and their corresponding polynomial algebras can be obtained. It has also been demonstrated how SUSYQM can be combined with these ideas to generate extended Lissajous models related to Jacobi exceptional orthogonal polynomials \cite{mar15}. 

All these new approaches have previously been restricted mainly to two and three dimensional superintegrable systems. The purpose of this chapter is to extend the recurrence approach for a class of higher-dimensional systems and construct higher rank polynomial algebras. We consider the following two models \cite{fh1, fh2}
\begin{eqnarray}
&&{\bf Model\quad 1}:\quad H_{dso}=\frac{p^2}{2}+\frac{\omega^2 r^2}{2}+\frac{c_1}{x^2_1+...+x^2_n}+\frac{c_2}{x^2_{n+1}+...+x^2_N},\label{chamil1}
\\&&
{\bf Model\quad 2}:\quad H_{KC}=\frac{1}{2}p^{2}-\frac{c_{0}}{r}+\frac{c_{1}}{r(r+x_{N})}+\frac{c_{2}}{r(r-x_{N})}, \label{chamil2}
\end{eqnarray}
where $ \vec{r}=(x_{1},x_{2},...,x_{N})$, $\vec{p}=(p_{1},p_{2},...,p_{N})$, $r^{2}=\sum_{i=1}^{N}x_{i}^{2}$, $p_{i}=-i \hbar \partial_{i}$ and $c_0$, $c_1$, $c_2$ are positive real constants. The model (\ref{chamil1}) is a family of $N$-dimensional superintegrable double singular oscillators and model (\ref{chamil2}) is $N$-dimensional superintegrable Kepler-Coulomb system with non-central terms in $N$-dimensional Euclidean space. In fact, both models appear to be minimally superintegrable with $N+1$ algebraically independent integrals of motion \cite{fh1, fh2}. In two recent papers, using a direct approach and ansatz to construct the integrals and the quadratic algebra, an algebraic derivation of the energy spectra have been presented \cite{fh1, fh2}. In this chapter, we will show using coupling constant metamorphosis \cite{kal6} that a constructive approach can be developed for these $N$-dimensional models and be used to simplify the calculation and analysis. 

The plan of the chapter is as follows. In section 2, higher order integrals of motion are constructed from ladder operators using separated eigenfunctions of the system (\ref{chamil1}) and the corresponding higher rank cubic algebra is presented. In section 3, higher order integrals and higher rank cubic algebras are constructed for the St\"{a}ckel equivalent system of (\ref{chamil2}) from coupling constant metamorphosis and ladder operators. The realizations in terms of deformed oscillators are obtained and applied to compute energy spectra of the two systems algebraically. Finally, in section 4, we provide some discussions on the results of this chapter and some open problems.

\section{Recurrence approach to $H_{dso}$}
In this section, we develop for the  model (\ref{chamil1}) the recurrence relations in order to generate higher order integrals of motion and the corresponding higher rank polynomial algebra. We show how in fact the integrals close into a cubic algebra with structure constants involving Casimir operators of certain Lie algebras and derive energy spectrum of the system algebraically. This provides also a poof for the superintegrability of this $N$-dimensional system.

\subsection{Separation of variables}
We recall \cite{fh2} in this subsection the separable solutions of (\ref{chamil1}) in double hyperspherical coordinates using the sum of two singular oscillators $H_{dso}=H_1+H_2$ of dimensions $n$ and $N-n$ respectively, where
\begin{eqnarray}
 &&H_1=\frac{1}{2}(p^2_1+...+p^2_n)+\frac{\omega^2}{2}r^2_1+\frac{c_1}{r^2_1},
 \\&&
 H_2=\frac{1}{2}(p^2_{n+1}+...+p^2_N)+\frac{\omega^2}{2}r^2_2+\frac{c_2}{r^2_2}.
 \end{eqnarray}
The Schrodinger equation of $H_1$ in $n$-dimensional hyperspherical coordinates is 
\begin{eqnarray}
-\frac{1}{2}\left[\frac{\partial^2}{\partial r^2_1}+\frac{n-1}{r_1}\frac{\partial}{\partial r_1}-\frac{1}{r^2_1}\Lambda^2(n)-\omega'^2 r^2_1-\frac{2c'_1}{r^2_1}\right]\psi_1(r_1,\Omega_{n-1})=E'_1\psi_1(r_1,\Omega_{n-1}),\label{ckpH1}
\end{eqnarray}
where $c'_1=\frac{c_1}{\hbar^2}$, $\omega'=\frac{\omega}{\hbar}$,  $E'_1=\frac{E_1}{\hbar^2}$ and $\Lambda^2(n)$ is the grand angular momentum operator. The wave function $\psi_1(r_1,\Omega_{n-1})$ is proportional to
\begin{eqnarray}
e^{-\frac{\omega' r^2_1}{2}}r_1^{\alpha_1+\frac{n}{2}}L^{\alpha_1}_{n_1}(\omega' r^2_1)y_1(\Omega_{n-1}),\label{ckpWf}
\end{eqnarray}
where $L^{\alpha}_n(x)$ is the $n$th order Laguerre polynomial \cite{and1}, 
$\alpha_1=2\delta_1+l_{n}+\frac{n-2}{2}$, 
$\delta_1=\left\{\sqrt{(\frac{1}{2}l_{n}+\frac{n-2}{4})^2+\frac{1}{2}c'_1}-\frac{n-2}{4}\right\}-\frac{1}{2}l_{n}$ and $n_1=\frac{E'_1}{2\omega'}-\left(\delta_1+\frac{1}{2}l_{n}+\frac{n}{4}\right)$. The wave function for $H_2$ has similar form. The energy spectrum of Hamiltonian (\ref{chamil1}) is
\begin{eqnarray}
E_{dso}=2\hbar \omega\left(p+1+\frac{\alpha_1+\alpha_2}{2}\right),
\end{eqnarray}
where the parameters $\alpha_2=2\delta_2+l_{N-n}+\frac{N-n-2}{2}$, $\delta_2=\left\{\sqrt{(\frac{1}{2}l_{N-n}+\frac{N-n-2}{4})^2+\frac{1}{2}c'_2}-\frac{N-n-2}{4}\right\}-\frac{1}{2}l_{N-n}$, $c'_2=\frac{c_2}{\hbar^2}$ and $p=n_1+n_2$, $n_2=\frac{E'_2}{2\omega'}-\left(\delta_2+\frac{1}{2}l_{N-n}+\frac{N-n}{4}\right)$, $E'_2=\frac{E_2}{\hbar^2}$.

\subsection{Recurrence formulas and algebra structure}
 
Consider the gauge transformations to $H_i$
 \begin{eqnarray}
 \tilde{H_i}=\mu^{-1}_i H_i\mu_i
 \end{eqnarray}
 with $\mu_1=r^{\frac{1-n}{2}}_1$, $\mu_2=r^{\frac{1-N+n}{2}}_2$.
We obtain the following gauge equivalent operators that possesses the same eigenvalues 
\begin{eqnarray}
\tilde{H_i}=-\frac{1}{2}\left\{\partial^2_{r_i}-\omega'^2 r^2_i-\frac{\beta_i}{r^2_i}\right\},\quad i=1,2.
\end{eqnarray}
Here $\beta_1=\frac{(n-1)(n-3)}{4}+2c'_1+\frac{J_{(2)}}{\hbar^2}$ and $\beta_2=\frac{(N-n-1)(N-n-3)}{4}+2c'_2+\frac{K_{(2)}}{\hbar^2}$ with
\begin{eqnarray}
&&J_{(2)}=\sum_{i<j}J^2_{ij},\quad J_{ij}=x_i p_j-x_j p_i, i, j= 1,2,....,n,
\\&&
 K_{(2)}=\sum_{i<j}K^2_{ij},\quad K_{ij}=x_i p_j-x_j p_i, i, j= n+1,....,N,
 \end{eqnarray}
 $J_{(2)}$ and $ K_{(2)}$ are related to $\Lambda^2(n)$ and $\Lambda^2(N-n)$ respectively.
 
  Let $Z=X_{n_1}X_{n_2}y_1(\Omega_{n-1})y_2(\Omega_{N-n-1})$ with
\begin{eqnarray}
&&X_{n_i}=e^{-\frac{\omega' r^2_i}{2}}r_i^{\alpha_i+\frac{1}{2}}L^{\alpha_i}_{n_i}(\omega' r^2_i), \quad i=1,2.
\end{eqnarray}
Then the wave functions of the gauge transformed $\tilde{H_i}$ are given simply by $Z=Z_1Z_2$, $Z_i= X_{n_i}y_i=\mu_i^{-1}\psi_i(r_i,\Omega), i=1, 2$. Alternatively, we can gauge transform $\tilde{H_i}$ back to get the initial Hamiltonian
\begin{eqnarray}
 H_{dso}=\mu_1\mu_2\tilde{H} \mu^{-1}_1\mu^{-1}_2,\quad \tilde{H}=\tilde{H_1}+\tilde{H_2}.
\end{eqnarray}
Thus we have the eigenvalues equations $\tilde{H_i} Z_i=\lambda_{r_i}Z_i=\tilde{E'_i} Z_i$ with
\begin{eqnarray}
\lambda_{r_i}=\omega'(2n_i+\alpha_i+1),\quad i=1,2.
\end{eqnarray} 
Hence the energy eigenvalues of $\tilde{H}Z=\tilde{E'}Z$ are given
\begin{eqnarray}
\tilde{E'}=\{2+2(n_1+n_2)+\alpha_1+\alpha_2\}\omega'.
\end{eqnarray}
Now we define the ladder operators
\begin{eqnarray}
\tilde{ D^\pm_i}(\omega', r_i)=-2\tilde{H_i}\mp 2\omega' r_i\partial_{r_i}+2\omega'^2 r^2_i\mp\omega',\quad i=1,2.
\end{eqnarray}
The action of the symmetry operators on the wave functions $Z_i=X_{n_i}y_i, i=1,2$ provides the following recurrence formulas 
\begin{eqnarray}
&&\tilde{D^+_i}(\omega', r_i)X_{n_i}y_i=-4\omega'(n_i+1)X_{n_i+1}y_i,\quad i=1,2,
\\&&
\tilde{D^-_i}(\omega', r_i)X_{n_i}y_i=-4\omega'(n_i+\alpha_i)X_{n_i-1}y_i,\quad i=1,2.
\end{eqnarray}
The following diagram indicates how $\tilde{D^\pm_i}$ change the quantum numbers of the wave functions.

\begin{tikzpicture}
\draw[dashed,color=gray] (-0.1,-0.1) grid (4.1,4.1);

    \draw[->] (0,0) -- (0,4.5) node[above] {$n_1$};
    \draw[ ->] (0,0) -- (4.5,0) node[right] {$\alpha_1$};
       
    \draw[-> ,blue] (2,2.01) -- (2,2.97)node[above left] {$\tilde{D^+_1}$};
    \draw[-> ,blue] (2,2) --(2,1.03) node[below left] {$\tilde{D^-_1}$};
    
    \draw[- ,blue] (1,2) -- (3,2);
        \node at (2,0)[below]{$\alpha_1$};
    \node at (0,2)[left]{$n_1$};
      
\end{tikzpicture}
\begin{tikzpicture}
\draw[dashed,color=gray] (-0.1,-0.1) grid (4.1,4.1);

    \draw[->] (0,0) -- (0,4.5) node[above] {$n_2$};
    \draw[ ->] (0,0) -- (4.5,0) node[right] {$\alpha_2$};
       
    \draw[-> ,blue] (2,2.01) -- (2,2.97)node[above left] {$\tilde{D^+_2}$};
    \draw[-> ,blue] (2,1.97) --(2,1.03) node[below left] {$\tilde{D^-_2}$};
    
    \draw[- ,blue] (1,2) -- (3,2);
        \node at (2,0)[below]{$\alpha_2$};
    \node at (0,2)[left]{$n_2$};
      
\end{tikzpicture}

Let us consider the suitable combination of the operators $\tilde{L_1}=\tilde{D^+_1}\tilde{ D^-_2}$, $\tilde{L_2}=\tilde{D^-_1}\tilde{ D^+_2}$, $\tilde{H}=\tilde{H_1}+\tilde{H_2}$ and $\tilde{B}=\tilde{H_1}-\tilde{H_2}$. The action of the operators on the wave functions are given by 
\begin{eqnarray}
&&\tilde{L_1}Z=16\omega'^2(n_1+1)(n_2+\alpha_2)X_{n_1+1}X_{n_2-1}y_1y_2,
\\&&
\tilde{L_2}Z=16\omega'^2(n_2+1)(n_1+\alpha_1)X_{n_1-1}X_{n_2+1}y_1y_2,
\\&&
\tilde{L_1}\tilde{L_2} Z=256\omega'^4 n_1(n_2+1)(n_1+\alpha_1)(n_2+\alpha_2+1)Z,
\\&&
\tilde{L_2}\tilde{L_1} Z=256\omega'^4 n_2(n_1+1)(n_2+\alpha_2)(n_1+\alpha_1+1)Z.
\end{eqnarray}
It follows that in operator they form the cubic algebra $C(3)$,
\begin{eqnarray}
&&[\tilde{L_1},\tilde{H}]=0= [\tilde{L_2},\tilde{H}],\label{ckpf1}
\\&&[\tilde{L_1},\tilde{B}]=-4\omega'\tilde{L_1} ,\qquad [\tilde{L_2},\tilde{B}]=4\omega'\tilde{L_2},\label{ckpf2}
\end{eqnarray} 
\begin{eqnarray}
&\tilde{L_1}\tilde{L_2}& =\left[(\tilde{B} +\tilde{H}-2\omega')^2-\frac{4\omega'^2}{\hbar^2}\left\{J_{(2)}+2c'_1\hbar^2+\frac{(n-2)^2}{4}\hbar^2\right\}\right] \nonumber\\&& \times\left[(\tilde{B} -\tilde{H}-2\omega')^2-\frac{4\omega'^2}{\hbar^2}\left\{K_{(2)}+2c'_2\hbar^2+\frac{(N-n-2)^2}{4}\hbar^2\right\}\right],\label{ckpf5}
\\&
\tilde{L_2}\tilde{L_1}&=\left[(\tilde{B} +\tilde{H}+2\omega')^2-\frac{4\omega'^2}{\hbar^2}\left\{J_{(2)}+2c'_1\hbar^2+\frac{(n-2)^2}{4}\hbar^2\right\}\right] \nonumber\\&& \times\left[(\tilde{B} -\tilde{H}+2\omega')^2-\frac{4\omega'^2}{\hbar^2}\left\{K_{(2)}+2c'_2\hbar^2+\frac{(N-n-2)^2}{4}\hbar^2\right\}\right].\label{ckpf6}
\end{eqnarray}

The first order integrals of motion $J_{ij}$ and $K_{ij}$ generate algebras isomorphic to the $so(n)$  and $so(N-n)$ Lie algebras respectively. Namely, 
\begin{eqnarray}
&[J_{ij},J_{kl}]&= i(\delta_{ik}J_{jl}+ \delta_{jl}J_{ik}-\delta_{il}J_{jk}-\delta_{jk}J_{il})\hbar, 
\end{eqnarray}
where $i, j, k, l=1, ..., n$ and 
\begin{eqnarray}
&[K_{ij},K_{kl}]&=i( \delta_{ik}K_{jl}+ \delta_{jl}K_{ik}-\delta_{il}K_{jk}-\delta_{jk}K_{il})\hbar, 
\end{eqnarray}
with $i, j, k, l=n+1, ..., N-n$. Moreover,
\begin{eqnarray}
&[J_{ij},\tilde{L_1}]=0=[J_{ij},\tilde{L_2}],\quad[K_{ij},\tilde{L_1}]=0=[K_{ij},\tilde{L_2}],&
\\&
[J_{ij},\tilde{H}]=0=[K_{ij},\tilde{H}],\quad [J_{ij},\tilde{B}]=0=[K_{ij},\tilde{B}].&
\end{eqnarray}
So the full symmetry algebra is a direct sum of the cubic algebra $C(3)$, $so(n)$ and $so(N-n)$ Lie algebras. Thus the $su(N)$ Lie algebra generated by the integrals of motion of the $N$-dimensional isotropic harmonic oscillators is deformed into higher rank cubic algebra $C(3) \oplus so(n) \oplus so(N-n)$ for Hamiltonian (\ref{chamil1}).

The recurrence method gives rise to higher order integrals of motion and higher rank polynomial algebra, more specifically 4th-order integrals and cubic algebra $C(3)\oplus so(n)\oplus so(N-n)$ involving Casimir operators of $so(n)$ and $so(N-n)$ Lie algebras. In our recent work \cite{fh2} we showed that the second order integrals of motion $A$, $B$ and $C$ generate the quadratic algebra $Q(3)\oplus so(n)\oplus so(N-n)$ with commutation relations $[A,B]=C$, $[A,C]=f_1(A, B, H, J_{(2)},  K_{(2)})$, $[B,C]=f_2(A, B, H, J_{(2)},  K_{(2)})$, where  $f_1$ and $f_2$ are quadratic polynomials in the generators $A$ and $B$ and the central elements $H$, $J_{(2)}$ and $K_{(2)}$. In order to derive the spectrum using the quadratic algebra $Q(3)$,  realizations of $Q(3)$ in terms of deformed oscillator algebra \cite{das2, das1} $\{\aleph, b^{\dagger}, b\}$ of the form
\begin{eqnarray}
[\aleph,b^{\dagger}]=b^{\dagger},\quad [\aleph,b]=-b,\quad bb^{\dagger}=\Phi (\aleph+1),\quad b^{\dagger} b=\Phi(\aleph),\label{ckpfh}
\end{eqnarray}
have been used. Where $\aleph $ is the number operator and $\Phi(x)$ is well behaved real function satisfying the constraints
\begin{eqnarray}
\Phi(0)=0, \quad \Phi(x)>0, \quad \forall x>0.\label{ckpbc}
\end{eqnarray}
It is non-trivial to obtain such a realization and find the structure function $\Phi(x)$. However in the recurrence approach presented in this paper, the cubic algebra $C(3)$ relations (\ref{ckpf1}) - (\ref{ckpf6}) already have, in fact, the form of deformed oscillator algebra (\ref{ckpfh}).

\subsection{Unirreps and energy spectrum}
We write the cubic algebra relations (\ref{ckpf1})-(\ref{ckpf6}) in the form of deformed oscillator (\ref{ckpfh}) by letting $\aleph=\frac{\tilde{B}}{4w'}$, $b^{\dagger}=\tilde{L_1}$ and $b=\tilde{L_2}$. We then readily obtain the structure function
\begin{eqnarray}
&\Phi(x,u,\tilde{H})&=\left[4\omega'(x+u)+\tilde{H}-2(1-\alpha_1)\omega'\right]\left[4\omega'(x+u)+\tilde{H}-2(1+\alpha_1)\omega'\right]\nonumber\\&&\times\left[4\omega'(x+u)-\tilde{H}-2(1-\alpha_2)\omega'\right]\left[4\omega'(x+u)-\tilde{H}-2(1+\alpha_2)\omega'\right],\label{cun1}
\end{eqnarray}
where $u$ is arbitrary constant. In order to obtain the $(p+1)$-dimensional unirreps, we should impose the following constraints on the structure function 
\begin{equation}
\Phi(p+1; u,\tilde{E'})=0,\quad \Phi(0;u,\tilde{E'})=0,\quad \Phi(x)>0,\quad \forall x>0,\label{cpro2}
\end{equation}
where $p$ is a positive integer. The solutions give the energy $\tilde{E'}$ and the arbitrary constant $u$. Let $\varepsilon_1=\pm 1$, $\varepsilon_2=\pm 1$. We have
\begin{eqnarray}
u=\frac{-\tilde{E'}+2\omega'(1+\varepsilon_1\alpha_1)}{4\omega'},\quad\tilde{E'}=(2+2p+\varepsilon_1\alpha_1+\varepsilon_2\alpha_2)\omega',
\end{eqnarray}
\begin{eqnarray}
\Phi(x) =256x\omega'^4(x+\varepsilon_1\alpha_1)(1+p-x)(1+p-x+\varepsilon_2\alpha_2).
\end{eqnarray}
The physical wave functions involve other quantum numbers and we have in fact degeneracy of $p+1$ only when these other quantum numbers would be fixed. The total number of degeneracies may be calculated by taking into account the further constraints on these quantum numbers. The results have been obtained in the gauge transformed Hamiltonian. However, the relations $D^{\pm}_i =\mu_i \tilde{D^{\pm}_1} \mu^{-1}_i, i=1,2$  provide the integrals of motion $L_1= D^{+}_1 D^{-}_2$ and $L_2= D^{-}_1 D^{+}_2$ of the initial Hamiltonian $H_{dso}$ and the algebraic derivation remain valid as gauge transformations preserve the spectrum.

\section{Recurrence approach to $H_{KC}$}
In this section, we construct recurrence relations to generate higher order integrals of motion and higher rank polynomial algebra from ladder operators and coupling constant metamorphosis for the Stackel equivalent of model (\ref{chamil2}).
 
\subsection{Saparation of variables}
The Schrodinger equation of the system (\ref{chamil2}) in the hyperparabolic coordinates reads 
\begin{eqnarray}
&H\psi(\xi, \eta, \Omega_{N-1})&=\left[-\frac{2}{\xi+\eta}\left[\Delta(\xi)+\Delta(\eta)-\frac{\xi+\eta}{4\xi\eta}\Lambda^2(\Omega_{N-1})\right]-\frac{2\beta_{0}}{\xi+\eta}\right.\nonumber\\&&\left.+\frac{2\beta_{1}}{\xi(\xi+\eta)}+\frac{2\beta_{2}}{\eta(\xi+\eta)}\right]\psi(\xi, \eta, \Omega_{N-1})=\varepsilon \psi(\xi, \eta, \Omega_{N-1}),\label{cfk1}
\end{eqnarray} 
where $\Lambda^2(N)$ is the grand angular momentum operator and 
\begin{eqnarray*}
&\Delta(\xi)& = \xi^{-\frac{N-3}{2}}\frac{\partial}{\partial \xi}  \xi^{\frac{N-1}{2}}\frac{\partial}{\partial \xi}, \quad \Delta(\eta) = \eta^{-\frac{N-3}{2}}\frac{\partial}{\partial \eta}  \eta^{\frac{N-1}{2}}\frac{\partial}{\partial \eta},\\& \beta_{0}&=\frac{c_{0}}{\hbar{^2}}, \quad \beta_{1}=\frac{c_{1}}{\hbar{^2}},\quad \beta_{2}=\frac{c_{2}}{\hbar{^2}},\quad \varepsilon=\frac{E}{\hbar^{2}}.
\end{eqnarray*}
We can write the equivalent system of (\ref{cfk1}) as 
\begin{eqnarray}
&H'\psi(\xi, \eta, \Omega_{N-1})&=\left[\Delta(\xi)+\Delta(\eta)-\frac{\beta_{1}}{\xi}-\frac{\beta_{2}}{\eta}+\frac{\omega'}{2}(\xi+\eta)-\frac{1}{4\xi}\Lambda^2(\Omega_{N-1})\right.\nonumber\\&&\left.-\frac{1}{4\eta}\Lambda^2(\Omega_{N-1})\right]\psi(\xi, \eta, \Omega_{N-1})=\varepsilon'\psi(\xi, \eta, \Omega_{N-1}).\label{cfk2}
\end{eqnarray}
The original energy parameter $\varepsilon$ now plays the role of model parameter (or coupling constant) $\omega'$ and the model parameter $-\beta_0$ plays the role of energy $\varepsilon'$. This change in the role of the parameters is called coupling constant metamorphosis. Moreover, the Hamiltonian (\ref{cfk2}) is related to the one in (\ref{cfk1}) by a St\"{a}ckel transformation and thus the two systems are St\"{a}ckel equivalent \cite{boy2, kal6}. 

After the change of variables $\xi=2r^2_1$, $\eta=2r^2_2$, the wave function in the new variables is proportional to
 \begin{eqnarray} 
e^{-\frac{\sqrt{-\omega'} r^2_i}{2}}r_i^{\alpha'_i+\frac{N-1}{2}}L^{\alpha'_i}_{n_i}(\sqrt{-\omega'} r^2_i)y_i(\Omega_{N-1}),\quad i=1,2,\label{cWf1} 
\end{eqnarray}
where $L^{\alpha}_{n}(x)$ is again the $n$th order Laguerre polynomial and 
$\alpha'_i=\delta_i+l_{N-2}+\frac{N-3}{2}$, $\delta_{i}=\left\{\sqrt{(I_{N-2}+\frac{N-3}{2})^2+4\beta_{i}}-\frac{N-3}{2}\right\}-I_{N-2}$, $n_{i}=-\frac{1}{2}\left(\delta_{i}+I_{N-2}+\frac{N-1}{2}\right)+\frac{\varepsilon'}{\omega'}$, $i=1, 2$.  The energy spectrum of system (\ref{chamil2}) is
 \begin{equation}
E=\frac{-c^{2}_{0}}{\hbar^{2}\left\{n_{1}+n_{2}+\frac{1}{2}(\delta_{1}+\delta_{2}+2I_{N-2}+N-1)\right\}^{2}}.\label{cen2}
\end{equation}

\subsection{Recurrence formula and algebra structure}

Let $H'=H_1'+H_2'$ and perform gauge rotations $\tilde{H_i}=\chi_i^{-1} H'_i \chi_i$, $\chi_i=r^{\frac{2-N}{2}}_i$, $i=1,2$ to get 
\begin{eqnarray}
&\tilde{H_i}&=\frac{1}{4}\left\{\partial^2_{r_i}+2\omega'r^2_i-\frac{\frac{(N-2)(N-4)}{4}+4 \beta_i+\frac{J_{(2)}}{\hbar^2}}{r^2_i}\right\},
\end{eqnarray}
where  
\begin{eqnarray}
J_{(2)}=\sum_{i<j}^{N-1}L_{ij}^{2},\quad L_{ij}=x_{i}p_{j}-x_{j}p_{i},\quad i, j=1,\dots, N-1,
\end{eqnarray}
and $J_{(2)}$ is related to $\Lambda^2(\Omega_{N-1})$.  
Let $Z$ be eigenfunction of the gauge rotated Hamiltonian $\tilde{H}=\tilde{H_1}+\tilde{H_2}$, $\tilde{H}Z=\tilde{\varepsilon} Z$. Writing $Z=X_{n_1}X_{n_2}y_1(\Omega_{N-1})y_2(\Omega_{N-1})=\chi^{-1}_i\psi(r_1, r_2, \Omega_{N-1})$ and using (\ref{cWf1}), we have 
\begin{eqnarray}
X_{n_i}=e^{-\frac{\sqrt{-\omega'} r^2_i}{2}}r_i^{\alpha'_i+\frac{1}{2}}L^{\alpha'_i}_{n_i}(\sqrt{-\omega'} r^2_i),\quad i=1,2.
\end{eqnarray}
$X_{n_i}y_i(\Omega_{N-1})$ are eigenfunctions of $\tilde{H_i}$ with eigenvalues
\begin{eqnarray}
\tilde{\varepsilon_i}=(n_i+\frac{1}{2}\alpha'_i+\frac{1}{2})\sqrt{-\omega'},\quad i=1,2.
\end{eqnarray}
Now we define ladder operators
\begin{eqnarray}
&&\tilde{ D^\pm_i}(\sqrt{-\omega'}, r_i)=4\tilde{H_i}\pm2\sqrt{-\omega'} r_i\partial_{r_i}+2\omega' r^2_i\pm\sqrt{-\omega'},\quad i=1,2.
\end{eqnarray}
The action of the symmetry operators on the function $X_{n_i}y_i(\Omega_{N-1})$ gives the recurrences formulas (for $i=1,2$)
\begin{eqnarray}
&&\tilde{ D^+_i}(\sqrt{-\omega'}, r_i)X_{n_i}y_i(\Omega_{N-1})=4(n_i+1)\sqrt{-\omega'}X_{n_i+1}y_i(\Omega_{N-1}), 
\\&&
\tilde{ D^-_i}(\sqrt{-\omega'}, r_i)X_{n_i}y_i(\Omega_{N-1})=4(n_i+\alpha'_i)\sqrt{-\omega'}X_{n_i-1}y_i(\Omega_{N-1}).
\end{eqnarray}
The following diagram indicates how $\tilde{D^\pm_i}, i=1,2$ change the quantum numbers.

\begin{tikzpicture}
\draw[dashed,color=gray] (-0.1,-0.1) grid (4.1,4.1);

    \draw[->] (0,0) -- (0,4.5) node[above] {$n_1$};
    \draw[ ->] (0,0) -- (4.5,0) node[right] {$\alpha_1$};
       
    \draw[-> ,blue] (2,2.01) -- (2,2.97)node[above left] {$\tilde{D^+_1}$};
    \draw[-> ,blue] (2,1.97) --(2,1.03) node[below left] {$\tilde{D^-_1}$};
    
    \draw[- ,blue] (1,2) -- (3,2);
        \node at (2,0)[below]{$\alpha_1$};
    \node at (0,2)[left]{$n_1$};
      
\end{tikzpicture}
\begin{tikzpicture}
\draw[dashed,color=gray] (-0.1,-0.1) grid (4.1,4.1);

    \draw[->] (0,0) -- (0,4.5) node[above] {$n_2$};
    \draw[ ->] (0,0) -- (4.5,0) node[right] {$\alpha_2$};
       
    \draw[-> ,blue] (2,2.01) -- (2,2.97)node[above left] {$\tilde{D^+_2}$};
    \draw[-> ,blue] (2,1.97) --(2,1.03) node[below left] {$\tilde{D^-_2}$};
    
    \draw[- ,blue] (1,2) -- (3,2);
        \node at (2,0)[below]{$\alpha_2$};
    \node at (0,2)[left]{$n_2$};
      
\end{tikzpicture}

Let us consider the higher order operators $\tilde{L_1}=\tilde{D^+_1}\tilde{ D^-_2}$, $\tilde{L_2}=\tilde{D^-_1}\tilde{ D^+_2}$, $\tilde{H}=\tilde{H_1}+\tilde{H_2}$ and $\tilde{B}=\tilde{H_1}-\tilde{H_2}$. Then the action of these operators provide us the following higher order integrals of motion for the gauge rotated Hamiltonian $\tilde{H}$ 
\begin{eqnarray}
&&\tilde{L_1}Z=-16\omega'(n_1+1)(n_2+\alpha'_2)X_{n_1+1}X_{n_2-1}y_1(\Omega_{N-1})y_2(\Omega_{N-1}) ,
\\&&
\tilde{L_2}Z=-16\omega'(n_2+1)(n_1+\alpha'_1)X_{n_1-1}X_{n_2+1}y_1(\Omega_{N-1})y_2(\Omega_{N-1}),
\\&&
\tilde{L_1}\tilde{L_2} Z=256\omega'^2 n_1(n_2+1)(n_1+\alpha'_1)(n_2+\alpha'_2+1)Z,
\\&&
\tilde{L_2}\tilde{L_1} Z=256\omega'^2 n_2(n_1+1)(n_2+\alpha'_2)(n_1+\alpha'_1+1)Z
\end{eqnarray}
and the cubic algebra $C(3)$ in operators form
\begin{eqnarray}
&[\tilde{L_1},\tilde{H}]=0=[\tilde{L_2},\tilde{H}],&\label{ckp5}
\\&[\tilde{L_1},\tilde{B}]=-2\sqrt{-\omega'}\tilde{L_1} ,\qquad [\tilde{L_2},\tilde{B}]=2\sqrt{-\omega'}\tilde{L_2},&\label{ckp6}
\end{eqnarray}
\begin{eqnarray}
&\tilde{L_1}\tilde{L_2}& =16\left[(\tilde{B}+\tilde{H}-\sqrt{-\omega'})^2+\frac{\omega'}{\hbar^2}\left\{J^2+4c'_1\hbar^2+\frac{(N-3)^2}{4}\hbar^2\right\}\right]\nonumber\\&&\quad\times\left[(\tilde{B}-\tilde{H}-\sqrt{-\omega'})^2+\frac{\omega'}{\hbar^2}\left\{J^2+4c'_2\hbar^2+\frac{(N-3)^2}{4}\hbar^2\right\}\right],\label{ckp9}
\end{eqnarray}
\begin{eqnarray}
&\tilde{L_2}\tilde{L_1}& =16\left[(\tilde{B}+\tilde{H}+\sqrt{-\omega'})^2+\frac{\omega'}{\hbar^2}\left\{J^2+4c'_1\hbar^2+\frac{(N-3)^2}{4}\hbar^2\right\}\right]\nonumber\\&&\quad\times\left[(\tilde{B}-\tilde{H}+\sqrt{-\omega'})^2+\frac{\omega'}{\hbar^2}\left\{J^2+4c'_2\hbar^2+\frac{(N-3)^2}{4}\hbar^2\right\}\right].\label{ckp10}
\end{eqnarray}
Applying the simple gauge rotations to $D'^{\pm}_i$, we can obtain the corresponding integrals of motion and cubic algebra of Hamiltonian $H'$.  

The first order integrals of motion $L_{ij}$ generate $so(N-1)$ Lie algebra
\begin{eqnarray}
&[L_{ij},L_{kl}]&= i(\delta_{ik}L_{jl}+ \delta_{jl}L_{ik}-\delta_{il}L_{jk}-\delta_{jk}L_{il})\hbar, 
\end{eqnarray}
where $i, j, k, l=1, ..., N-1.$ Moreover,
\begin{eqnarray}
[L_{ij},\tilde{L_1}]=0=[L_{ij},\tilde{L_2}],\quad[L_{ij},\tilde{H}]=0=[L_{ij},\tilde{B}].
\end{eqnarray}
Thus the full symmetry algebra is a direct sum of the cubic algebra $C(3)$ and $so(N-1)$ Lie algebra (i.e., $C(3)\oplus so(N-1))$.

\subsection{Unirreps and Energy spectrum}

The cubic algebra (~\ref{ckp5})-(~\ref{ckp10}) 
 already has the form of the deformed oscillator algebra with $\aleph=\frac{\tilde{B}}{2\gamma}$, $b^{\dagger}=\tilde{L_1}$ and $b=\tilde{L_2}$. The structure function is
\begin{eqnarray}
\Phi(x;u,\tilde{H})&=&16\left[2(x+u)\gamma+\tilde{H}-(1-\alpha_1)\gamma\right]\left[2(x+u)\gamma+\tilde{H}-(1+\alpha_1)\gamma\right]\nonumber\\&&\times\left[2(x+u)\gamma-\tilde{H}-(1-\alpha_2)\gamma\right]\left[2(x+u)\gamma-\tilde{H}-(1+\alpha_2)\gamma\right],\nonumber\\&&
\end{eqnarray}
where $\gamma=\sqrt{-\omega'}$ and $u$ is arbitrary constant. We should impose the following constraints on the structure function in order for the representations to be finite dimensional
\begin{equation}
\Phi(p+1; u,\tilde{\varepsilon})=0,\quad \Phi(0;u,\tilde{\varepsilon})=0,\quad \Phi(x)>0,\quad \forall x>0,\label{cpro2}
\end{equation}
where $p$ is a positive integer. The solutions of the constraints give the energy $\tilde{\varepsilon}$, the arbitrary constant $u$ and the structure function (i.e. the $(p+1)$-dimensional unirreps)  
\begin{eqnarray}
u=\frac{-\tilde{\varepsilon}+\gamma(1+\alpha'_1)}{2\gamma},\quad\tilde{\varepsilon}=\frac{\gamma}{2}(2+2p+k_1 \alpha'_1+k_2 \alpha'_2),
\end{eqnarray}
\begin{eqnarray}
\Phi(x) &=&16x\gamma^4(\alpha'_1+x)\left[2+2p-2x-(1-k_1)\alpha'_1+(1+k_2)\alpha'_2\right]\nonumber\\&&\times\left[2+2p-(1-k_1)\alpha'_1-(1-k_2)\alpha'_2\right],
\end{eqnarray}
where $k_1=\pm 1$, $k_2=\pm 1$.
By equivalence of spectra between Hamiltonians related by the gauge transformations, we have 
\begin{eqnarray}
\varepsilon'=\frac{\gamma}{2}(2+2p+k_1 \alpha'_1+k_2 \alpha'_2).\label{cEng}
\end{eqnarray}
The coupling constant metamorphosis provides $\varepsilon'\leftrightarrow-\beta_0$ and $\varepsilon\leftrightarrow\omega'$ and give a correspondence between $\gamma=\sqrt{-\omega'}$ and the energy $\varepsilon$ of the original Hamiltonian $H_{KC}$. From (\ref{cEng}) we obtain 
\begin{eqnarray}
E=\frac{-c^2_0}{\hbar^2\left(p+1+\frac{k_1\alpha'_1+k_2\alpha'_2}{2}\right)^2}.\label{cen4}
\end{eqnarray}
Making the identifications $p=n_1+n_2$, $k_1= 1$, $k_2= 1$, then (\ref{cen4}) coincides with the physical spectra (\ref{cen2}).

We remark that the recurrence method gives 4th-order integrals of motion and higher rank cubic algebra for the Stackel equivalent system. This algebra has already a factorized form and much easier to handle than the quadratic algebra $Q(3)$ obtained in \cite{fh1} by using ansatz and the direct approach. 

\section{Conclusion}
The main results of this chapter is extending the recurrence approach, that is a constructive approach to the $N$-dimensional superintegrable Kepler-Coulomb systems with non-central terms and the double singular oscillators of type $(n,N-n)$ to obtain algebraic derivations of their spectra. For both cases, we obtained 4th-order integrals of motion and corresponding higher rank cubic algebras. One interesting feature of these algebras is the fact they admit a factorized form which enable to simplify the calculation of the realizations in terms of deformed oscillator algebras and the corresponding unirreps.

Let us point out that the classical analogs of the recurrence relations \cite{mar6} have been developed and exploited to generate superintegrable systems with higher order integrals and polynomial Poisson algebras \cite{mar5, mar11}. Recently another classical method based on subgroup coordinates \cite{kal8, kal9, des1} has been introduced. One interesting open problem, that we study in a future paper, would be to extend these approaches for the classical analog of the two models considered in this chapter. Moreover, the application of the co-algebra approach \cite{bal4, rig1, pos1} to these new models also remain to be investigated.


\chapter{Superintegrable monopole systems }\label{ch7}

{\bf \large{Acknowledgement}}
\\This chapter is based on the work that was published in  Ref. \cite{fh4}. I have incorporated text of that paper \cite{fh4}. In this chapter, we introduce a Hartmann system in the generalized Taub-NUT space with Abelian monopole interaction. This quantum system includes well known Kaluza-Klein monopole and MIC-Zwanziger monopole as special cases.  It is shown that the corresponding Schr\"{o}dinger equation of the Hamiltonian is separable in both spherical and parabolic coordinates. We obtain the integrals of motion of this superintegrable model and construct the quadratic algebra and Casimir operator. This algebra can be realized in terms of a deformed oscillator algebra and has finite dimensional unitary representations (unirreps) which provide energy spectra of the system. This result coincides with the physical spectra obtained from the separation of variables.

\section{Introduction}
Dirac first explored the existence of monopoles in the quantum mechanical interest and the quantization of electric charge \cite{dir1}. Later the Kepler problems involving additional magnetic monopole interaction was independently discovered by McIntosh and Cisneros \cite{mci1} and  Zwanziger \cite{zwa1} which is known as MICZ-Kepler problem. The MICZ-Kepler problem discusses the existence of the Runge-Lenz vector in addition to the angular momentum vector and a large dynamical symmetry $so(4)$ algebra \cite{bar4, jac1}. The generalized Dirac monopole exhibits a hidden dynamical algebra \cite{men1}. The MICZ-Kepler problems have been generalized using many approaches to higher dimensions \cite{men1, yan1, mad3, ner1, mad5}. These monopole systems are separable in hyperspherical, spheroidal and parabolic coordinates \cite{men1, yan1, mad3, ner1, mad5}. 

One important class of monopole models is Kaluza-Klein monopoles. Kaluza and Klein  introduced a five-dimensional theory with one dimension curled up to form a circle in the context of unification theory \cite{kalu1, kle1}. The complete algebraic description of Kaluza-Klein monopole allows a dynamical symmetry of the quantum motions \cite{gro1, gib1, feh2, cor1}. Models in space with Taub-NUT (Taub-Newman-Unit-Tambrino) metric have attracted much attention because the geodesic of the Taub-NUT metric describes appropriately the motion of well-separated monopole-monopole interactions ( see e.g. \cite{gro1, gib1, feh2, cor1, man1, ati1, gro2, cot1, mar8}). This Taub-NUT metric is well known to admit the Kepler-type symmetry  and provides non-trivial generalization of the Kepler problems. Iwai and his collaborators published a series of papers  \cite{iwa1, iwa2, iwa3, iwa4} about the reduction system to admit a Kepler and harmonic oscillator type symmetry via generalized Taub-NUT metric. The generalized MICZ-Kepler problems  \cite{mad6} represent the intrinsic Smorodinsky-Winternitz system \cite{fri1, eva2} with monopole in 3D Euclidean space. Supersymmetry could be constructed in the generalized MICZ-Kepler system \cite{ran1}. The MICZ-Kepler problem was also considered in $S^3$ \cite{gri1}.

Quadratic algebra is a useful tool to obtain energy spectrum of superintegrable systems in the viewpoint of classical and quantum mechanics \cite{gra1}. General quadratic algebras involving three generators generated by second-order integrals of motion and their realizations in terms of deformed oscillator algebra have been investigated in Ref. \cite{das2}. Most of the applications of the quadratic algebra and representation theory have been on systems with scalar potential interactions \cite{kal4, kal5, tan1, mil1, isa1, gen1, fh1, fh2}. In this chapter, we introduce a new superintegrable system in a Taub-NUT space with Abelian monopole interaction. 

The contents of this chapter are organized as follows: Section 2 introduces a new Kepler monopole system in a Taub-NUT space which includes Kaluza-Klein and MICZ monopole as special cases. It is remarked that this system covers a class of dynamical systems of interest. In section 3, the Schrodinger equation of the Hamiltonian in the Taub-NUT space is solved in both spherical and parabolic coordinates. In section 4, we construct second-order integrals of motion which show the superintegrability of the model in the parabolic coordinates. We obtain quadratic algebra and Casimir operator generated by integrals, and realize these algebras in terms of deformed oscillator algebra. This enables us to obtain the energy spectra of the system algebraically. Finally in section 5, we discuss the results and some open problems.

\section{Kepler monopole system}
Let us consider the generalized Taub-NUT metric in $\mathbb{R}^3$
\begin{eqnarray}
ds^2=f(r)d\textbf{r}^2+g(r)(d\psi+\textbf{A}. d\textbf{r} )^2,\label{dmc1}
\end{eqnarray}
where 
\begin{eqnarray}
&& f(r)=\frac{a}{r}+b, \qquad g(r)=\frac{r(a+br)}{1+c_1 r+dr^2},\label{dfg1}
\\&&
A_1=\frac{-y}{r(r+z)}, \quad A_2=\frac{x}{r(r+z)}, \quad A_3=0,
\end{eqnarray}
$r=\sqrt{x^2+y^2+z^2}$ and the three dimensional Euclidean line element $d\textbf{r}^2=dx^2+dy^2+dz^3$, $a$, $b$, $c_1$, $d$ are constants. Here $\psi$ is the additional angular variable which describes the relative phase and its coordinate is cyclic with period $4\pi$ \cite{ gro1, cor1}. The functions $f(r)$ and $g(r)$ in the metric represent gravitational effects and $A_i$ is identified the monopole interaction. 

We consider the Hamiltonian system associated with (\ref{dmc1}) 
\begin{eqnarray}
H=\frac{1}{2}\left[\frac{1}{f(r)}\left\{p^2+\frac{c_0}{2r}+\frac{c_2}{2r(r+z)}+\frac{c_3}{2r(r-z)}+c_4\right\}+\frac{Q^2}{g(r)}\right],\label{dk1}
\end{eqnarray}
where $c_0$, $c_2$, $c_3$, $c_4$ are constants and the operators
\begin{eqnarray}
p_i=-i(\partial_i-i A_i Q), \quad Q=-i\partial_\psi
\end{eqnarray}
satisfying the following commutation relations
\begin{eqnarray}
[p_i,p_j]=i\epsilon_{ijk}B_k Q, \quad [p_i,Q]=0,\quad \textbf{B}=\frac{\textbf{r}}{r^3}.
\end{eqnarray} 
The system with Hamiltonian (\ref{dk1}) is generalized Hartmann system \cite{har2} in a curved Taub-NUT space with abelian monopole interaction. The Hartmann system is a deformed Coulomb interaction in 3D Euclidean space. This new system (\ref{dk1}) is referred to as Kepler monopole system. It contains the Kaluza-Klein \cite{gib1, mar8} and MICZ monopoles \cite{mci1, zwa1} as special cases:  it is Kaluza-Klein monopole system when $a=1$, $b=1$, $c_1=2$, $d=1$ and MICZ monopole system when $a=0$, $b=1$, $c_1=-2$, $d=1$, $c_2=0$, $c_3=0$.

The Kepler monopole system allows the following suitable total angular momentum operator $\textbf{L}$ and the Runge-Lenz operator \textbf{M} which can be constructed into the form
\begin{eqnarray}
\textbf{L}=\textbf{r}\times \textbf{p}-\frac{\textbf{r}}{r}Q, \quad \textbf{M}=\frac{1}{2}(\textbf{p}\times \textbf{L}-\textbf{L}\times \textbf{p})-\frac{\textbf{r}}{r}(aH-\frac{c_1}{2} Q^2).
\end{eqnarray}
The operators $L$ and $M$ commute with the Kepler monopole system (\ref{dk1}) when $c_0=c_2=c_3=c_4=0$ and verify its maximally superintegrability.
These operators close to an $o(4)$ or $o(3,1)$ dynamical symmetry algebra in the quantum state with fixed energy and eigenvalue of operator $Q$:
\begin{eqnarray}
[L_i, L_j]=i\epsilon_{ijk}L_k,\quad [L_i, M_j]=i\epsilon_{ijk}M_k, \quad [M_i, M_j]=i\epsilon_{ijk}L_k(\frac{c_1 Q^2}{2}-aH).
\end{eqnarray}
It is $o(4)$ algebra for $\frac{c_1 Q^2}{2}-aH>0$  and $o(3,1)$ algebra for $\frac{c_1 Q^2}{2}-aH<0$.

In the next section, we examine model (\ref{dk1}) in the spherical and parabolic coordinate systems for separation of variables.

\section{Separation of variables}
The Hamiltonian (\ref{dk1}) with monopole interaction is multiseparable and allows the separation of variables for the corresponding Schr\"{o}dinger equations in spherical and parabolic coordinates.

\subsection{Spherical coordinates}
Let us consider the spherical coordinates 
\begin{eqnarray}
&&x=r \sin\theta\cos\phi,\quad y=r\sin\theta\sin\phi,\quad z=r\cos\theta,
\end{eqnarray}
where $r>0$, $0\leq\theta\leq\pi$ and $0\leq \phi\leq 2\pi$.
In terms of these coordinates, the Taub-NUT metric (\ref{dmc1}) takes on the form 
\begin{eqnarray}
ds^2=f(r)(dr^2+r^2d\theta^2+r^2\sin^2\theta d\phi^2)+g(r)(d\psi+\cos\theta d\phi)^2,\label{dmc2}
\end{eqnarray}
\begin{eqnarray}
A_1=-\frac{1}{r}\tan\frac{\theta}{2}\sin\phi,\quad A_2=\frac{1}{r}\tan\frac{\theta}{2}\cos\phi,\quad A_3=0,
\end{eqnarray}
and the Schrodinger equation $H\Psi=E\Psi$ of the model (\ref{dk1}) takes the following form 
\begin{eqnarray}
&&\frac{-r}{2(a+br))}\left[\frac{\partial^2}{\partial r^2}+\frac{2}{r}\frac{\partial}{\partial r}-\frac{c_0}{2r}-\frac{c_2}{4r^2\cos^2\frac{\theta}{2}}-\frac{c_3}{4r^2\sin^2\frac{\theta}{2}}-c_4\right.\nonumber\\&&\left.+\frac{1}{r^2}\left(\frac{\partial^2}{\partial\theta^2}+\cot\theta\frac{\partial}{\partial\theta}+\frac{1}{\sin^2\theta}\frac{\partial^2}{\partial\phi^2}\right)+\left(\frac{1}{r^2\cos^2\frac{\theta}{2}}+\frac{c_1}{r}+d\right)\frac{\partial^2}{\partial\psi^2}\right.\nonumber\\&&\left.-\frac{1}{r^2\cos^2\frac{\theta}{2}}\frac{\partial}{\partial\phi}\frac{\partial}{\partial\psi}   \right]\Psi(r,\theta,\phi,\psi) =E\Psi(r,\theta,\phi,\psi).\label{dkp2}
\end{eqnarray}
For the separation of (\ref{dkp2}), the ansatz
\begin{eqnarray}
\Psi(r,\theta,\phi,\psi)=R(r)\Theta(\theta)e^{i(\nu_1\phi+\nu_2\psi)},
\end{eqnarray}
leads readily to the following radial and angular ordinary differential equations   
\begin{eqnarray}
&&\left[\frac{d^2}{d r^2}+\frac{2}{r}\frac{d}{d r}+\alpha+\frac{\beta}{r}-\frac{k_1}{r^2}\right ]R(r)=0,\label{dkp3}
\\
&&\left[\frac{d^2}{d\theta^2}+\cot\theta\frac{d}{d\theta}+\left\{k_1-\frac{c_2+(\nu_1-2\nu_2)^2}{2(1+\cos\theta)}-\frac{c_3+\nu_1^2}{2(1-\cos\theta)}\right\}\right]\Theta(\theta)=0,\label{dkp4}
\end{eqnarray}
where $\alpha=2b E-d \nu_2^2-c_4$, $\beta=2a E-c_1\nu_1^2-\frac{c_0}{2}$ and $k_1$ is separable constant.
We now turn to (\ref{dkp4}), which can be converted, by setting $z=\cos\theta$ and $\Theta(z)=(1+z)^{a}(1-z)^{b} Z(z)$, to
\begin{eqnarray}
&&(1-z^2)Z''(z)+\{2a-2b-(2a+2b+2)z\}Z'(z)\nonumber \\&&\qquad +\{k_2-(a+b)(a+b+1)\}Z(z)=0,\label{dpr5}
\end{eqnarray}
where $2a=\delta_{1}+\nu_1$, $2b=\delta_{2}+\nu_1$ and
\begin{eqnarray}
\delta_{1}=\sqrt{c_2+(\nu_1-2\nu_2)^2}-\nu_1,\quad \delta_{2}=\sqrt{c_3+\nu_1^2}-\nu_1. \label{dpr1}
 \end{eqnarray}
Comparing (\ref{dpr5}) with the Jacobi differential equation  
\begin{equation}
(1-x^{2})y''+\{\beta_1-\alpha_1-(\alpha_1+\beta_1+2)x\}y'+\lambda(\lambda+\alpha_1+\beta_1+1)y=0,\label{dJd1}
\end{equation}
we obtain the  separation constant 
\begin{equation}
k_1=(l+\frac{\delta_{1}+\delta_{2}}{2})(l+\frac{\delta_{1}+\delta_{2}}{2}+1),\label{dpr2} 
\end{equation}
where $l=\lambda+\nu_1$. 
Hence solutions of (\ref{dkp4}) are given in terms of the Jacobi polynomials as
\begin{eqnarray}
\Theta(\theta)&\equiv &\Theta_{l \nu_1}(\theta; \delta_{1}, \delta_{2})
= F_{l \nu_1}(\delta_{1}, \delta_{2})(1+\cos\theta)^{\frac{(\delta_{1}+\nu_1)}{2}}(1-\cos\theta)^{\frac{(\delta_{2}+\nu_1)}{2}}\nonumber\\&&\quad\times P^{(\delta_{2}+\nu_1, \delta_{1}+\nu_1)}_{l-\nu_1}(\cos\theta),\label{djp1}
\end{eqnarray}
where $P^{(\alpha, \beta)}_{\lambda}$ denotes Jacobi polynomial, $F_{l \nu_1}(\delta_{1}, \delta_{2})$ is the normalized constant and $l\in \mathbb{N}$. 

Let us now turn to the radial equation (\ref{dkp3}), which can be converted, by setting  
 $z=\varepsilon r$, $R(z)=z^{l+\frac{\delta_{1}+\delta_{2}}{2}} e^{-\frac{z}{2}}R_1(z)$ and $\alpha =\frac{-\varepsilon^2}{4}$, to
\begin{equation}
z\frac{d^2R_1(z)}{dz^2}+\{(2l+\delta_{1}+\delta_{2}+2)-z\}\frac{dR_1(z)}{dz}-(\frac{\delta_{1}+\delta_{2}}{2}+l+1)-\frac{\beta}{\varepsilon})R_1(z)=0.\label{dan6}
\end{equation}
Set  
\begin{eqnarray}
n=\frac{\beta}{\varepsilon}-\frac{\delta_{1}+\delta_{2}}{2}.\label{dan7}
\end{eqnarray}
Then (\ref{dan6}) can be expressed as
\begin{equation}
z\frac{d^2R_1(z)}{dz^2}+\{(2l+\delta_{1}+\delta_{2}+2)-z\}\frac{dR_1(z)}{dz}-(-n+l+1)R_1(z)=0.\label{dan8}
\end{equation}
This is the confluent hypergeometric equation. Hence we can write the solution of (\ref{dkp3}) in terms of the confluent hypergeometric function as  
\begin{eqnarray}
&R(r)&\equiv R_{nl}(r;\delta_{1}, \delta_{2})=F_{nl}(\delta_{1},\delta_{2})(\varepsilon r)^{l+\frac{\delta_{1}+\delta_{2}}{2}} e^{\frac{-\varepsilon r}{2}}\nonumber\\&&
\quad \times {}_1 F_1(-n+l+1, 2l+\delta_{1}+\delta_{2}+2; \varepsilon r),\label{dan9}
\end{eqnarray}
where $F_{nl}(\delta_{1},\delta_{2})$ is the normalized constant.
In order to have a discrete spectrum the parameter $n$ needs to be positive integer. From (\ref{dan7}) we have  
\begin{equation}
\varepsilon=\frac{\beta}{(n+\frac{\delta_{1}+\delta_{2}}{2})}
\end{equation} and hence the energy spectrum is given by
\begin{equation}
\frac{2aE-c_1\nu_2^2-\frac{c_0}{2} }{2\sqrt{c_4-2b E+d\nu_2^2}}=n+\frac{\delta_1+\delta_2}{2},\qquad n=1, 2, 3,\dots\label{den2}
\end{equation}

\subsection{Parabolic Coordinates }
The parabolic coordinate system has the form
\begin{eqnarray}
&&x=\sqrt{\xi\eta}\cos\phi,\quad y=\sqrt{\xi\eta}\sin\phi,\\&&
z=\frac{1}{2}(\xi-\eta),\quad r=\frac{1}{2}(\xi+\eta)
\end{eqnarray}
with $\xi, \eta >0$ and $0\leq\phi\leq 2\pi$. 
In terms of the coordinates, the Taub-NUT metric (\ref{dmc1}) takes the form
\begin{eqnarray}
&&ds^2=f(r)\left[(\xi+\eta)(d\xi^2+d\eta^2)+\xi\eta d\phi^2\right]+g(r)\left[ d\psi+\left(1-\frac{\xi-\eta}{\xi+\eta} d\phi\right) \right]^2,\label{dmc3}
\end{eqnarray}
\begin{eqnarray}
A_1=\frac{-2 \sqrt{\eta}}{\sqrt{\xi}}\frac{\sin\phi}{\xi+\eta}, \quad A_2=\frac{-2 \sqrt{\eta}}{\sqrt{\xi}}\frac{\cos\phi}{\xi+\eta}, \quad A_3=0
\end{eqnarray}
and the Schrodinger equation of the system (\ref{dk1}) is
\begin{eqnarray}
&&\frac{-1}{2\{2a+b(\xi+\eta)\}}\left[4\xi\frac{\partial^2}{\partial\xi^2}+ 4\eta\frac{\partial^2}{\partial\eta^2}+4\frac{\partial}{\partial\xi} +4\frac{\partial}{\partial\eta}-c_0-\frac{c_2}{\xi}-\frac{c_3}{\eta}\right.\nonumber\\&&\left.-c_4(\xi+\eta)+\frac{\xi+\eta}{\xi\eta}\frac{\partial^2}{\partial\phi^2}-\frac{4}{\xi}\frac{\partial}{\partial\phi}\frac{\partial}{\partial\psi}+\left\{\frac{4}{\xi}+c_1+\frac{d}{4}(\xi+\eta)\right\}\frac{\partial^2}{\partial\psi^2}\right]\nonumber\\&&\times \Psi(\xi,\eta,\phi,\psi)=E\Psi(\xi,\eta,\phi,\psi).\label{dkp1}
\end{eqnarray}
By making the Ansatz,
\begin{eqnarray}
\Psi(\xi,\eta,\phi,\psi)=\Theta_1(\xi)\Theta_2(\eta)e^{i(\nu_1\phi+\nu_2\psi)},
\end{eqnarray}
(\ref{dkp1}) becomes  
\begin{eqnarray}
&&\left[\partial_{\xi}(\xi\partial_{\xi})+\frac{\alpha}{4}\xi+\frac{\beta}{4}-\frac{c_2+(\nu_1-2\nu_2)^2}{4\xi}\right ]\Theta_1(\xi)=k_2 \Theta_1(\xi),\label{dfk4}
\\
&&\left[\partial_{\eta}(\eta\partial_{\eta})+\frac{\alpha}{4}\eta+\frac{\beta}{4}-\frac{c_3+\nu_1^2}{4\eta}\right]\Theta_2(\eta)=-k_2 \Theta_2(\eta),\label{dfk5}
\end{eqnarray}
where $\alpha=2b E-d\nu_2^2-c_4$, $\beta=2a E-c_1\nu_2^2-\frac{c_0}{2}$ and $k_2$ is separable constant. Putting $z_{1}=\varepsilon \xi $ in (\ref{dfk4}), $z_{2}=\varepsilon \eta $ in (\ref{dfk5}) and $\Theta_i(z_i)=z_i^\frac{\delta_i+\nu_1}{2}e^{-\frac{z_i}{2}}F_i(z_i)$, $\alpha=-\varepsilon^2$, these two equations represent
\begin{equation}
z_{i}\frac{d^2F_{i}(z_i)}{dz^2_{i}}+\{(\delta_{i}+\nu_1+1)-z_{i})\}\frac{dF_{i}(z_i)}{dz_{i}}-(\frac{\delta_{i}+\nu_1+1}{2}-\frac{\beta}{4\varepsilon}+\frac{k_i}{\varepsilon})F_{i}(z_i)=0,\label{dfk6}
\end{equation}
where $i=1,2$, $k_2=-k_1$ and 
\begin{eqnarray}
\delta_1=\sqrt{c_2+(\nu_1-2\nu_2)^2}-\nu_1,\quad \delta_2=\sqrt{c_3+\nu_1^2}-\nu_1.\label{ddt1}
\end{eqnarray}
Let us now denote 
\begin{eqnarray}
n_{i}=-\frac{1}{2}(\delta_{i}+\nu_1+1)+\frac{\beta}{4\varepsilon}-\frac{k_i}{\varepsilon}, \quad i=1,2.\label{dfk8}
\end{eqnarray}
Then (\ref{dfk6}) can be identified with the Laguerre differential equation. Thus we have the normalized wave function
 \begin{eqnarray} 
\Psi(\xi,\eta,\phi,\psi)&=&U_{n_{1}n_{2}\nu_1}(\xi,\eta,\phi,\psi; \delta_{1},\delta_{2})\nonumber\\&=&\frac{\hbar \varepsilon^2}{\sqrt{-8c_{0}}}f_{n_{1}\nu_1}(\xi;\delta_{1})f_{n_{2}\nu_1}(\eta;\delta_{2})\frac{e^{i(\nu_1\phi+\nu_2\psi)}}{\sqrt{2\pi}}, 
\end{eqnarray}
where
\begin{eqnarray*}
 &f_{n_{i} \nu_1}(t_{i};\delta_{i})& \equiv f_{i}(t_{i})=\frac{1}{\Gamma(\nu_1+\delta_{i}+1)}\sqrt{\frac{\Gamma(n_{i}+\nu_1+\delta_{i}+1)}{ n_{i}!}} 
 (\varepsilon t_{i})^{(\nu_1+\delta_{i})/2} e^{-\varepsilon t_{i}/2}\nonumber\\&& \times {}_1F_{1}(-n_{i}, \nu_1+\delta_{i}+1; \varepsilon t_{i}),\label{dfk7}   
\end{eqnarray*}
 $i=1, 2$ and $ t_{1}\equiv\xi , t_{2}\equiv\eta $. We look for the discrete spectrum and thus $n_{1}$ and $n_{2}$ are both positive integers. The expression for the energy of the system in terms of $n_{1}$ and $n_{2}$  can be found by using $\alpha=-\varepsilon^2$ in (\ref{dfk8}) to be 
\begin{equation}
\frac{2a E-c_1\nu_2^2-\frac{c_0}{2}}{2\sqrt{c_4-2bE+d\nu_2^2}}=n_1+n_2+\frac{\delta_1+\delta_2}{2}+\nu_1+1.\label{den3}
\end{equation}
We can relate the quantum numbers in (\ref{den2}) and (\ref{den3}) by the following relation
\begin{eqnarray}
n_{1}+n_{2}+\nu_1=n-1,
\end{eqnarray}
where $n_{1}, n_{2}=0, 1, 2,\dots.$

\section{Kepler monopole in Taub-NUT space and algebra structure}
In this section, we construct integrals of motion for the superintegrable monopole system (\ref{dk1}), their quadratic algebra and Casimir operator. The realization of the algebra in terms of deformed oscillator algebra which generate a finite dimensional unitary representation (unirrep) to degenerate energy spectrum of the model (\ref{dk1}) is presented.

\subsection{Integrals of motion and quadratic algebra}
The Hamiltonian system (\ref{dk1}) has the following algebraically independent integrals of motion 
\begin{eqnarray}
&&A=L^2+\frac{c_3 \xi}{4\eta}+\frac{c_2\eta}{4\xi}, \label{dk2}
\\&&
B=M_3+\frac{\xi-\eta}{\xi+\eta}\left\{\frac{c_3\xi^2-c_2\eta^2}{2\xi^2\eta-2\xi\eta^2}+\frac{c_0}{4}\right\}\label{dk3},
\\&& L_3,
\end{eqnarray}
where $L_3=\sqrt{\xi\eta}\cos\phi p_2-\sqrt{\xi\eta}\sin\phi p_1-\frac{\xi+\eta}{\xi-\eta}Q$ and $M_3=\frac{1}{2}\{(p_1L_2-p_2L_1)-(L_1p_2-L_2p_1)\}-\frac{z}{r}(aH-\frac{c_1}{2}Q^2).$
The integral of motion $A$ is associated with the separation of variables in spherical coordinates and $B$ is associated with the separation of variables in parabolic coordinates system. The Hamiltonian (\ref{dk1}) is minimally superintegable as it allows five integrals of motion including $H$ and the superintegrability can be verified by proving the commutation relations
\begin{eqnarray}
&[A, L_3]=0=[A,H], \quad [B, L_3]=0=[B,H],&\\&[A,Q]=0=[B,Q], \quad [H, Q]=0=[H, L_3], \\&[Q,L_3]=0.&
\end{eqnarray}
For convenience we present a diagram of the above commutation relations
\begin{eqnarray}
\begin{xy}
(10,0)*+{Q}="f"; (50,0)*+{L_{3}}="k"; (0,30)*+{A}="a"; (60,30)*+{B}="b"; (30,60)*+{H}="h";  
"f";"k"**\dir{--}; 
"h";"b"**\dir{--};
"h";"f"**\dir{--};
"h";"a"**\dir{--};
"h";"k"**\dir{--};
"h";"f"**\dir{--};
"a";"f"**\dir{--}; 
"a";"k"**\dir{--};
"b";"f"**\dir{--};
"b";"k"**\dir{--};
\end{xy}
\end{eqnarray}
The diagram shows that $Q$ and $L_3$ are central elements. 

We now construct a new integral of motion $C$ of the system from (\ref{dk2}) and (\ref{dk3}) via commutator 
\begin{eqnarray}
[A,B]=C.
\end{eqnarray}
Here $C$ is a cubic function of momenta. By direct computation we can show that the integrals of motion $A$, $B$  and central elements $H$, $Q$, $L_3$ satisfy the following quadratic algebra $Q(3)$, 
\begin{eqnarray}
&[A,B]&=C.\label{dk4}
\\&[A,C]&=2\{A,B\}-4aHQL_3+2c_1Q^3L_3+c_0QL_3+(c_2+c_3)B\nonumber\\&&\quad+a(c_2-c_3)H-\frac{1}{2}c_1(c_2-c_3)Q^2-\frac{1}{4}c_0(c_2-c_3)\label{dk5},
\\&[B,C]&=-2B^2+8bAH+2a^2H^2-2(ac_1+2b)HQ^2-4bHL_3^2-4dAQ^2\nonumber\\&&\quad+\frac{1}{2}(c_1^2+4d)Q^4+2dQ^2L_3^2-4c_4A+(4b-ac_0)H\nonumber\\&&\quad+\frac{1}{2}(c_0 c_1+4c_4-4d)Q^2+2c_4L_3^2+\frac{1}{8}(c_0^2-16c_4)\label{dk6}.
\end{eqnarray}
The Casimir operator of $Q(3)$ in terms of central elements is given by
\begin{eqnarray}
K&=&4a^2H^2L_3^2+4a^2H^2Q^2+a^2(c_2+c_3)H^2-4ac_1HQ^4-4(2b+ac_1)HQ^2L_3^2\nonumber\\&&-(2ac_0-2bc_2+ac_1c_2-2bc_3+ac_1c_3)HQ^2-2(ac_0-bc_2-bc_3)HL_3^2\nonumber\\&&-\frac{1}{2}(4bc_2+ac_0c_2+4bc_3+ac_0c_3-4bc_2c_3)H+4b(c_2-c_3)HQL_3+c_1^2Q^6 \nonumber\\&& +(c_1^2+4d)Q^4L_3^2+\frac{1}{4}(4c_0c_1 +c_1^2c_2 +c_1^2c_3 -4c_2d -4c_3d)Q^4+(c_0c_1+4c_4\nonumber\\&&-c_2d-c_3d)Q^2L_3^2-2d(c_2-c_3)L_3Q^3 +\frac{1}{4}(c_0^2+c_0c_1c_2+c_0c_1c_3-4c_2c_4\nonumber\\&&-4c_3c_4+4c_2d+4c_3d-4c_2c_3d)Q^2-2c_4(c_2-c_3)QL_3+\frac{1}{4}(c_0^2-4c_2c_4\nonumber\\&&-4c_3c_4)L_3^2 +\frac{1}{16}(c_0^2c_2+c_0^2c_3+16c_2c_4+16c_3c_4-16c_2c_3c_4)\label{dk7}.
\end{eqnarray}
This is a main step in the application of the deformed oscillator algebra approach which relies on the quadratic algebra $Q(3)$ and the Casimir operator $K$. In order to derive the spectrum of the system we realize the quadratic algebra $Q(3)$ in terms of deformed oscillator algebras \cite{das1, das2} $\{\aleph, b^{\dagger}, b\}$ of the form
\begin{eqnarray}
[\aleph,b^{\dagger}]=b^{\dagger},\quad [\aleph,b]=-b,\quad bb^{\dagger}=\Phi (\aleph+1),\quad b^{\dagger} b=\Phi(\aleph).\label{dkpfh}
\end{eqnarray}
Here $\aleph $ is the number operator and $\Phi(x)$ is well behaved real function satisfying 
\begin{eqnarray}
\Phi(0)=0, \quad \Phi(x)>0, \quad \forall x>0.\label{dkpbc}
\end{eqnarray}
It is non-trivial to obtain such a realization and find the structure function $\Phi(x)$. The realization of $Q(3)$ is of the form $A=A(\aleph)$, $B=b(\aleph)+b^{\dagger}\rho(\aleph)+\rho(\aleph)b$, where
\begin{eqnarray}
A(\aleph)&=&\left\{(\aleph+u)^2-\frac{1}{4}-\frac{c_2+c_3}{4}\right\},
\\
b(\aleph)&=&\left[4aHQL_3-2c_1Q^3L_3-c_0QL_3-a(c_2-c_3)H+\frac{1}{2}c_1(c_2-c_3)Q^2\right.\nonumber\\&&\left.+\frac{1}{4}c_0(c_2-c_3)\frac{1}{4}\right]\frac{1}{4(\aleph+u)^2-1} ,
\\
\rho(\aleph)&=&\frac{1}{3. 2^{20}(\aleph+u)(1+\aleph+u)\{1+2(\aleph+u)^2\}},
\end{eqnarray}
and $u$ is a constant to be determined from constraints on the structure function.
\subsection{Unirreps and energy spectrum} 
We now construct the structure function $\Phi(x)$ from the realizations of the quadratic algebra ((\ref{dk4})-(\ref{dk6})) and the Casimir operator (\ref{dk7}) as follows
\begin{eqnarray}
\Phi(x;u, H)&=&12288 \left[c_0^2 + c_0 (-8aH + 4 c_1 Q^2) + 4 [2 H \{2 a^2 H + b(1 - 2 (x+u))^2\}\right.\nonumber \\&&\left. - c_4 (1 - 2 (x+u))^2- \{4 a c_1 H + d(1 - 2 (x+u))^2\} Q^2 + c_1^2 Q^4]\right]\nonumber\\&&\times\left[c_2^2 -2 c_2 \{c_3 + (1 - 2 (x+u))^2 + 4 L_3 Q\}+ \{c_3 - (-1 + 2L_3 \right. \nonumber\\&&\left.+ 2 (x+u)) (-1 + 2 (x+u) - 2 Q)\} \{c_3 + (1 + 2 L_3 - 2 (x+u)) \right. \nonumber\\&&\left.\times(-1 + 2 (x+u) + 2 Q)\}\right].
\end{eqnarray}
We need to use an appropriate Fock space to obtain finite dimensional unirreps. Thus the action of the structure function on the Fock basis $|n, E\rangle$ with $\aleph|n, E\rangle =n|n,E\rangle$ and using the eigenvalues of $H$, $Q$ and $L_3$, the structure function becomes the following factorized form:
\begin{eqnarray}
\Phi(x;u, E)&=&-3145728 (c_4 - 2 bE + d q_2^2)[x+u-\frac{1}{2}(1-m_1+m_2)]\nonumber\\&&\times[x+u-\frac{1}{2}(1+m_1-m_2)][x+u-\frac{1}{2}(1-m_1-m_2)]\nonumber\\&&\times [x+u-\frac{1}{2}(1+m_1+m_2)]\left[x+u-\left(\frac{1}{2}+\frac{c_0-4aE+2c_1 q_2^2}{4\sqrt{c_4-2bE+dq_2^2}}\right)\right]\nonumber\\&&\times \left[x+u-\left(\frac{1}{2}-\frac{c_0-4aE+2c_1 q_2^2}{4\sqrt{c_4-2bE+dq_2^2}}\right)\right],\label{dstf1}
\end{eqnarray}
where $m_1^2=c_2+(q_1-q_2)^2$, $m_2^2=c_3+(q_1+q_2)^2$, $q_1$  and $q_2$ are the eigenvalues of $L_3$ and $Q$ respectively.

For finite dimensional unitary representations we should impose the following constraints on the structure function (\ref{dstf1}), 
\begin{equation}
\Phi(p+1; u,E)=0,\quad \Phi(0;u,E)=0,\quad \Phi(x)>0,\quad \forall x>0,\label{dpro2}
\end{equation}
where $p$ is a positive integer. These constraints give $(p+1)$-dimensional unitary representations and their solution gives the energy $E$ and the arbitrary constant $u$. Thus all the possible structure function and energy spectra for $
\epsilon_1=\pm 1$, $\epsilon_2=\pm 1$,
\\Set-1:
\begin{eqnarray}
u=\frac{1}{2}(1 +\epsilon_1 m_1 +\epsilon_2 m_2), \quad \frac{2aE-c_1 q^2_2-\frac{c_0}{2}}{\sqrt{c_4-2bE+dq_2^2}}= 2 + 2 p  + \epsilon_1 m_1 + \epsilon_2 m_2 ,  
\end{eqnarray}
\begin{eqnarray}
\Phi(x)&=&3145728 x(x +\epsilon_1 m_1)(x +\epsilon_2 m_2)(x+\epsilon_1 m_1+\epsilon_2 m_2)(1+p-x)\nonumber\\&&\times(1+ p+x +\epsilon_1 m_1+\epsilon_2 m_2 )s_1^2.
\end{eqnarray}
\\Set-2:
\begin{eqnarray}
&&u=\frac{1}{2}-\frac {2aE-c_1 q^2_2-\frac{c_0}{2}}{2\sqrt{c_4-2bE+dq_2^2}}, \quad \frac{2aE-c_1 q^2_2-\frac{c_0}{2}}{\sqrt{c_4-2bE+dq_2^2}}= 2 + 2 p  + \epsilon_1 m_1 + \epsilon_2 m_2,   
\end{eqnarray}
\begin{eqnarray}
\Phi(x)&=& \frac{1572864 (1 + p)}{a}[x-1-p][1 + p- x+\epsilon_1 m_1 ][1+p-x+\epsilon_2 m_2 ]\nonumber\\&& \times[1+ p- x+\epsilon_1 m_1 +\epsilon_2 m_2 ][2+ 2 p- x+\epsilon_1 m_1 +\epsilon_2 m_2 ]\left[-2a(c_4 + d q_2^2)\right.\nonumber\\&&\left. + b\{c_0 + 2c_1 q_2^2 + 2 (2+2 p+\epsilon_1 m_1 +\epsilon_2 m_2 )s_1\}\right],
\end{eqnarray}
\begin{eqnarray*}
\text{where} &s^2_1&=c_4+d q_2^2+\frac{b}{2a^2}\left[b \eta_1^2-a c_0 - 2 a c_1 q_2^2\right.\\&&\left.\quad +\sqrt{\eta_1^2 \{b^2 \eta_1^2 - 2 a b (c_0 + 2 c_1 q_2^2) + 4 a^2 (c+4 + d q_2^2)\}}\right], \\& \eta_1&=2p + 2 + \epsilon_1 m_1 +\epsilon_2 m_2.
\end{eqnarray*}
The Structure functions are positive for the constraints $\varepsilon_1=1$, $\varepsilon_2=1$ and $m_1, m_2>0$. Using formula (\ref{ddt1}) and $\nu_2=q_2$, we can write $m_1=\sqrt{c_2+(\nu_1-2\nu_2)^2}-\nu_1$ and $m_2=\sqrt{c_3+\nu_1^2}-\nu_1$. Making the identification $p=n_1+n_2$, the energy spectrum coincides with the physical spectra (\ref{den3}). The physical wave functions involve other quantum numbers and we have in fact degeneracy of $p+1$ only when these other quantum numbers would be fixed. The total number of degeneracies may be calculated by taking into account the further constraints on these quantum numbers.

\section{Conclusion}
In this chapter, we have introduced a new superintegrable monopole system in the Taub-NUT space whose wave functions are given in terms of a product of  Laguerre and Jacobi polynomials. By construction, algebraically independent integrals of motion of the Hamiltonian (\ref{dk1}) make it a superintegrable system with monopole interactions. We have presented the quadratic algebra and corresponding Casimir operator generated by the integrals. The realization of this algebra in terms of deformed oscillator enables us to provide the finite dimensional unitary representations and the degeneracy of the energy spectrum of the monopole system.


 \chapter{ Deformed Kepler system and Yang-Coulomb monopole}\label{ch8}

{\bf \large{Acknowledgement}}
\\This chapter is based on the work that was published in  Ref. \cite{fh5}. I have incorporated text of that paper \cite{fh5}. In this chapter, we construct the integrals of motion for the 5D deformed Kepler system with non-central potentials in $su(2)$ Yang-Coulomb monopole field. We show that these integrals form a higher rank quadratic algebra $Q(3; L^{so(4)}, T^{su(2)})\oplus so(4)$, with structure constants involving the Casimir operators of $so(4)$ and $su(2)$ Lie algebras. We realize the quadratic algebra in terms of the deformed oscillator and construct its finite-dimensional unitary representations. This enable us to derive the energy spectrum of the system algebraically. Furthermore we show that the model is multiseparable and allows separation of variables in the hyperspherical and parabolic coordinates. We also show the separability of its 8D dual system (i.e. the 8D singular harmonic oscillator) in the Euler and cylindrical coordinates.

\section{Introduction}
Quantum mechanical systems in Dirac monopole \cite{dir1} or Yang's non-abelian $su(2)$ monopole \cite{yan1} fields have remained a subject of international research interest. In \cite{tru1} the structural similarity between  $su(2)$ gauge system and the 5D Kepler problem was discussed in the scheme of algebraic constraint quantization. Various 5D Kepler systems with $su(2)$ monopole (or Yang-Coulomb monopole) interactions have been investigated by many authors \cite{mad1, mad2, mad3, ple2, ple3, mad4, ple4, kal1, ner1, mad5, kar1, bel1, bel2}. Superintegrability  of certain monopole systems has been shown in \cite{mad6, mad10, ran1, sal1, mar7, fh4, fh6}.

There exists established connection between  2, 3, 5 and 9-dimensional hydrogen-like atoms and 2, 4, 8 and 16-dimensional harmonic oscillators, respectively
\cite{lev2, dav1, kle2, le1, van1, van2}.  In \cite{mar10} one of the present authors proved the superintegrability of the 5D Kepler system deformed with non-central terms in Yang-Coulomb monopole (YCM) field. This was achieved by relating this system to an 8D singular harmonic oscillator via the Hurwitz transformation \cite{hur1, kus1}. An algebraic calculation of the spectrum of the 8D dual system was performed, which enabled the author to deduce the spectrum of the 5D deformed Kepler YCM system via duality between the two models.

However, algebraic structure of the 5D deformed Kepler YCM system and a direct derivation of its spectrum via the symmetry algebra have remained an open problem. Such algebraic approach is expected to provide deeper understanding to the degeneracies of the energy spectrum and the connection of the wavefunctions with special functions and orthogonal polynomials.

Symmetry algebras generated by integrals of motion of superintegrable systems are in general polynomial algebras with structure constants involving Casimir operators of certain Lie algebras \cite{gra1, das2, kal4, kal5, tan1, mil1, gen1, isa1, fh1, fh2, fh3}.
The purpose of this chapter is to obtain the algebraic structure in the 5D deformed Kepler YCM system  and apply it to give a direct algebraic derivation of the energy spectrum of the model. We construct the integrals of motion and show that they form a higher rank quadratic algebra $Q(3; L^{so(4)}, T^{su(2)})\oplus so(4)$ with structure constants involving the Casimir operators of $so(4)$ and $su(2)$ Lie algebras. This quadratic algebra structure enable us to give an algebraic derivation of the energy spectrum of the model. Furthermore, we show that the model is separable in the hyperspherical and parabolic coordinates. We also show the separability of its 8D dual system (i.e. the 8D singular harmonic oscillator) in Euler and cylindrical coordinates.

\section{Kepler system in Yang-Coulomb monopole field}

The 5D Kepler-Coulomb system with Yang-Coulomb monopole interaction is defined by the Hamiltonian \cite{mad1, mar10}
\begin{eqnarray}
H=\frac{1}{2}\left(-i\hbar \frac{\partial}{\partial x_{j}} -\hbar A_{j}^{a} T_{a}\right)^{2}  +\frac{\hbar^{2}}{2r^{2}} \hat{T}^2 -\frac{c_{0}}{r},\label{eH-undeformed}
\end{eqnarray}
where $j=0, 1, 2, 3, 4$ and $\{T_a, a=1, 2, 3\}$ are the $su(2)$ gauge group generators which satisfy the  commutation relations
\begin{eqnarray}
[T_a, T_b]=i\epsilon_{abc}T_c.
\end{eqnarray}
$\hat{T}^2=T^a T^a$ is the Casimir operator of $su(2)$ and $A^a_j$ are the components of the monopole potential $\textbf{A}^a$, which can be written as
\begin{eqnarray}
A_j^a=\frac{2i}{r(r+x_0)}\tau^a_{jk}x_k.
\end{eqnarray}
Here $\tau^a$ are the $5\times 5$ matrices
\begin{eqnarray}
\tau^1 =\frac{1}{2}
\begin{pmatrix}
0 & 0 & 0 \\
0 & 0 & -i\sigma^1 \\
0 & i\sigma^1 & 0
\end{pmatrix},\quad
\tau^2 =\frac{1}{2}
\begin{pmatrix}
0 & 0 & 0 \\
0 & 0 & i\sigma^3 \\
0 & -i\sigma^3 & 0
\end{pmatrix},\quad
\tau^3 =\frac{1}{2}
\begin{pmatrix}
0 & 0 & 0 \\
0 & \sigma^2 & 0 \\
0 & 0 &  \sigma^2
\end{pmatrix},
\end{eqnarray}
which satisfy $[\tau^a, \tau^b]=i\epsilon_{abc}\tau^c$ and $\sigma^i$ are the Pauli matrices
\begin{eqnarray}
\sigma^1 =
\begin{pmatrix}
0 & 1 \\
1 & 0
\end{pmatrix},\quad
\sigma^2 =
\begin{pmatrix}
0 & -i \\
i & 0
\end{pmatrix},\quad
\sigma^3 =
\begin{pmatrix}
1 & 0 \\
0 & -1
\end{pmatrix}.
\end{eqnarray}
The vector potentials $\textbf{A}^a$ are orthogonal to each other,
\begin{eqnarray}
\textbf{A}^a\cdot \textbf{A}^b=\frac{1}{r^2}\frac{r-x_0}{r+x_0}\delta_{ab}
\end{eqnarray}
and to the vector $\textbf{x}=(x_0, x_1, x_2, x_3, x_4)$.

The Kepler-Coulomb model Hamiltonian has the following integrals of motion \cite{yan1}
\begin{eqnarray}
 L_{jk}=(x_j\pi_k-x_k\pi_j)-r^2\hbar F^a_{jk}T_a,\quad
 M_{k}=\frac{1}{2}(\pi_jL_{jk}+L_{jk}\pi_j)+c_0\frac{x_k}{r},\label{ear1}
\end{eqnarray}
which satisfy $[H, L_{jk}]=0=[H, M_{k}]$. Here
\begin{eqnarray}
F^a_{jk}=\partial_j A^a_k-\partial_k A^a_j+\epsilon_{abc}A^b_j A^c_k.\label{ef1}
\end{eqnarray}
is the Yang-Mills field tensor and $\pi_j=-i\hbar \frac{\partial}{\partial x_{j}} -\hbar A_{j}^{a} T_{a}$ which obey the commutation relations
\begin{eqnarray}
[\pi_j, x_k]=-i\hbar\delta_{jk}, \quad [\pi_j,\pi_k]=i\hbar^2 F^a_{jk} T_a.
\end{eqnarray}
The integrals $L_{jk}$ and $M_k$ close to the  $so(6)$ algebra  \cite{lan1}
\begin{eqnarray}
&&[L_{ij}, L_{mn}]=i\hbar\delta_{im}L_{jn}-i\hbar\delta_{jm}L_{in}-i\hbar\delta_{in}L_{jm}+i\hbar\delta_{jn}L_{im},\nonumber\\
&&[L_{ij}, M_{k}]=i\hbar\delta_{ik}M_{j}-i\hbar\delta_{jk}M_{i}, \quad [M_{i}, M_{k}]=-2i\hbar H L_{ik}.
\end{eqnarray}
The $so(6)$  symmetry algebra is very useful in deriving the energy spectrum of the system algebraically.

The Casimir operators of $so(6)$ are given by \cite{bar5, mad8}
\begin{eqnarray}
\hat{K}_1=\frac{1}{2}D_{\mu\nu}D_{\mu\nu}, \quad \hat{K}_2=\epsilon_{\mu\nu\rho\sigma\tau\lambda} D_{\mu\nu}D_{\rho\sigma}D_{\tau\lambda},\quad
 \hat{K}_3=\frac{1}{2}D_{\mu\nu}D_{\nu\rho}D_{\rho\tau}D_{\tau\mu},
\end{eqnarray}
where $\mu, \nu= 0,1,2,3,4$ and
\begin{eqnarray}
D =
\begin{pmatrix}
L_{jk} & -(-2H)^{1/2}M_k \\
(-2H)^{1/2}M_k & 0
\end{pmatrix}.
\end{eqnarray}
The eigenvalues of the operators $\hat{K}_1, \hat{K}_2$ and $\hat{K}_3$ are  \cite{per1}
\begin{eqnarray}
&&K_1=\mu_1(\mu_1+4)+\mu_2(\mu_2+2)+\mu_3^2,\quad
K_2=48(\mu_1+2)(\mu_2+1)\mu_3,\nonumber \\
&& K_3=\mu_1^2(\mu_1+4)^2+6\mu_1(\mu_1+4)+\mu_2^2(\mu_2+2)^2+\mu_3^4-2\mu_3^2,
\end{eqnarray}
respectively, where $\mu_1$, $\mu_2$ and $\mu_3$ are positive integers or half-integers and $\mu_1\geq\mu_2\geq\mu_3$. One can represent the operators $\hat{K}_1, \hat{K}_2, \hat{K}_3$ in the form \cite{mad8}
\begin{eqnarray}
&&\hat{K}_1=-\frac{c_0}{2\hbar^2 H}+2\hat{T}^2-4, \quad\quad \hat{K}_2=48\left(-\frac{\mu_0}{2\hbar^2 H}\right)^{1/2}\hat{T}^2,\nonumber \\
&& \hat{K}_3=K_1^2+6K_1-4K_1 \hat{T}^2-12\hat{T}^2+6\hat{T}^4.
\end{eqnarray}
Denote by $T$ the eigenvalue of $\hat{T}^2$. Then,
\begin{eqnarray}
&&K_1-2T(T+1)=\mu_1(\mu_1+4),\nonumber  \\
&&\mu_2^2(\mu_2+2)^2+\mu_3^4-2\mu_3^2=2T^2(T+1)^2,
\end{eqnarray}
The energy levels of the Yang-Coulomb monopole system are given by
\begin{eqnarray}
\epsilon^{T}_n=-\frac{c_0}{2\hbar^2(\frac{n}{2}+2)^2},
\end{eqnarray}
where use has been made of $\mu_1=\frac{n}{2}$ with $n$ being nonnegative integer.

\section{Kepler system with non-central potentials and Yang-Coulomb monopole interaction}
Let us now consider the 5D Kepler system deformed with non-central potentials in Yang-Coulomb monopole field. The Hamiltonian is given by \cite{yan1, mad1, mar10}
\begin{eqnarray}
&&H=\frac{1}{2}\left(-i\hbar \frac{\partial}{\partial x_{j}} -\hbar A_{j}^{a} T_{a}\right)^{2}  +\frac{\hbar^{2}}{2 r^{2}} \hat{T}^2 -\frac{c_{0}}{r} + \frac{c_{1}}{r(r+x_{0})} + \frac{c_{2}}{r(r-x_{0})},\nonumber\\&&\label{eKP1}
\end{eqnarray}
where $c_1$ and $c_2$ are positive real constants. We can construct the integrals of motion of (\ref{eKP1}) by deforming the integrals (\ref{ear1}) of (\ref{eH-undeformed}),
\begin{eqnarray}
&&A= \tilde{L}^2+ \frac{2 r c_{1}}{(r+x_{0})} + \frac{2 r c_{2}}{(r-x_{0})},\nonumber\\
&&B=M_k+ \frac{c_{1}(r-x_{0})}{\hbar^{2} r(r+x_{0})}  + \frac{c_{2}(r+x_{0})}{\hbar^{2} r(r-x_{0})},\label{eIntegralsAB}
\end{eqnarray}
where
\begin{eqnarray}
\tilde{L}^2=\sum_{j<k}{L}^2_{jk}, \quad j, k =0, 1, 2, 3, 4.
\end{eqnarray}
It can be checked that these integrals satisfy the commutation relations
\begin{eqnarray}
[H,A]=0=[H,B].
\end{eqnarray}
In addition, still some components of the first order integrals of motion $L_{jk}$ for $j, k = 1, 2, 3, 4$ of the Hamiltonian (\ref{eH-undeformed}) commute as well as with the Hamiltonian (\ref{eKP1}). So the 5D deformed Kepler YCM system (\ref{eKP1}) is minimally superintegrable as it has 6 algebraically independent integrals of motion including $H$. Define
\begin{eqnarray}
\hat{L}^2=\sum_{j<k}{L}^2_{jk}, \quad j, k = 1, 2, 3, 4.
\end{eqnarray}
$\hat{L}^2$ is a Casimir operator of the $so(4)$ Lie algebra and is also a central element of the commutation algebra.
The integrals $A, B$ (\ref{eIntegralsAB}) have the following differential operator realization:
\begin{eqnarray}
&A&=\hbar^2 \left[-r^{2}\frac{\partial^2}{\partial x_j^2} +x_{j}x_{k} \frac{\partial^{2}}{\partial x_{j} \partial x_{k}}+4 x_{j} \frac{\partial}{\partial x_{j}} \frac{2 r}{r+x_{0}} \hat{T}^2  \right.\nonumber\\
&&\quad\left. + 2 i r^{2} A_{j}^{a} T_{a} \frac{\partial}{\partial x_{j}}+ \frac{2 r c_{1}}{(r+x_{0})} + \frac{2 r c_{2}}{(r-x_{0})}\right],\nonumber\\
&B&=\hbar^2 \left[ x_{0} \frac{\partial^{2}}{\partial x_j^2 }- x_j \frac{ \partial^{2}}{\partial x_{0} \partial x_j}+i(r-x_{0})A_{j}^{a} T_{a} \frac{\partial}{\partial x_{j}} -2 \frac{\partial}{\partial x_{0}} \right.\nonumber\\
&&\quad\left.+ \frac{ (r-x_{0})}{r(r+x_{0})}\hat{T}^2 +\frac{c_{0}}{\hbar^{2}}\frac{x_{0}}{r}\right]+ \frac{c_{1}(r-x_{0})}{\hbar^{2} r(r+x_{0})}  + \frac{c_{2}(r+x_{0})}{\hbar^{2} r(r-x_{0})}.\label{eDifferentialAB}
\end{eqnarray}
These differential expressions are associated with the multiseparability of the Hamiltonian, as will be seen later. However, let us point out that in general systems with monopole interactions are not separable \cite{ber1}.

\section{Algebra structure, unirreps and energy spectrum}
By direct computation, we can show that the integrals $A$, $B$ (\ref{eIntegralsAB}) and the central elements $H$, $\hat{L}^2$, $\hat{T}^2$ close to form the following quadratic algebra
$Q(3; L^{so(4)}, T^{su(2)})$,
\begin{eqnarray}
&&[A,B]=C,\label{eq1}
\\&&[A,C]=2 \{A,B\} +8 B -2c_{0}(c_{1}-c_{2})-4 c_{0} \hat{T}^2,\label{eq2}
\\&&[B,C]=-2 B^{2} +8 H A - 4 \hat{L}^2 H + (16 -4c_{1}-4c_{2})H +2 c_{0}^{2}.\label{eq3}
\end{eqnarray}
The Casimir operator is a cubic combination of the generators, given explicitly by
\begin{eqnarray}
\hat{K}&=&C^2-2\{A,B^2\}-4B^2-2\{4c_0(c_2-c_1)-4c_0 \hat{T}^2\}B+8HA^2 \nonumber\\&&+2\{(16-8c_1-8c_2)H-4\hat{L}^2H+2c_0^2\}A. \label{eC1}
\end{eqnarray}
Using the differential realization (\ref{eDifferentialAB}) of the integrals $A$ and $B$,  we can show that the Casimir operator (\ref{eC1}) takes the form,
\begin{eqnarray}
\hat{K}&=&-8 H (\hat{T}^2)^{2} +16 \hat{L}^2 H -8(c_{1}-c_{2}) \hat{T}^2 H-2\{(c_{1}-c_{2})^{2} \nonumber\\
&&+8 (2-c_{1}-c_{2})\}H + 4 c_{0}^{2}\hat{L}^2 +4c_{0}^{2}(c_{1} + c_{2}-1).
\end{eqnarray}
Notice that the first order integrals of motion $\{L_{ij},~ i, j=1,2,3,4\}$ generate the $so(4)$ algebra
\begin{eqnarray}
[L_{ij}, L_{mn}]=i\hbar(\delta_{im}L_{jn}-\delta_{jm}L_{in}-\delta_{in}L_{jm}+\delta_{jn}L_{im}).
\end{eqnarray}
So the full dynamical symmetry algebra of the Hamiltonian (\ref{eKP1}) is a direct sum \\ $Q(3; L^{so(4)}, T^{su(2)})\oplus so(4)$ of the quadratic algebra $Q(3; L^{so(4)}, T^{su(2)})$ and the $so(4)$ Lie algebra, with structure constants involving the Casimir operators $\hat{L}^2, \hat{T}^2$ of $so(4), su(2)$.

In order to obtain the energy spectrum of the system, we now construct a realization of the quadratic algebra $Q(3; L^{so(4)}, T^{su(2)})$ in terms of the deformed oscillator algebra of the form
\cite{das1, das2},
\begin{eqnarray}
[\aleph,b^{\dagger}]=b^{\dagger},\quad [\aleph,b]=-b,\quad bb^{\dagger}=\Phi (\aleph+1),\quad b^{\dagger} b=\Phi(\aleph).\label{ekpfh}
\end{eqnarray}
Here $\aleph $ is the number operator and $\Phi(x)$ is well behaved real function satisfying
\begin{eqnarray}
\Phi(0)=0, \quad \Phi(x)>0, \quad \forall x>0.\label{ekpbc}
\end{eqnarray}
It is non-trivial to obtain such a realization and to find the structure function $\Phi(x)$. After long computations, we get
\begin{eqnarray}
A=(\aleph+u)^2-\frac{9}{4},\quad
B=b(\aleph)+b^{\dagger}\rho(\aleph)+\rho(\aleph)b,
\end{eqnarray}
where
\begin{eqnarray}
b(\aleph)&=&\frac{c_0(c_1-c_2)+c_0 \hat{T}^2}{(\aleph+u)^2-\frac{1}{4}},\nonumber\\
\rho(\aleph)&=&\frac{1}{3. 2^{20}(\aleph+u)(1+\aleph+u)\{1+2(\aleph+u)^2\}},
\end{eqnarray}
and $u$ is a constant to be determined from the constraints on the structure function $\Phi$. Using (\ref{eq1}-\ref{eq3}) and  (\ref{eC1}), we find
\begin{eqnarray}
&\Phi(x;u,H)&=98304 [2 c_0^2 + H \{1 - 2 (x+u)\}^2] [4 c_1^2 +
   4 c_2^2 + \{1 - 2 (x+u)\}^2 \nonumber \\&&\times\{4(x+u)(x+u-1)-4\hat{L}^2-3\} -
   4 c_1 [2 c_2 + \{1 - 2 (x+u)\}^2 \nonumber \\&& - 4\hat{T}^2] + 16 (\hat{T}^2)^2 -
   4 c_2 [\{1 - 2 (x+u)\}^2 + 4 \hat{T}^2]].
   \end{eqnarray}
A set of appropriate quantum numbers can be defined in same way as in \cite{tru1, ras1}. We can use the subalgebra chains $so(4)\supset so(3)\supset so(2)$ and $so(3)\supset so(2)$ for $\hat{L}^2$ and $\hat{T}^2$ respectively. The Casimir operators $\hat{J}^2_{(\alpha)}$ of the related chains can be written as \cite{ras1}
\begin{eqnarray}
\hat{J}_{(\alpha)}^2= \sum^\alpha_{i<j} J^2_{ij}, \quad \alpha = 2, 3, 4.
\end{eqnarray}
We need to use an appropriate Fock space in order to obtain finite-dimensional unitary irreducible representations (unirreps) of $Q(3; L^{so(4)}, T^{su(2)})$. Let $|n,E>\equiv |n, E, l_4, l_3, l_2>$ donote the Fock basis states, where $l_4, l_3, l_2$ are quantum numbers of $\hat{J}_{(\alpha)}^2,~ \alpha=4, 3, 2$, respectively.
Then the eigenvalues of the Casimir operators $\hat{L}^2$ and $\hat{T}^2$ are $\hbar^2 l_4(l_4+2)$ and $\hbar^2 T(T+1)$, respectively.  By acting the structure function on the Fock states $|n, E\rangle$ with $\aleph|n, E\rangle =n|n,E\rangle$ and using the eigenvalues of $H$, $\hat{L}^2$ and $\hat{T}^2$, we get
\begin{eqnarray}
&\Phi(x;u,E)&=[ x+u -(\frac{1}{2}-\frac{c_{0}}{\sqrt{-2E}})][ x+u-(\frac{1}{2}+\frac{c_{0}}{\sqrt{-2E}})]\nonumber\\&& \times  [x+u-\frac{1}{2}(1+m_{1}+m_{2})][x+u-\frac{1}{2}(1+m_{1}-m_{2})]\nonumber\\&& \times [x+u-\frac{1}{2}(1-m_{1}+m_{2})][x+u-\frac{1}{2}(1-m_{1}-m_{2})],\label{epp1}
\end{eqnarray}
where $m_{1}^{2}=1+2c_{1} +\hbar^2 l_4(l_4+2) + 2 \hbar^2 T(T+1) $, $ m_{2}^{2}=1+2c_{2} +\hbar^2 l_4(l_4+2) - 2 \hbar^2 T(T+1)$.
For the unirreps to be finite dimensional, we impose the following constraints on the structure function (\ref{epp1}),
\begin{eqnarray}
\Phi(p+1; u,E)=0,\quad \Phi(0;u,E)=0,\quad \Phi(x)>0,\quad \forall x>0,\label{epro2}
\end{eqnarray}
where $p$ is a positive integer. These constraints give $(p+1)$-dimensional unirreps and their solution gives the energy $E$ and constant $u$. The energy spectrum is
\begin{eqnarray}
E=-\frac{c_{0}^{2}}{2( p+1+\frac{m_{1}+m_{2}}{2})^{2}}.\label{een1}
\end{eqnarray}
The total number of degeneracies depends on $p+1$ only when the other quantum numbers would be fixed. These quantum numbers do not increase the total number of degeneracies.

\section{Separation of variables}
In this section we show that the Hamiltonian (\ref{eKP1}) is multiseparatble and allows separation of variables in the hypersherical and parabolic coordinates. We also show that the dual system of (\ref{eKP1}) is separable in the Euler and cylindrical coordinates.

\subsection{Hyperspherical coordinates}
We define the hyperspherical coordinates $r\in [0,\infty)$, $\theta\in[0,\pi]$, $\alpha\in[0,2\pi)$, $\beta\in[0,\pi]$ and $\gamma\in[0,4\pi)$ in the space $\mathbb{R}^5$ by
\begin{eqnarray}
&&x_0=r\cos\theta,\nonumber
\\&&
x_1+i x_2=r\sin\theta\cos\frac{\beta}{2}e^{i(\alpha+\gamma)/2},\nonumber
\\&&
x_3+i x_4=r\sin\theta\sin\frac{\beta}{2}e^{i(\alpha-\gamma)/2}.
\end{eqnarray}
In this coordinate system the differential elements of length, volume and the Laplace operator can be expressed \cite{mad5} as
\begin{eqnarray}
&&dl^2=dr^2+r^2d\theta^2+\frac{r^2}{4}\sin^2\theta(d\alpha^2+d\beta^2+d\gamma^2+2\cos\theta d\alpha d\gamma),\nonumber\\&&
dV=\frac{r^4}{8}\sin^3\theta\sin\beta drd\theta d\alpha d\beta d\gamma,\nonumber\\&&
\Delta=\frac{1}{r^{4}}\frac{\partial}{\partial r} ( r^{4} \frac{\partial}{ \partial r}) + \frac{1}{r^{2}\sin^{3}\theta}\frac{\partial}{ \partial \theta}( \sin^{3}\theta\frac{\partial}{ \partial \theta})-\frac{4\hat{\mathcal{L}}^2}{r^2\sin^2\theta},
\end{eqnarray}
with $\hat{\mathcal{L}}^2=L_1^2+L_2^2+L_3^2 $ and
\begin{eqnarray}
&&L_1=i\left(\cos\alpha\cot\beta\frac{\partial}{\partial\alpha}+\sin\alpha\frac{\partial}{\partial\beta}-\frac{\cos\alpha}{\sin\beta}\frac{\partial}{\partial\gamma}\right),\nonumber\\&&
L_2=-i\left(\sin\alpha\cot\beta\frac{\partial}{\partial\alpha}-\cos\alpha\frac{\partial}{\partial\beta}-\frac{\sin\alpha}{\sin\beta}\frac{\partial}{\partial\gamma}\right),\nonumber\\&&
L_3=i\frac{\partial}{\partial\alpha}.
\end{eqnarray}
With the help of the identity\cite{mad2}
\begin{eqnarray}
iA^a_j\frac{\partial}{\partial x_j}=\frac{2}{r(r+x_0)}L_a,
\end{eqnarray}
with
\begin{eqnarray}
&& L_{1}=\frac{i}{2}(D_{41}+D_{32}), \qquad L_{2}=\frac{i}{2}(D_{13}+D_{42}),\nonumber\\
&& L_{3}=\frac{i}{2}(D_{12}+D_{34}),\quad   D_{jk}=-x_{j}\frac{\partial}{\partial x_{k}}+x_{k} \frac{\partial}{\partial x_{j}},
\end{eqnarray}
the Schr\"{o}dinger equation $H\psi=E\psi$ of (\ref{eKP1}) can be written as
\begin{eqnarray}
\left[\Delta_{r \theta}- \frac{\hat{\mathcal{L}}^2+c_{2}}{\hbar^2 r^{2}\sin^{2}\frac{\theta}{2}}-\frac{\hat{J}^{2}+c_{1}}{\hbar^2 r^{2}\cos^{2}\frac{\theta}{2}} \right]\psi + \frac{2}{\hbar^{2}}(E+\frac{c_{0}}{r})\psi =0, \label{ekp2}
\end{eqnarray}
where
\begin{eqnarray}
\Delta_{r \theta}=\frac{1}{r^{4}}\frac{\partial}{\partial r} ( r^{4} \frac{\partial}{ \partial r}) + \frac{1}{r^{2}\sin^{3}\theta}\frac{\partial}{ \partial \theta}( \sin^{3}\theta\frac{\partial}{ \partial \theta}),
\end{eqnarray}
$\hat{J}^2=J_aJ_a$ with $J_a=L_a+T_a$, $a=1, 2, 3$. The operators $L_a$ and $J_a$ satisfy the commutation relations
\begin{eqnarray}
[ L_{a},L_{b}]=i \epsilon_{abc} L_{c},\quad [ J_{a},J_{b}]=i \epsilon_{abc} J_{c}.
\end{eqnarray}
Equation (\ref{ekp2}) is separable in the hyperspherical coordinates using the eigenfunctions of $\hat{\mathcal{L}}^2$, $\hat{T}^2$ and $\hat{J}^2$ with the eigenvalues $\hbar^2 L(L+1)$, $\hbar^2 T(T+1)$ and $\hbar^2 J(J+1)$, respectively. This is seen as follows. We make the separation ansatz \cite{mad2}
\begin{eqnarray}
\psi &=& \Phi(r,\theta) D^{JM}_{LTm't'}(\alpha,\beta,\gamma,\alpha_{T},\beta_{T},\gamma_{T}),
\end{eqnarray}
where
\begin{eqnarray}
 D^{JM}_{LTm't'}(\alpha,\beta,\gamma,\alpha_{T},\beta_{T},\gamma_{T})&= &\sqrt{\frac{(2L+1)(2T+1)}{4 \pi^{4}}} \sum_{M=m+t}
 C_{L,m;T,t}^{JM}\nonumber\\&&\times D_{mm'}^{L}(\alpha,\beta,\gamma) D_{tt'}^{T}(\alpha_{T},\beta_{T},\gamma_{T}),
\end{eqnarray}
and $C_{L,m;T,t}^{JM}$ are the Clebseh-Gordon coefficients that arise in angular momentum coupling and appear as the expansion coefficients of total angular momentum eigenstates in uncoupled tensor product basis \cite{con1}; $D^j_{mm'}$ are the $su(2)$ Wigner D-functions of dimension $2j+1$ with $j=0, 1/2, 1, 3/2, 2,\dots $ and $m=-j, -j+1,\dots, j$ \cite{wig1}.
Substituting the ansatz into (\ref{ekp2}) leads to the differential equation for the the function $\Phi(r,\theta)$
\begin{eqnarray}
&&\left[\Delta_{r \theta}- \frac{L(L+1)+\frac{c_{2}}{\hbar^{2}}}{r^{2}\sin^{2}\frac{\theta}{2}}-\frac{J(J+1)+\frac{c_{1}}{\hbar^{2}}}{r^{2}\cos^{2}\frac{\theta}{2}}+ \frac{2}{\hbar^{2}}(E+\frac{c_{0}}{r}) \right]\Phi(r,\theta) =0.\nonumber\\&& \label{ekpp1}
\end{eqnarray}
Setting the function $\Phi(r,\theta)=R(r)F(\theta)$, (\ref{ekpp1}) is separated into the ordinary differential equations
\begin{eqnarray}
&&\left[\frac{d^2}{d\theta^2}+3\cot\theta\frac{d}{d\theta} - \frac{2\{L(L+1)+\frac{c_{2}}{\hbar^{2}}\}}{1-\cos\theta}
-\frac{2\{J(J+1)+\frac{c_{1}}{\hbar^{2}}\}}{1+\cos\theta}+\Lambda\right] F(\theta)=0, \nonumber\\
&&\label{ekpp3}
\\&&
\left[\frac{d^2}{d r^2} +\frac{4}{r}\frac{d}{d r} + \frac{2}{\hbar^{2}}(E+\frac{c_{0}}{r})-\frac{\Lambda}{r^2}\right]R(r)=0,\label{ekpp4}
\end{eqnarray}
where $\Lambda$ is the separation constant. Solutions of (\ref{ekpp3}) and (\ref{ekpp4}) in terms of the Jacobi and confluent hypergeometric polynomials \cite{and1} are as follows
\begin{eqnarray}
&&F(\theta)=F_{\lambda J L}(\theta;\delta_1,\delta_2)(1+\cos\theta)^{(\delta_{1}+J)/2}(1-\cos\theta)^{(\delta_{2}+L)/2}\nonumber\\&&\quad\quad\quad \times P_{\lambda -J-L}^{( \delta_{2}+L, \delta_{1}+J)}(\cos\theta),\nonumber
\\&&
R(r)=R_{n \lambda}(r:\delta_1,\delta_2)e^{-\frac{\kappa r}{2}} (\kappa r)^{\lambda + \frac{\delta_{1}+\delta_{2}}{2}} {}_1 F_1(-n, 4 +2 \lambda + \delta_{1}+\delta_{2};\kappa r),
\end{eqnarray}
where
\begin{eqnarray}
&&\Lambda=(\lambda+\frac{\delta_1+\delta_2}{2})(\lambda+\frac{\delta_1+\delta_2}{2}+3),\quad \lambda\in \mathbb{N},\nonumber
\\&&\delta_{1}=-1+\sqrt{\frac{4c_{1}}{\hbar^{2}}+(2J+1)^{2}}-J,\quad
\delta_{2}=-1+\sqrt{\frac{4c_{2}}{\hbar^{2}}+(2L+1)^{2}}-L,\nonumber
\\&&
-n= -\frac{2c_{0}}{\hbar^2 \kappa}+ \frac{\delta_{1}+\delta_{2}}{2} +\lambda+2, \quad E=\frac{-\kappa^2\hbar^2}{8}.
\end{eqnarray}
Hence the energy spectrum
\begin{eqnarray}
E=-\frac{c_{0}^{2}}{2 \hbar^{2}( n+\lambda+2+ \frac{\delta_{1}+\delta_{2}}{2})^{2}}.\label{een2}
\end{eqnarray}
This physical spectrum coincides with (\ref{een1}) obtained from algebraic derivation by the identification $p= n+\lambda+1$, $\delta_1=m_1$, $\delta_2=m_2$ and $\hbar=1$.

\subsection{Parabolic coordinates}
The parabolic coordinates are defined by
\begin{eqnarray}
&&x_0=\frac{1}{2}(\mu-\nu),\nonumber
\\&&
x_1+i x_2=\sqrt{\mu\nu}\cos\frac{\beta}{2}e^{i(\alpha+\gamma)/2},\nonumber
\\&&
x_3+i x_4=\sqrt{\mu\nu}\sin\frac{\beta}{2}e^{i(\alpha-\gamma)/2},
\end{eqnarray}
with $\mu,\nu\in[0,\infty)$. The differential elements of length, volume and Laplace operator in this coordinates can be expressed as
\begin{eqnarray}
&&dl^2=\frac{\mu+\nu}{4}\left(\frac{d\mu^2}{\mu}+\frac{du^2}{\nu}\right)+\frac{\mu\nu}{4}(d\alpha^2+d\beta^2+d\gamma^2+2\cos\theta d\alpha d\gamma),\nonumber\\&&
dV=\frac{\mu\nu}{32}(\mu+\nu)\sin\beta d\mu d\nu d\alpha d\beta d\gamma,\nonumber\\&&
\Delta=\frac{4}{\mu+\nu}\left[ \frac{1}{\mu}\frac{\partial}{\partial \mu} ( \mu^{2} \frac{\partial}{ \partial \mu})+\frac{1}{\nu}\frac{\partial}{\partial \nu} ( \nu^{2} \frac{\partial}{ \partial \nu})\right]-\frac{4}{\mu\nu}\hat{\mathcal{L}}^2.
\end{eqnarray}
The Schr\"{o}dinger equation $H\psi=E\psi$ of the Hamiltonian (\ref{eKP1}) in this coordinates becomes
\begin{eqnarray}
\left[ \Delta_{\mu \nu}-  \frac{ 4(\hat{J}^{2}+c_1)}{\hbar^2 \mu (\mu +\nu)} -  \frac{ 4(\hat{\mathcal{L}}^2+c_2)}{\hbar^2\nu(\nu+\mu)}+ \frac{2}{\hbar^{2}}(E+\frac{c_{0}}{\mu+\nu})\right]\psi =0,\label{ekp3}
\end{eqnarray}
where
\begin{eqnarray}
\Delta_{\mu \nu}=  \frac{4}{\mu+\nu}\left[ \frac{1}{\mu}\frac{\partial}{\partial \mu} ( \mu^{2} \frac{\partial}{ \partial \mu})+\frac{1}{\nu}\frac{\partial}{\partial \nu} ( \nu^{2} \frac{\partial}{ \partial \nu})\right].
\end{eqnarray}
Making the ansatz of the form \cite{mad2}
\begin{eqnarray}
&&\psi = f_{1}(\mu) f_{2}(\nu) D_{LTm't'}^{JM}(\alpha,\beta,\gamma,\alpha_{T},\beta_{T},\gamma_{T}),
\end{eqnarray}
in (\ref{ekp3}), the wave functions are separated and lead to the following ordinary differential equations
\begin{eqnarray}
&&\frac{1}{\mu} \frac{d}{d\mu}(\mu^2 \frac{ d f_{1}}{d\mu})+ \left[ \frac{E}{2\hbar^{2}} \mu  -  \frac{ J(J+1)+\frac{c_1}{\hbar^2}}{\mu}+\frac{c_0}{2\hbar^2}+\frac{\hbar}{2}\tilde{\Lambda} \right]f_{1}=0,\label{ekp4}
\\&&
\frac{1}{\nu} \frac{d}{d\nu}( \nu^2 \frac{ d f_{2}}{d\nu})  +  \left[ \frac{E}{2\hbar^{2}} \nu  -  \frac{ L(L+1)+\frac{c_2}{\hbar^2}}{\nu}+\frac{c_0}{2\hbar^2}-\frac{\hbar}{2}\tilde{\Lambda}\right]f_{2}=0,\label{ekp5}
\end{eqnarray}
where $\tilde{\Lambda}$ is the separation constant.
Solutions of (\ref{ekp4}) and (\ref{ekp5}) are given by the confluent hypergeometric polynomials \cite{and1},
\begin{eqnarray}
&&f_{1}=e^{-\frac{\kappa\mu}{2}}( \kappa \mu)^{\delta_{1}+J}F(-n_{1}, \delta_{1}+J +2 , \kappa \mu ),\nonumber
\\&&
f_{2}=e^{-\frac{\kappa\nu}{2}}( \kappa \nu)^{\delta_{2}+L}F(-n_{2}, \delta_{2}+L +2 , \kappa \nu ),
\end{eqnarray}
where
\begin{eqnarray}
&&\delta_{1}=-1+\sqrt{\frac{4c_{1}}{\hbar^{2}}+(2J+1)^{2}}-J,\quad
\delta_{2}=-1+\sqrt{\frac{4c_{2}}{\hbar^{2}}+(2L+1)^{2}}-L,\nonumber
\\&&
-n_{1}=\frac{1}{2}(\delta_{1}+J)+1-\frac{\hbar}{2\kappa}\tilde{\Lambda} -\frac{c_{0}}{2\kappa \hbar^{2}},\quad E=\frac{-\hbar^2\kappa^2}{2}, \nonumber
\\&&
-n_{2}=\frac{1}{2}(\delta_{2}+L)+1+\frac{\hbar}{2\kappa}\tilde{\Lambda} -\frac{c_{0}}{2\kappa \hbar^{2}}.
\end{eqnarray}
Set
\begin{eqnarray}
n_{1}+n_{2}=\frac{c_0}{\kappa\hbar^2}-\frac{1}{2}(\delta_{1}+\delta_{2}+J+L)-2.
\end{eqnarray}
The energy spectrum
\begin{eqnarray}
E=-\frac{c_{0}^{2}}{2\hbar^{2}\{n_1+n_2+\frac{1}{2}(\delta_1+\delta_2+J+L)+2\}^{2}}.
\end{eqnarray}
Making the identification $p= n_1+n_2+\frac{J+L}{2}+1$, $\delta_1=m_1$, $\delta_2=m_2$ and $\hbar=1$, the energy spectrum becomes (\ref{een1}).

\subsection{Euler spherical coordinates}
The 5D Kepler system with non-central terms and Yang-Coulomb monopole is dual to the 8D singular oscillator via Hurwitz transformation \cite{mar10}.
The symmetry algebra structure and energy spectrum of the 8D singular oscillator has been studied in \cite{mar10}. In this and next subsections
we show the separability this dual system in the Euler spherical and cylindrical coordinates, and compare our results for the spectrum with those obtained in \cite{mar10}.

The Hamiltonian of the 8D singular oscillator reads
\begin{eqnarray}
&&H=-\frac{\hbar^{2}}{2}\sum_{i=0}^7 \frac{ \partial^{2}}{\partial u_i^{2}} + \frac{ \omega^{2}}{2}\sum_{i=0}^7 u_i^{2} +\frac{ \lambda_{1}}{u_{0}^{2}+u_{1}^{2}+u_{2}^{2}+u_{3}^{2}}+\frac{ \lambda_{2}}{u_{4}^{2}+u_{5}^{2}+u_{6}^{2}+u_{7}^{2}}.\nonumber\\&&\label{eEP1}
\end{eqnarray}
In the Euler 8D spherical coordinates \cite{kar1}
\begin{eqnarray}
&&u_{0}+iu_{1}=u \cos\frac{\theta}{2}\sin\frac{ \beta_{T}}{2} e^{-i\frac{(\alpha_{T}-\gamma_{T})}{2}},\quad
u_{2}+iu_{3}=u \cos\frac{\theta}{2}\cos\frac{ \beta_{T}}{2} e^{i\frac{(\alpha_{T}+\gamma_{T})}{2}},\nonumber
\\&&
u_{4}+iu_{5}=u \sin\frac{\theta}{2}\sin \frac{ \beta_{K}}{2} e^{i\frac{(\alpha_{K}-\gamma_{K})}{2}},\quad
u_{6}+iu_{7}=u\sin\frac{\theta}{2}\cos\frac{ \beta_K}{2} e^{-i\frac{(\alpha_{K}+\gamma_{K})}{2}},
\end{eqnarray}
where $0\leq u<\infty$, $0\leq \theta\leq\pi$, we have
\begin{eqnarray}
\sum_{i=0}^7 \frac{ \partial^{2}}{\partial u_i^{2}}&=&\frac{1}{u^{7}} \frac{\partial}{\partial u}( u^{7} \frac{\partial}{\partial u})+  \frac{4}{u^{2}\sin^3\theta}\frac{ \partial}{\partial \theta}( \sin^{3}\theta \frac{\partial }{ \partial \theta}) - \frac{4}{u^{2}\cos^{2}\frac{\theta}{2}}\hat{T}^2 - \frac{4}{u^{2}\sin^{2}\frac{\theta}{2}}\hat{K}^{2},\nonumber\\
\end{eqnarray}
with
\begin{eqnarray}
 \hat{T}^2 =-\left[\frac{\partial^{2}}{\partial \beta_{T}^{2}}+ \cot\beta_{T}\frac{\partial}{\partial \beta_{T}}+ \frac{1}{\sin^{2}\beta_{T}}\left( \frac{ \partial}{\partial \alpha_{T}^{2}} -2 \cos\beta_{T} \frac{\partial^{2}}{\partial \alpha_{T} \partial \gamma_{T}}+  \frac{\partial^{2}}{ \partial \gamma_{T}^{2}}\right)\right],
\end{eqnarray}
\begin{eqnarray}
 \hat{K}^{2} =-\left[\frac{\partial^{2}}{\partial \beta_{K}^{2}}+ \cot\beta_{K}\frac{\partial}{\partial \beta_{K}}+ \frac{1}{\sin^{2}\beta_{K}}\left( \frac{ \partial}{\partial \alpha_{K}^{2}} -2 \cos \beta_{K} \frac{\partial^{2}}{\partial \alpha_{K} \partial \gamma_{K}}+  \frac{\partial^{2}}{ \partial \gamma_{K}^{2}}\right)\right].
\end{eqnarray}
For the separation of the wavefunctions in the form
\begin{eqnarray}
\psi= R(u)F(\theta) D_{tt'}^{T}(\alpha_{T},\beta_{T},\gamma_{T})  D_{kk'}^{K}(\alpha_{K},\beta_{K},\gamma_{K}),
\end{eqnarray}
with
\begin{eqnarray}
\hat{T}^2D_{tt'}^{T}=T(T+1)D_{tt'}^{T},
\quad
\hat{K}^{2}D_{kk'}^{K}=K(K+1)D_{kk'}^{K},\label{eWG1}
\end{eqnarray}
the Schr\"odinger equation $H\Psi=\epsilon \Psi$  leads to the ordinary differential equations
\begin{eqnarray}
&&\left[\frac{d^2}{d u^2}+ \frac{7}{u}\frac{d}{d u}- \frac{\Gamma}{u^{2}}+\frac{2 \epsilon}{\hbar^{2}}-\frac{\omega^{2}}{\hbar^{2}}u^{2}\right]R(u)=0,\label{erd1}
\\&&
\left[\frac{1}{\sin^{3}\theta} \frac{d}{d \theta}(\sin^{3}\theta\frac{d }{d \theta})- \frac{T(T+1)+\frac{\lambda_{1}}{2\hbar^{2}}}{  \cos^{2}\frac{\theta}{2}}- \frac{K(K+1)+\frac{\lambda_{2}}{2\hbar^{2}}}{  \sin^{2}\frac{\theta}{2}} + \frac{\Gamma}{4}\right] F(\theta) =0,\nonumber\\&&\label{eag1}
\end{eqnarray}
where $\Gamma$ is the separation of constant.
The solution of (\ref{eag1}) in terms of Jacobi polynomials \cite{and1} as follows
\begin{eqnarray}
&&F(\theta)=F_{\lambda T K}(\theta;\delta_1,\delta_2)(1+\cos\theta)^{(\delta_{1}+T)/2}(1-\cos\theta)^{(\delta_{2}+K)/2}\nonumber\\&&\quad\quad\quad \times P_{\lambda -T-K}^{( \delta_{2}+K, \delta_{1}+T)}(\cos\theta),
\end{eqnarray}
where
\begin{eqnarray}
&&\Gamma=4(\lambda+\frac{\delta_1+\delta_2}{2})(\lambda+\frac{\delta_1+\delta_2}{2}+3),\quad \lambda\in \mathbb{N},\nonumber
\\&&\delta_{1}=-1+\sqrt{\frac{2\lambda_{1}}{\hbar^{2}}+(2T+1)^{2}}-T,\nonumber
\\&&
\delta_{2}=-1+\sqrt{\frac{2\lambda_{2}}{\hbar^{2}}+(2K+1)^{2}}-K.
\end{eqnarray}
The solution of (\ref{erd1}) in terms of the confluent hypergeometric functions \cite{and1}
\begin{eqnarray}
R(u)=R_{n \lambda}(u;\delta_1,\delta_2)e^{-\frac{\kappa u^2}{2}} (\kappa u^2)^{\lambda + \frac{\delta_{1}+\delta_{2}}{2}} {}_1 F_1(-n, 4 +2 \lambda + \delta_{1}+\delta_{2};\kappa u^2),
\end{eqnarray}
where
\begin{eqnarray}
-n=\frac{\delta_1+\delta_2}{2}+\lambda+2-\frac{\epsilon}{2\kappa\hbar^2},\quad \omega^2=\kappa^2\hbar^2.
\end{eqnarray}
Hence
\begin{eqnarray}
\epsilon=2 \hbar^2 \kappa ( n +\frac{\delta_{1}+\delta_{2}}{2}+ \lambda +2 ).
\end{eqnarray}
Using the relations between the parameters of the generalized Yang-Coulomb monopole and the 8D singular oscillator,
\begin{eqnarray}
c_0=\frac{\epsilon}{4}, \quad E=\frac{-\omega^2}{8}, \quad 2c_i=\lambda_i,\quad i=1, 2,\label{eduality}
\end{eqnarray}
we obtain
\begin{eqnarray}
E=-\frac{c_0^2}{2\hbar^2\left(n+\lambda + 2+\frac{\delta_1+\delta_2}{2}\right)^2},
\end{eqnarray}
which coincides with (\ref{een2}).

\subsection{Cylindrical coordinates}

Consider the 8D cylindrical coordinates
\begin{eqnarray}
&&u_{0}+iu_{1}= \rho_{1} \sin\frac{ \beta_{T}}{2} e^{-i\frac{(\alpha_{T}-\gamma_{T})}{2}},\quad
u_{2}+iu_{3}= \rho_{1} \cos\frac{ \beta_{T}}{2}e^{i\frac{(\alpha_{T}+\gamma_{T})}{2}},\nonumber
\\&&
u_{4}+iu_{5}= \rho_{2}\sin\frac{ \beta_{K}}{2}e^{i\frac{(\alpha_{K}-\gamma_{K})}{2}},\quad
u_{6}+iu_{7}= \rho_{2}\cos\frac{ \beta_{K}}{2}e^{-i\frac{(\alpha_{K}+\gamma_{K})}{2}},
\end{eqnarray}
where $0\leq \rho_1, \rho_2<\infty$. Then we have in this coordinate system,
\begin{eqnarray}
\sum_{i=0}^7 \frac{ \partial^{2}}{\partial u_i^{2}}= \frac{1}{\rho_{1}^{3}} \frac{\partial}{\partial \rho_{1}}( \rho_{1}^{3} \frac{\partial}{\partial \rho_{1}})  + \frac{1}{\rho_{2}^{3}} \frac{\partial}{\partial \rho_{2}}( \rho_{2}^{3} \frac{\partial}{\partial \rho_{2}})-\frac{4}{\rho_{1}^{2}}\hat{T}^2-\frac{4}{\rho_{2}^{2}}\hat{K}^{2},
\end{eqnarray}
\begin{eqnarray}
 \hat{T}^2
  =-\left[\frac{\partial^{2}}{\partial \beta_{T}^{2}}+ \cot\beta_{T} \frac{1}{\sin^{2}\beta_{T}}\left( \frac{ \partial}{\partial \alpha_{T}^{2}} -2 \cos\beta_{T} \frac{\partial^{2}}{\partial \alpha_{T} \partial \gamma_{T}}+  \frac{\partial^{2}}{ \partial \gamma_{T}^{2}}\right)\right],
\end{eqnarray}
\begin{eqnarray}
 \hat{K}^{2} =-\left[\frac{\partial^{2}}{\partial \beta_{K}^{2}}+ \cot\beta_{K} \frac{1}{\sin^{2}\beta_{K}}\left( \frac{ \partial}{\partial \alpha_{K}^{2}} -2 \cos \beta_{K} \frac{\partial^{2}}{\partial \alpha_{K} \partial \gamma_{K}}+  \frac{\partial^{2}}{ \partial \gamma_{K}^{2}}\right)\right].
\end{eqnarray}
Making the ansatz
\begin{eqnarray}
\psi = f_{1} f_{2} D_{tt'}^{T}(\alpha_{T},\beta_{T},\gamma_{T})  D_{kk'}^{K}(\alpha_{K},\beta_{K},\gamma_{K})
\end{eqnarray}
and change of variables $x_{i}= a^{2} \rho_{i}^{2}$, $a =\sqrt{\frac{\omega}{\hbar}}$, the Schr\"odinger equation $H\psi=\epsilon \psi$ is converted to
\begin{eqnarray}
&&x_{1} \frac{d^{2}f_{1}}{dx_{1}^{2}}+2 \frac{df_{1}}{dx_{1}}-\left[\frac{ T(T+1)+\frac{\lambda_1}{2\hbar^2}}{x_{1}}+\frac{x_{1}}{4}-\frac{\epsilon_{1}}{2\hbar \omega}\right]f_{1}=0,\label{ehh1}
\\&&
x_{2} \frac{d^{2}f_{2}}{dx_{2}^{2}}+2 \frac{df_{2}}{dx_{2}}-\left[\frac{ K(K+1)+\frac{\lambda_2}{2\hbar^2}}{x_{2}}+\frac{x_{2}}{4}-\frac{\epsilon_{2}}{2\hbar \omega}\right]f_{2}=0.\label{ehh2}
\end{eqnarray}
where $ \epsilon_{1}+\epsilon_{2}=\epsilon$. Setting $f_{i}= e^{-\frac{x_{i}}{2}} x_{i}^{\frac{\delta_i+z_i}{2}} W(x_{i})$, $ i=1, 2$, then (\ref{ehh1}) and (\ref{ehh2}) become
\begin{eqnarray}
x_{i} \frac{ d^{2}W(x_{i})}{dx_{i}^{2}} + ( \gamma_{i} -x_{i}) \frac{dW(x_{i})}{dx_{i}}- \alpha_{i} W(x_{i}) =0,\label{epk1}
\end{eqnarray}
where $ \alpha_i=\frac{\gamma_i}{2}-\frac{\epsilon_i}{2\hbar \omega}, \gamma_{i}=\delta_i+z_i+2$, $z_1=T$, $z_2=K$ and
\begin{eqnarray}
\delta_{i}=-1+\sqrt{\frac{2\lambda_{i}}{\hbar^{2}}+(2z_i+1)^{2}}-z_i,\quad i=1, 2.
\end{eqnarray}
Solutions of (\ref{epk1}) in terms of confluent hypergeometric polynomials \cite{and1} are given by
\begin{eqnarray}
 f_{n_{i},\gamma_{i}} = ( a \rho_{i})^{\gamma_{i}-2} e^{-\frac{a^{2} \rho_{i}^{2}}{2}} {}_1 F_1(-n_{i}, \gamma_{i}, a^{2} \rho_{i}^{2}),
\end{eqnarray}
where
\begin{eqnarray}
n_{i}=\frac{\epsilon_{i}}{2\hbar \omega}-\frac{1}{2}(\delta_i+z_i+2), \quad i=1,2.
\end{eqnarray}
We thus found the energy spectrum of the  8D singular oscillator
\begin{eqnarray}
\epsilon=2\hbar\omega\left(n_1+n_2+\frac{\delta_1+\delta_2}{2}+\frac{T+K}{2}+2 \right).
\end{eqnarray}
Using the relations (\ref{eduality}) between the parameters of the generalized Yang-Coulomb monopole and the 8D singular oscillator, we obtain
\begin{eqnarray}
E=-\frac{c_0^2}{2\hbar^2\left(n_1+n_2+\frac{\delta_1+\delta_2}{2}+\frac{T+K}{2}+2\right)^2},
\end{eqnarray}
which coincides with (\ref{een1}) by making the identification $p= n_1+n_2+\frac{T+K}{2}+1$, $\delta_1=m_1$, $\delta_2=m_2$ and $\hbar=1$.

\section{Conclusion}
One of the results of this chapter is the determination of the higher rank quadratic algebra structure $Q(3; L^{so(4)},T^{su(2)})\oplus so(4)$ in the 5D deformed Kepler system with non-central terms and Yang-Coulomb monopole interaction. The structure constants of the quadratic algebra $Q(3; L^{so(4)},T^{su(2)})$ contain Casimir operators of the $so(4)$ and $su(2)$ Lie algebras. The realization of this algebra in terms of deformed oscillator enable us to provide the finite dimensional unitary representations and the degeneracy of the energy spectrum of the model. We also connected these results with method of separation of variables, solution in terms of orthogonal polynomials and the dual under Hurwitz transformation which is a 8D singular oscillator.

It would be interesting to extend these results to systems in curved spaces, in particular to the 8D pseudospherical (Higgs) oscillator and its dual system involving monopole  \cite{bel2}. Moreover, higher dimensional superintegrable models with monopole interactions are still relatively unexplored area. One of the open problems is to construct the non-central deformations of the known higher-dimensional models \cite{men1, kri1} and their symmetry algebras.


 \chapter{Constructive approach on superintegrable monopole systems}\label{ch9}

{\bf \large{Acknowledgement}}
\\This chapter is based on the work that was published in  Ref. \cite{fh6}. I have incorporated text of that paper \cite{fh6}. In this chapter, we revisit the MIC-harmonic oscillator in flat space with monopole interaction and derive the polynomial algebra satisfied by the integrals of motion and its energy spectrum using the ad hoc recurrence approach. We introduce a superintegrable monopole system in generalized Taub-NUT space. The Schr\"{o}dinger  equation of this model is solved in spherical coordinates in the framework of St\"{a}ckel transformation. It is shown that wave functions of the quantum system can be expressed in terms of the product of Laguerre and Jacobi polynomials. We construct ladder and shift operators based on the corresponding wave functions and obtain the recurrence formulas. By applying these recurrence relations, we construct higher order algebraically independent integrals of motion. We show the integrals form a polynomial algebra. We construct the structure functions of the polynomial algebra and obtain the degenerate energy spectra of the model.

\section{Introduction}

The connection between quantum models and magnetic monopoles, integrable and superintegrable systems is well-known. In this paper, we show that the same connection applies to the harmonic oscillator with Abelian monopole using the recurrence approach. To our knowledge the recurrence approach had not previously been applied to Hamiltonian systems with magnetic monopole interactions. 

 In classical and quantum mechanical systems, constructive approach is one of the powerful tools to derive integrals of motion. The first- and second-order ladder operators have been used by several authors to construct integrals of motion and their corresponding polynomial algebras \cite{jau1, fri2, boy1, eva3, mar6}. There are many distinct approaches for the constructions of integrals of motion using higher order ladder operators (see e.g. \cite{kre1, adl1, jun1, dem1, mar3, rag1, mar9, mar13, mar14}). In fact, there is a close connection between recurrence approach and special functions and orthogonal polynomials \cite{kal7}. The operator version of recurrence relations, their algebraic relations \cite{cal2, cal3} and connection to Lissajous models related to Jacobi exceptional orthogonal polynomials were investigated \cite{mar15}. Recently the authors in the present paper applied coupling constant metamorphosis to systems amenable to the ladder operator method \cite{fh3}. However most these previous studies have been restricted to systems with scalar potentials.
 
Superintegrable systems with non-scalar potentials such as spin \cite{win1, win2, nik1}, magnetic field and magnetic monopole \cite{wu1, jac1, dho1, lab1, mad5} have recently attracted much interest. In \cite{mar8}, a quantum superintegrable system in the field of Kaluza-Klein magnetic monopole was studied. We are interested superintegrable monopole system in space with Taub-NUT metric. The geodesic of the Taub-NUT metric appropriately describes the motion of well-separated monopole-monopole interactions ( see e.g. \cite{cor1, iwa3, iwa4, gro2, cot1, gib1}). Recently we introduced a Kepler quantum monopole system in a generalized Taub-NUT space which includes the Kaluza-Klein and MIC-Zwanziger monopole systems as special cases \cite{fh4}.

The purpose of the present chapter is twofold: Firstly we revisit the MIC-harmonic oscillator in the field of magnetic monopole in flat space \cite{lab1} by means of a somewhat ad hoc recurrence approach. We construct the integrals of motion and (polynomial) algebraic relations satisfied by them. This enables us to present an algebraic derivation of the energy spectrum of the system. Secondly we introduce a new MIC-harmonic oscillator type Hartmann system with monopole interaction in a generalized Taub-NUT space. We construct its integrals of motion using recurrence relations based on wave functions. We show the integrals satisfy a higher-order polynomial algebra and apply this algebraic structure to derive the energy spectrum.

\section{MIC-harmonic oscillator with monopole in flat space}
The problem of the accidental degeneracies in the spectrum of a harmonic oscillator in the field of magnetic monopole was investigated in \cite{lab1, mci1}. In this section we revisit this model using a somewhat ad hoc recurrence method.

Consider the Hamiltonian with monopole interaction 
\begin{eqnarray}
H=\frac{1}{2}\left[\textbf{p}^2+\frac{c_0 r^2}{2}+\frac{Q^2}{r^2}\right],\label{fham1}
\end{eqnarray}
where $p_i=-i\partial_i-A_i Q$; $A_1=\frac{-y}{r(r+z)}$, $A_2=\frac{x}{r(r+z)}$,  and $A_3=0$ are the 3 components of the vector potential associated with the magnetic monopole; $c_0$ and $Q$ are constants. This system is the well-known MIC-harmonic oscillator \cite{mci1}.
Setting $Q^2=\lambda$ and $\frac{c_0}{2}=\omega^2$, the Hamiltonian becomes the one in \cite{lab1}.
The total angular momentum of the system reads
\begin{eqnarray}
\textbf{L}=\textbf{r}\times \textbf{p}-Q\frac{\textbf{r}}{r}.
\end{eqnarray}
The eigenvalues of $\textbf{L}^2$ are, as usual, of the form $l(l+1)$ with $l=|Q|, |Q|+1, |Q|+2,\dots$. Let 
\begin{eqnarray}
T=-\frac{1}{4}(\textbf{r}.\textbf{p}-\textbf{p}.\textbf{r}),\quad
S=-\frac{1}{2\omega}\left(\frac{1}{2}\textbf{p}^2+\frac{Q^2}{2r^2}-\frac{\omega^2 r^2}{2}\right).
\end{eqnarray}
Then $T$, $S$ and $\frac{1}{2\omega}H$ satisfy the $O(2,1)$ commutation relations and eigenstates of $H$ belong to irreducible $O(2,1)$ representation spaces \cite{lab1}. The eigenvalues of $H$ is of the form $2\omega(d^{+}_l+n)$,  where $n=0, 1, 2,\dots$ and $d^{+}_l=\frac{1}{2}\{1+ (l+\frac{1}{2})\}$. As pointed out in \cite{lab1}, a complete set of quantum number is obtained by simultaneously diagonalizing $H$, $\textbf{L}^2$, $L_3$. The action of $H$, $\textbf{L}^2$ and $L_3$ on the basis vectors $|n,l,m\rangle$ is given by
\begin{eqnarray}
&H|n,l,m\rangle&=2\omega(d^{+}_l+n)|n,l,m\rangle,\quad n=0, 1, 2,\dots,\label{fh1}
\\&
\textbf{L}^2|n,l,m\rangle&= l(l+1)|n,l,m\rangle, \quad l=|Q|, |Q|+1,\dots,\label{fl1}
\\&
L_3|n,l,m\rangle&= m|n,l,m\rangle,\quad m=-l, -l+1,\dots, l-1, l. \label{fl2}
\end{eqnarray}
The physical energy spectrum of $H$ is found to be 
\begin{eqnarray}
E_{k}=\omega(k+\frac{3}{2}), \quad k=2n+l.\label{fpe1}
\end{eqnarray}
Introduce \cite{lab1}
\begin{eqnarray}
a_i=\frac{1}{\sqrt{2}}(u_i+\frac{i}{\omega}\dot{u_i})\quad \text{and}\quad a^\dagger_i=\frac{1}{\sqrt{2}}(u_i+\frac{i}{\omega}\dot{u_i}),
\end{eqnarray}
where $u_i=\frac{\epsilon_{ijk}}{\sqrt{2}}(L_j r_k+r_k L_j)$ and $\dot{u_i}=\frac{\epsilon_{ijk}}{\sqrt{2}}(L_j v_k+v_k L_j)$, $i,j,k=1,2,3$. They also satisfy the commutation relations  $[H, a^\dagger_i]=\omega a^\dagger_i$ and $[H, a_i]=-\omega a_i$. Let 
\begin{eqnarray}
A=\frac{1}{2\omega}(H+\omega B-\omega), \quad H_{\pm}\equiv S\pm iT,
\end{eqnarray}
where $B=\sqrt{\textbf{L}^2+\frac{1}{4}}$ which is a well-defined operator \cite{lab1}.

 We now construct ladder operators
 \begin{eqnarray}
A X^{+}=A a^\dagger_3-H_{+}a_3 \quad \text{and}\quad X^{-}A= a_3 A-a^\dagger_3 H_{-}.
\end{eqnarray}
Then on the basis vectors $|n,l,m\rangle$, 
\begin{eqnarray}
&H_{+}|n,l,m\rangle&=\sqrt{(n+1)(n+l+\frac{3}{2})} |n+1,l,m\rangle,
\\&
H_{-}|n,l,m\rangle &=\sqrt{n(n+l+\frac{1}{2})} |n-1,l,m\rangle,
\\
&a_3|n,l,m\rangle & =c_0(n, l-1, m) |n,l-1,m\rangle +c_1(n-1, l+1,m)\nonumber\\&&\quad\times |n-1,l+1,m\rangle,
\\&
a^\dagger_3|n,l,m\rangle &=c^{*}_0(n, l, m) |n,l+1,m\rangle +c^{*}_1(n, l,m)|n+1,l-1,m\rangle,\nonumber\\&&
\\
&AX^{+}|n,l,m\rangle &=(l+\frac{3}{2})c^{*}_0(n,l,m)|n,l+1,m\rangle,
\\&
X^{-}A|n,l,m\rangle &=(l+\frac{1}{2})c_0(n,l-1,m)|n,l-1,m\rangle,  
\end{eqnarray}
where
\begin{eqnarray*}
&&c_0(n,l,m)=-i\sqrt{\frac{(2n+2l+3)(l-m+1)(l+m+1)(l-Q+1)(l+Q+1)}{\omega(2l+1)(2l+3)}},
\\&&
c_1(n,l,m)=i\sqrt{\frac{2(n+1)(l-m)(l+m)(l-Q)(l+Q)}{\omega(2l-1)(2l+1)}}.
\end{eqnarray*}

\subsection{Integrals of motion, algebra structure and unirreps}
We now take the combinations 
\begin{eqnarray}
D_1 = H_{+}(X^{-}A)^2(B-2), \quad D_2=(B-2)(AX^{+})^2 H_{-} 
\end{eqnarray}
whose action on the basis vectors show the raising and lowering of quantum numbers while preserving energy $E$. We have 
\begin{eqnarray}
D_1|n,l,m\rangle &=&(l-\frac{3}{2})(l-\frac{1}{2})(l+\frac{1}{2})\sqrt{(n+1)(n+l-\frac{1}{2})} c_0(n,l-1,m)\nonumber\\&&\times c_0(n,l-2,m)|n+1,l-2,m\rangle,
\\
D_2|n,l,m\rangle &=&(l+\frac{1}{2})(l+\frac{3}{2})(l+\frac{5}{2})\sqrt{n(n+l+\frac{1}{2})} c^{*}_0(n-1,l,m)\nonumber\\&&\times c^{*}_0(n-1,l+1,m)|n-1,l+2,m\rangle.  
\end{eqnarray}
We can also obtain the action of the operators $D_1D_2$ and $D_2D_1$ on the basis vectors. Then together with (\ref{fh1}), (\ref{fl1}) and (\ref{fl2}), we can conclude that on the operator level,
\begin{eqnarray}
&&[D_1, H]=0=[D_2, H], \quad [D_1, L_3]=0=[D_2, L_3],\label{fkffh1}
\\&&
[B, D_1]= -2D_1,\quad \qquad [B, D_2]= 2D_2,\label{fkffh2}
\end{eqnarray}
\begin{eqnarray}
&D_1D_2&=\frac{B(B+2)}{16384\omega^6 }[2B-2L_3-1][2B-2L_3+1][2B+2L_3-1]\nonumber\\&&\quad\times [2B+2L_3+1][2B-2Q-1][2B-2Q+1][2B+2Q-1]\nonumber\\&&\quad\times[2B+2Q+1][H+\omega B-\omega]^2[H-\omega B-\omega][H+\omega B+\omega],\nonumber\\&&\quad\label{fkffh3}
\end{eqnarray}
\begin{eqnarray}
&D_2D_1&=\frac{(B-2)B}{16384\omega^6 }[2B-2L_3-3][2B-2L_3-1][2B+2L_3-3]\nonumber\\&&\quad\times [2B+2L_3-1][2B-2Q-3][2B-2Q-1][2B+2Q-3]\nonumber\\&&\quad\times[2B+2Q-1][H+\omega B-3\omega]^2[H-\omega B+\omega][H+\omega B-\omega].\nonumber\\&&\quad\label{fkffh4}
\end{eqnarray}
Thus $D_1$, $D_2$ and $B$ form a higher-order polynomial algebra with central elements $H$ and $L_3$.
In order to derive the spectrum using the polynomial algebra, we realize this algebra in terms of deformed oscillator algebra \cite{das1, das2} $\{\aleph, b^{\dagger}, b\}$ of the form
\begin{eqnarray}
[\aleph,b^{\dagger}]=b^{\dagger},\quad [\aleph,b]=-b,\quad bb^{\dagger}=\Phi (\aleph+1),\quad b^{\dagger} b=\Phi(\aleph).\label{fkpfh}
\end{eqnarray}
Here $\aleph $ is the number operator and $\Phi(x)$ is well behaved real function satisfying 
\begin{eqnarray}
\Phi(0)=0, \quad \Phi(x)>0, \quad \forall x>0.\label{fkpbc}
\end{eqnarray}
We rewrite ((\ref{fkffh1})-(\ref{fkffh4})) in the form of deformed oscillator (\ref{fkpfh}) by letting $\aleph=\frac{B}{2}$, $b=D_1$ and $b^{\dagger}=D_2$. We then obtain the structure function
\begin{eqnarray}
&\Phi(x;u,E)&=\frac{(2x+u)(2x+u-2)}{16384\omega^6}[E+\omega(2x+u-3)]^2[E-\omega(2x+u-1)]\nonumber\\&&\times [E+\omega(2x+u-1)][2(u+2x)-2L_3-3][2(u+2x)-2L_3-1]\nonumber\\&&\times [2(u+2x)+2L_3-3][2(u+2x)+2L_3-1][2(u+2x)-2Q-3]\nonumber\\&&\times [2(u+2x)-2Q-1][2(u+2x)+2Q-3][2(u+2x)+2Q-1],\nonumber\\&&
\end{eqnarray}
where $u$ is arbitrary constant. In order to obtain the $(p+1)$-dimensional unirreps, we should impose the following constraints on the structure function
\begin{equation}
\Phi(p+1; u,E)=0,\quad \Phi(0;u,E)=0,\quad \Phi(x)>0,\quad \forall x>0,\label{fpro2}
\end{equation}
where $p$ is a positive integer. These constraints give $(p+1)$-dimensional unitary representations and their solution gives the energy $E$ and the arbitrary constant $u$. We have the following possible constant $u$ and energy spectra $E$, for the constraints $\varepsilon_1=\pm 1$, $\varepsilon_2=\pm 1$, $\varepsilon_3=\pm 1$:
\begin{eqnarray}
&&u=\frac{1}{2}(1+2\varepsilon_1 m), \quad E=\frac{\varepsilon_2 \omega}{2}[2+\varepsilon_3(1+4p)+2\varepsilon_1 m];\label{fe1}
\\&&
u=\frac{1}{2}(1+2\varepsilon_1 Q), \quad E=\frac{\varepsilon_2 \omega}{2}[2+\varepsilon_3(1+4p)+2\varepsilon_1 Q] ;
\\&&
u=\frac{1}{\omega}(\omega+\varepsilon_1 E), \quad E=\frac{\varepsilon_2 \omega}{2}(3+2p+2\varepsilon_1 m);\label{fe2}
\\&&
u=\frac{1}{\omega}(\omega+\varepsilon_1 E), \quad E=\frac{\varepsilon_2 \omega}{2}(3+2p+2\varepsilon_1 Q).
\end{eqnarray}
Making the identification $p=n$, $l=m$, $\varepsilon_1=1$, $\varepsilon_2=1$, $\varepsilon_3=1$, the energy spectra (\ref{fe1}) and (\ref{fe2}) coincide with the physical spectra (\ref{fpe1}). The physical wave functions involve other quantum numbers and we have in fact the degeneracy of $p$ only when these other quantum numbers would be fixed.

\section{MIC-harmonic oscillator with monopole in generalized Taub-NUT space}
Let us consider the generalized Taub-NUT metric in $\mathbb{R}^3$ 
\begin{eqnarray}
ds^2=f(r)dl^2+g(r)(d\psi+A_i d\textbf{r})^2,\label{fmc1}
\end{eqnarray}
where 
\begin{eqnarray}
&&f(r)=a r^2+b, \quad g(r)=\frac{r^2(ar^2+b)}{1+c_1 r^2+d r^4},\label{ffg1}
\\&&
A_1=\frac{-y}{r(r+z)}, \quad A_2=\frac{x}{r(r+z)}, \quad A_3=0,
\end{eqnarray}
$r=\sqrt{x^2+y^2+z^2}$ and the three dimensional Euclidean line element $dl^2=dx^2+dy^2+dz^3$, $a$, $b$, $c_1$, $d$ are constants. Here $\psi$ is the additional angular variable which describes the relative phase and its coordinate is cyclic with period $4\pi$ \cite{cor1, gro1}. The functions $f(r)$ and $g(r)$ in the metric represent gravitational effects and $A_i$ are components of the potential associated with the magnetic monopole field.

We consider the Hamiltonian associated with (\ref{fmc1}) 
\begin{eqnarray}
H=\frac{1}{2}\left[\frac{1}{f(r)}\left\{\textbf{p}^2+\frac{c_0 r^2}{2}+c_4\right\}+\frac{Q^2}{g(r)}\right]\label{fkf1},
\end{eqnarray}
where $c_0$ and $c_4$ are constants and the operators 
\begin{eqnarray}
p_i=-i(\partial_i-iA_i Q), \quad Q=-i\partial_\psi
\end{eqnarray}
satisfying the following commutation relations
\begin{eqnarray}
[p_i,p_j]=i\epsilon_{ijk}M_k Q, \quad [p_i,Q]=0,\quad \textbf{M}=\frac{\textbf{r}}{r^3}.
\end{eqnarray} 
The system with Hamiltonian (\ref{fkf1}) is a Hartmann system in a curved Taub-NUT space with abelian monopole interaction. This new system is referred to as MIC-harmonic oscillator monopole system. In this section, we solve the Schr\"{o}dinger St\"{a}ckel equivalent of the system (\ref{fkf1}) in spherical coordinates, derive the recurrence relations and construct higher order integrals and the corresponding higher order polynomial algebra.

\subsection{Separation of variables}
Let us consider the spherical coordinates 
\begin{eqnarray}
&&x=r \sin\theta\cos\phi,\quad y=r\sin\theta\sin\phi,\quad z=r\cos\theta,
\end{eqnarray}
where $r>0$, $0\leq\theta\leq\pi$ and $0\leq \phi\leq 2\pi$.
In terms of these coordinates, the Taub-NUT metric (\ref{fmc1}) takes on the form 
\begin{eqnarray}
ds^2=f(r)(dr^2+r^2d\theta^2+r^2\sin^2\theta d\phi^2)+g(r)(d\psi+\cos\theta d\phi)^2,\label{fmc2}
\end{eqnarray}
\begin{eqnarray}
A_1=-\frac{1}{r}\tan\frac{\theta}{2}\sin\phi,\quad A_2=\frac{1}{r}\tan\frac{\theta}{2}\cos\phi,\quad A_3=0.
\end{eqnarray}
The Schr\"{o}dinger equation of the model (\ref{fkf1}) is
\begin{eqnarray} 
&H\Psi(r,\theta,\phi,\psi)&=\frac{-1}{2(a r^2+b)}\left[\frac{\partial^2}{\partial r^2}+\frac{2}{r}\frac{\partial}{\partial r}-\frac{c_0 r^2}{2}-c_4\right.\nonumber\\&&\left.+\frac{1}{r^2}\left(\frac{\partial^2}{\partial\theta^2}+\cot\theta\frac{\partial}{\partial\theta}+\frac{1}{\sin^2\theta}\frac{\partial^2}{\partial\phi^2}\right)+\left(\frac{1}{r^2\cos^2\frac{\theta}{2}}+c_1+d r^2\right)\right.\nonumber\\&&\left.\times\frac{\partial^2}{\partial\psi^2}-\frac{1}{r^2\cos^2\frac{\theta}{2}}\frac{\partial}{\partial\phi}\frac{\partial}{\partial\psi}\right] \Psi(r,\theta,\phi,\psi) =E\Psi(r,\theta,\phi,\psi).\label{fkf2}
\end{eqnarray}
We can write $\Psi(r,\theta,\phi,\psi)=\chi(r,\theta)e^{i(\nu_1\phi+\nu_2\psi)}$. Then we obtain the equivalent system of (\ref{fkf2}) as
\begin{eqnarray} 
&H'\chi(r,\theta)e^{i(\nu_1\phi+\nu_2\psi)}&=\left[\frac{\partial^2}{\partial r^2}+\frac{2}{r}\frac{\partial}{\partial r}+\left(2aE-d\nu_2^2-\frac{c_0 }{2}\right)r^2\right.\nonumber\\&&\left.+\frac{1}{r^2}\left(\frac{\partial^2}{\partial\theta^2}+\cot\theta\frac{\partial}{\partial\theta}-\frac{\nu_1^2}{\sin^2\theta}\right)-\frac{\nu_2^2}{r^2\cos^2\frac{\theta}{2}}+\frac{\nu_1\nu_2}{r^2\cos^2\frac{\theta}{2}}\right]\nonumber\\&&\times\chi(r,\theta)e^{i(\nu_1\phi+\nu_2\psi)} =E'\chi(r,\theta)e^{i(\nu_1\phi+\nu_2\psi)},\label{fkff2}
\end{eqnarray}
where $E'=c_4+c_1\nu_2^2-2bE$. The original energy parameter $E$ now plays as the role of model parameter and the model parameter $c_4+c_1\nu_2^2-2bE$ plays the role of energy $E'$. This change in the role of the parameters is called coupling constant metamorphosis. Moreover, the model (\ref{fkff2}) is related to the one in (\ref{fkf2}) by St\"{a}ckel transformation and thus the two systems are St\"{a}ckel equivalent \cite{boy2, kal6}.

By making the Ansatz,
\begin{eqnarray}
\Psi(r,\theta,\phi,\psi)=R(r)\Theta(\theta)e^{i(\nu_1\phi+\nu_2\psi)},
\end{eqnarray}
(\ref{fkff2}) becomes the radial and angular ordinary differential equations
\begin{eqnarray}
&&\left[\frac{\partial^2}{\partial r^2}+\frac{2}{r}\frac{\partial}{\partial r}-E'+(2aE-d\nu_2^2-\frac{c_0 }{2}) r^2 -\frac{k_1}{r^2}\right ]R(r)=0,\label{fkf5}
\\
&&\left[\frac{\partial^2}{\partial\theta^2}+\cot\theta\frac{\partial}{\partial\theta}+\left\{k_1-\frac{(\nu_1-2\nu_2)^2}{2(1+\cos\theta)}-\frac{\nu_1^2}{2(1-\cos\theta)}\right\}\right]\Theta(\theta)=0,\label{fkf6}
\end{eqnarray}
where $k_1$ is separable constant. 

We now turn to (\ref{fkf6}), which can be converted, by setting $z=\cos\theta$ and $\Theta(z)=(1+z)^{a}(1-z)^{b} Z(z)$, to
\begin{eqnarray}
&&(1-z^2)Z''(z)+\{2a-2b-(2a+2b+2)z\}Z'(z)\nonumber \\&&\qquad +\{k_1-(a+b)(a+b+1)\}Z(z)=0,\label{fpr5}
\end{eqnarray}
where $2a=\nu_1-2\nu_2$, $2b=\nu_1$.
Comparing (\ref{fpr5}) with the Jacobi differential equation 
\begin{equation}
(1-x^{2})y''+\{\beta_1-\alpha_1-(\alpha_1+\beta_1+2)x\}y'+\lambda(\lambda+\alpha_1+\beta_1+1)y=0,\label{fJd1}
\end{equation}
we obtain the  separation constant 
\begin{equation}
k_1=(l-\nu_2)(l-\nu_2+1),\label{fpr2} 
\end{equation}
where $l=\lambda+\nu_1$. 
Hence the solutions of (\ref{fpr5}) are given in terms of the Jacobi polynomials as
\begin{eqnarray}
\Theta(\theta)&\equiv &\Theta_{l \nu_1}(\theta; \nu_{1}, \nu_{2})
= F_{l \nu_1}(\nu_{1}, \nu_{2})(1+\cos\theta)^{\frac{(\nu_1-2\nu_2)}{2}}(1-\cos\theta)^{\frac{\nu_1}{2}}\nonumber\\&&\quad\times P^{(\nu_1, \nu_1-2\nu_2)}_{l-\nu_1}(\cos\theta),\label{fjp1}
\end{eqnarray}
where $P^{(\alpha, \beta)}_{\lambda}$ denotes a Jacobi polynomial \cite{mag1}, $F_{l \nu_1}(\nu_{1}, \nu_{2})$ is the normalized constant and $l\in \mathbb{N}$.

The radial equation (\ref{fkf5}) can be converted, by setting  
 $z=\varepsilon r^2$, $R(z)=z^{\frac{1}{2}( l-\nu_2)} e^{-\frac{z}{2}}R_1(z)$ and $\varepsilon^2=\frac{c_0}{2}-2aE+d\nu_2^2$, to
\begin{eqnarray}
&&z\frac{d^2R_1(z)}{dz^2}+\left\{(l-\nu_2+\frac{3}{2})-z\right\}\frac{dR_1(z)}{dz}-\left\{\frac{1}{2}(l-\nu_2+\frac{3}{2})+\frac{E'}{4\varepsilon}\right\}R_1(z)=0.\nonumber\\&&\label{fan11}
\end{eqnarray}
Set  
\begin{eqnarray}
 n=\frac{\nu_{2}}{2}-\frac{E'}{4\varepsilon}-\frac{l}{2}-\frac{3}{4}.\label{fan12}
\end{eqnarray}
Then (\ref{fan11}) can be identified with the Laguerre differential equation. Hence we can write the solution of (\ref{fkf5}) in terms of the confluent hypergeometric polynomial as  
\begin{eqnarray}
&R(r)&\equiv R_{nl}(r;\nu_{1}, \nu_{2})=F_{nl}(\nu_{1},\nu_{2})(\varepsilon r^2)^{\frac{1}{2}(l-\nu_2)} e^{\frac{-\varepsilon r^2}{2}}\nonumber\\&&
\quad \times {}_1 F_1(-n, l-\nu_2+\frac{3}{2}; \varepsilon r^2),\label{fan14}
\end{eqnarray}
where $F_{nl}(\nu_{1},\nu_{2})$ is the normalized constant.
In order to have a discrete spectrum the parameter $n$ needs to be positive integer. From (\ref{fan12}) 
\begin{equation}
\varepsilon=\frac{-E'}{4n+2l-2\nu_{2}+3}\label{fen1}
\end{equation} and hence the energy spectrum of the system (\ref{fkf1}) is given by
\begin{equation}
\frac{2bE-c_1\nu_2^2-c_4 }{\sqrt{\frac{c_0}{2}-2a E+d\nu_2^2}}=4n+2l-2\nu_2+3,\quad n=1, 2, 3,\dots\label{fen1}
\end{equation}

\subsection{Ladder operators and recurrence approach}
The solutions of the eigenfunction for the equation $H'\Psi=E'\Psi$ of the form $\Psi(r,\theta,\phi,\psi)=\psi^{l-\nu_2+\frac{1}{2}}_{n}\Theta^{( \nu_1,\nu_1-2\nu_2)}_{l-\nu_1} e^{i(\nu_1\phi+\nu_2\psi)}$ are given by
\begin{eqnarray}
&&\psi^{l-\nu_2+\frac{1}{2}}_{n}=e^{\frac{-\varepsilon r^2}{2}}r^{l-\nu_2} L^{l-\nu_2+\frac{1}{2}}_n
(\varepsilon r^2),
\\&&
\Theta^{(\nu_1, \nu_1-2\nu_2)}_{l-\nu_1}=\sin^{\nu_1}{\frac{\theta}{2}}\cos^{\nu_1-2\nu_2}{\frac{\theta}{2}}P^{( \nu_1, \nu_1-2\nu_2)}_{l-\nu_1}(\cos\theta),
\end{eqnarray} 
where $L^\alpha_n$ is the $n$th order Laguerre polynomial, $P^{(\beta,\gamma)}_\lambda$ is the Jacobi polynomial \cite{mag1},  $n=\frac{\nu_2}{2}-\frac{E'}{4\varepsilon}-\frac{3}{4}-\frac{l}{2}$ and $\varepsilon^2=\frac{c_0}{2}-2aE+d\nu_2^2$. The energy eigenvalues of the equation $H'\Psi=E'\Psi$ is
\begin{eqnarray}
E'=-\varepsilon(4n+2l-2\nu_2+3).
\end{eqnarray}
Let us now construct the ladder operators based on radial part of the separated solutions using differential identities for Laguerre functions \cite{mag1}
\begin{eqnarray}
&&K^{+}_{l-\nu_2+\frac{1}{2},n}=\frac{1}{r}(B-Q+1)\partial_{r}-\frac{H'}{2}+\frac{1}{r^2}(B-Q+1)(B-Q-\frac{1}{2}),
\nonumber\\&&
K^{-}_{l-\nu_2+\frac{1}{2},n}=-\frac{1}{r}(B-Q-1)\partial_{r}-\frac{H'}{2}-\frac{1}{r^2}(B-Q-1)(B-Q+\frac{1}{2})
\end{eqnarray}
and the shift operators based on the angular functions using Jacobi function identities \cite{mag1}
\begin{eqnarray}
&J^{+}_{l-\nu_1} &=-2(B-Q+\frac{1}{2})\sin\theta \partial_\theta-2(B-Q+\frac{1}{2})^2\cos\theta -2Q(L_3-Q),\nonumber
\\
&J^{-}_{l-\nu_1} &=2(B-Q-\frac{1}{2})\sin\theta \partial_\theta-2(B-Q-\frac{1}{2})^2\cos\theta -2Q(L_3-Q).
\end{eqnarray}
Here $B=\sqrt{\textbf{L}^2+\frac{1}{4}}$ as in section 2.
The action of the operators on the corresponding wave functions provide the following recurrence formulas
\begin{eqnarray}
&&K^{+}_{l-\nu_2+\frac{1}{2},n}\psi^{l-\nu_2+\frac{1}{2}}_{n}=-2\varepsilon^2 \psi^{l-\nu_2+\frac{5}{2}}_{n-1},\nonumber
\\&&
K^{-}_{l-\nu_2+\frac{1}{2},n}\psi^{l-\nu_2+\frac{1}{2}}_{n}=- 2(n+1)(n+l-\nu_2+\frac{1}{2}) \psi^{l-\nu_2-\frac{3}{2}}_{n+1},\nonumber
\\
&&J^{+}_{l-\nu_1} \Theta^{(\nu_1, \nu_1-2\nu_2)}_{l-\nu_1}=-2(l-\nu_1+1)(l-\nu_1-2\nu_2+1)\Theta^{(\nu_1, \nu_1-2\nu_2)}_{l-\nu_1+1},\nonumber
\\&&
J^{-}_{l-\nu_1} \Theta^{(\nu_1, \nu_1-2\nu_2)}_{l-\nu_1}=-2l(l-2\nu_2)\Theta^{(\nu_1, \nu_1-2\nu_2)}_{l-\nu_1-1}.
\end{eqnarray}

\subsection{Integrals of motion, algebra structure and spectrum}
Let us now consider the suitable operators 
\begin{eqnarray}
D_1=K^{+}_{l-\nu_2+\frac{1}{2},n}J^{+}_{l-\nu_1+1}J^{+}_{l-\nu_1} B, \quad D_2=BJ^{-}_{l-\nu_1}J^{-}_{l-\nu_1-1}K^{-}_{l-\nu_2+\frac{1}{2},n}.
\end{eqnarray}
The explicitly action of the operators $D_i, i=1, 2$ on the wave functions is given by
\begin{eqnarray}
D_1\Psi(r,\theta,\phi,\psi)&=&-8\varepsilon^2l(l+1)(l-\nu_1+1)(l-\nu_1-2\nu_2+1)(l-\nu_1+2)\nonumber\\&&\times(l-\nu_1-2\nu_2+2) \psi^{l-\nu_2+\frac{5}{2}}_{n-1}\Theta^{(\nu_1, \nu_1-2\nu_2)}_{l-\nu_1+2}e^{i(\nu_1\phi+\nu_2\psi)},
\\
D_2\Psi(r,\theta,\phi,\psi)&=&-8l(l-1)(l-2\nu_2)(l-2\nu_2-1)(l-\frac{3}{2})(n+1)\nonumber\\&&\times(n+l-\nu_2+\frac{1}{2}) \psi^{l-\nu_2-\frac{3}{2}}_{n+1}\Theta^{(\nu_1, \nu_1-2\nu_2)}_{l-\nu_1-2}e^{i(\nu_1\phi+\nu_2\psi)}.
\end{eqnarray}
We can also obtain the action of the operators $D_1D_2$ and $D_2D_1$ on the wave functions. It follows in the operator from construction they form an algebraically independent set of differential operators and there has a common feature of superintegrable systems to close polynomially symmetry algebra. Direct computation shows that they form higher order polynomial algebra 
\begin{eqnarray}
&&[D_1, H']=0=[D_2, H'],\label{fff1}
\\&& [B, D_1]= 2D_1, \quad [B, D_2]=-2D_2,\label{fff2}
\end{eqnarray}
\begin{eqnarray}
&D_1D_2&=\frac{B-2}{256}(2B-5)(2B-3)^2(2B-2L_3-4Q-1)\nonumber\\&&\times (2B-2L_3-1)(2B-4Q-3)(2B-4Q-1)\nonumber\\&&\times(2B-2L_3-4Q-3)(2B-1)(2B-2L_3-3)\nonumber\\&&\times [H'-2\varepsilon(B-Q-1)][H'+2\varepsilon(B-Q-1)],\label{fff3}
\end{eqnarray}
\begin{eqnarray}
&D_2D_1&=\frac{B}{256}(2B-1)(2B+1)^2 (2B-2L_3-4Q+3)\nonumber\\&&\times (2B-2L_3+3)(2B-4Q+1)(2B-4Q+3)\nonumber\\&&\times(2B-2L_3-4Q+1)(2B+3)(2B-2L_3+1)\nonumber\\&&\times [H'-2\varepsilon(B-Q+1)][H'+2\varepsilon(B-Q+1)].\label{fff4}
\end{eqnarray}
We rewrite ((\ref{fff1})-(\ref{fff4})) in the form of deformed oscillator algebra (\ref{fkpfh}) by letting $\aleph=\frac{B}{2}$,  $b^{\dagger}=D_1$ and $b=D_2$. We then obtain structure function
\begin{eqnarray}
&\Phi(x;u,H')&=\frac{(2x+u-2)}{256}[2(2x+u)-3]^2[2(2x+u)-2L_3-1]\nonumber\\&&\times[2(2x+u)-1] [2(2x+u)-2L_3-3][2(2x+u)-5]\nonumber\\&&\times[2(2x+u)-2L_3-4Q-3][H'-2\varepsilon\{(2x+u)-Q-1\}]\nonumber\\&&\times[2(2x+u)-4Q-1][H'+2\varepsilon\{(2x+u)-Q-1\}]\nonumber\\&&\times [2(2x+u)-4Q-3][2(2x+u)-2L_3-4Q-1] ,
\end{eqnarray}
where $u$ is an arbitrary constant to be determined. We should impose the following constraints on the structure function in order to obtain a finite dimensional unirreps, 
\begin{equation}
\Phi(p+1; u,E')=0,\quad \Phi(0;u,E')=0,\quad \Phi(x)>0,\quad \forall x>0,\label{fpro2}
\end{equation}
where $p$ is a positive integer. These constraints give $(p+1)$-dimensional unitary representations and their solutions give the energy $E'$ and the arbitrary constant $u$. We have all the possible energy spectra and structure functions as 
\begin{eqnarray}
u=\frac{\varepsilon_1 E'+2\varepsilon(1+\nu_2)}{2\varepsilon}, \quad E'=-\varepsilon(4p-2\nu_1+2\varepsilon_2\nu_2+3)
\end{eqnarray}
\begin{eqnarray}
&\Phi(x)&=2\varepsilon^2[2x-2p+(\varepsilon_2-1)\nu_1-2][2x-2p+(\varepsilon_2-1)\nu_1-2\nu_2-2]\nonumber\\&&\times[4x-(4p+2\varepsilon_2\nu_1+2\nu_2+3)(1+\varepsilon_1)][2x-2p+\varepsilon_2\nu_1-2\nu_2-1]\nonumber\\&&\times[4x-(4p+2\varepsilon_2\nu_1+2\nu_2+3)(1-\varepsilon_1)][2x-2p+\varepsilon_2\nu_1-2\nu_2-2]\nonumber\\&&\times[2x-2p+\varepsilon_2-2][2x-2p+\varepsilon_2\nu_1-3][2x-2p+\varepsilon_2\nu_1-1]\nonumber\\&&\times[4p-2x+2\varepsilon_2\nu_1+5][2x-2p+\varepsilon_2\nu_1-2]^2\nonumber\\&&\times[2x-2p+(\varepsilon_2-1)\nu_1-2\nu_2-1],
\end{eqnarray}
where $\varepsilon_1=\pm 1$, $\varepsilon_2=\pm 1$. The coupling constant metamorphosis provides $E'\leftrightarrow c_4+c_1\nu_2^2-2bE$ and $\varepsilon^2\leftrightarrow \frac{c_0}{2}-2aE+d\nu_2^2$. Hence we have the energy  of the original Hamiltonian
\begin{eqnarray}
\frac{2bE-c_1\nu_2^2-c_4 }{\sqrt{\frac{c_0}{2}-2a E+d\nu_2^2}}=4p-2\nu_1+2\varepsilon_2 \nu_2+3.\label{fen2}
\end{eqnarray}
Making the identifications $p=n$, $-\nu_1=l$, $\varepsilon_1=1$, $\varepsilon_2=1$, then (\ref{fen2}) coincides with the physical spectra (\ref{fen1}).

\section{Conclusion}
One of the main results of this chapter is the construction via recurrence method of the higher order integrals of motion and higher order polynomial algebra for the MIC-harmonic oscillator systems with monopole interactions in both flat space and curved Taub-NUT space. The method is systematic and well constructed based on wave functions of the systems. To our knowledge this is the first application of the recurrence approach in superintegrable monopole systems.

Let us point out that superintegrable systems with monopole interactions and their polynomial algebras are largely unexplored area \cite{mar10}. It is interesting to generalize the results to systems with non-abelian monopole interactions.


 \chapter{Superintegrable systems from exceptional orthogonal polynomials}\label{ch10}

{\bf \large{Acknowledgement}}
\\This chapter is based on the work that was published in  Ref. \cite{fh8}. I have incorporated text of that paper \cite{fh8}. In this chapter, we introduce an extended Kepler-Coulomb quantum model in spherical coordinates. The Schr\"{o}dinger equation of this Hamiltonian is solved in these coordinates and it is shown that the wave functions of the system can be expressed in terms of Laguerre, Legendre and exceptional Jacobi polynomials (of hypergeometric type). We construct ladder and shift operators based on the corresponding wave functions and obtain their recurrence formulas. These recurrence relations are used to construct higher-order, algebraically independent integrals of motion to prove superintegrability of the Hamiltonian. The integrals form a higher rank polynomial algebra. By constructing the structure functions of the associated deformed oscillator algebras we derive the degeneracy of energy spectrum of the superintegrable system.

\section{Introduction}
Many families of exceptional orthogonal polynomials have been successfully used to construct new superintegrable systems, higher order integrals of motion and higher order polynomial algebras \cite{pos3, mar17, mar18, mar13, mar14}. In this chapter, we use the recurrence approach to extend the three parameter Kepler-Coulomb system \cite{kal11}.  

The exceptional orthogonal polynomials (EOP) were first explored in \cite{gom1, gom2}. These polynomials form complete, orthogonal systems extending the classical orthogonal polynomials of Hermite, Laguerre and Jacobi. More recently much research has been done extending the theory of  EOPs in various directions in mathematics and physics, in particular, exactly solvable quantum mechanical problems for describing bound states \cite{gom3, dut1, gra1, gra2, leva1, oda1, que1, que2, ses1} and scattering states \cite{ho1, yad1, yad2, yad3}, diffusion equations and random processes \cite{ho2, ho3, cho1}, quantum information entropy \cite{dut2}, exact solutions to Dirac equation \cite{sch1}, Darboux transformations \cite{que3, oda1, oda2, que1, gom4, sas1, ho4} and finite-gap potentials \cite{gom5}. Recent progress has been made constructing systems relating superintegrability and supersymmetric quantum mechanics with exceptional orthogonal polynomials \cite{pos3, mar15}.

The research for superintegrable systems with second-order integrals in conformally flat spaces started in the mid sixties \cite{fri2}. Over the last decade the topic of superintegrability has become an attractive area of research as these systems possess many desirable properties and can be found throughout various subjects in mathematical physics. For a detailed list of references on superintegrability, we refer the reader to the review paper \cite{mil1}. One systematic approach to superintegrability is to derive spectra of 2D superintegrable systems based on quadratic and cubic algebras involving three generators \cite{das2, mar19, mar20}.  In particular, the method of realization in the deformed oscillator algebras \cite{das1} has been effective for obtaining finite dimensional unitary representations  \cite{mar19, mar12}. In fact, this approach was extended to classes of higher order polynomial algebras with three generators \cite{isa1}  as well as higher rank polynomial algebras of superintegrable systems in higher dimensional spaces \cite{fh1, fh2}. However, it is quite involved to apply the direct approach to obtain the corresponding polynomial algebras, Casimir operators and deformed oscillator algebras.

These difficulties can be overcome using a constructive approach based on eigenfunctions of the models. This approach is a useful tool to construct well-defined integrals of motion in classical and quantum mechanical problems. Many papers were devoted to construct integrals of motion and their corresponding higher order symmetry algebras based on lower-(first and second) ones \cite{jau1, fri2, boy1, eva3, mar6} and higher-order ladder operators \cite{mar13, mar14, kre1, adl1, jun1, dem1, mar3, rag1, mar9} in various aspects. In fact, the constructive approach has shown a close connection with special functions and (exceptional) orthogonal polynomials \cite{pos3, mar15, kal7, cal1, cal2, cal3, fh3, fh6}.

In this chapter, we introduce a new exactly solvable Hamiltonian system in 3D, which is a singular deformation of the Coulomb potential. Its wave functions are given as products of Laguerre, Legendre and exceptional Jacobi polynomials. We show that the system is superintegrable by constructing integrals of the motion using the recurrence relation approach. The symmetry algebra enables us to give an algebraic derivation for the energy spectrum.

\section{Extended Kepler-Coulomb system}
Consider the generalization of the three parameter Kepler-Coulomb Hamiltonian \cite{kal11} in spherical coordinates 
\begin{eqnarray}
&&H=\frac{1}{2}{\bf p}^2 -\frac{\alpha}{2r}+\frac{1}{2r^2\sin^2\theta}\left[\frac{\gamma^2-\frac{1}{4}}{4\sin^2\frac{\phi}{2}}+\frac{\delta^2-\frac{1}{4}}{4\cos^2\frac{\phi}{2}} +\frac{2(1-b\cos\phi)}{(b-\cos\phi)^2}\right],\label{Nh1}
\end{eqnarray}
where $p_i=-i\partial_i$, $b=\frac{\delta+\gamma}{\delta-\gamma}$, $\gamma\neq \delta$ and $\alpha, \gamma, \delta$ are three real constants. 
The Schr\"{o}dinger equation $H\Psi(r,\theta,\phi)=E\Psi(r,\theta,\phi)$ of (\ref{Nh1}) can be expressed as 
\begin{eqnarray}
&&\left[\frac{\partial^2}{\partial r^2}+\frac{2}{r}\frac{\partial}{\partial r}+\frac{\alpha}{r}+2E+\frac{1}{r^2}\left\{\frac{\partial^2}{\partial\theta^2}+\cot\theta\frac{\partial}{\partial\theta}  \right\}\nonumber\right.\\&&\left.+ \frac{1}{r^2\sin^2\theta}\left\{ \frac{\partial^2}{\partial\phi^2} -\frac{\gamma^2-\frac{1}{4}}{4\sin^2\frac{\phi}{2}} -\frac{\delta^2-\frac{1}{4}}{4\cos^2\frac{\phi}{2}}-\frac{2(1-b\cos \phi)}{(b-\cos\phi)^2}\right\} \right]\Psi(r,\theta,\psi)=0.
\end{eqnarray}
The separation of variable of the Hamiltonian (\ref{Nh1}) for the wave equation $H\Psi=E\Psi$ by the ansatz
\begin{eqnarray}
\Psi(r,\theta,\phi)=R(r)\Theta(\theta)Z(\phi)
\end{eqnarray}
provides the following radial and angular ordinary differential equations
\begin{eqnarray}
&&\left[\frac{d^2}{dr^2}+\frac{2}{r}\frac{d}{dr}+\frac{\alpha}{r}+2E-\frac{k_2}{r^2}\right]R(r)=0, \label{Ra1}
\\&&
\left[\frac{d^2}{ d\theta^2}+\cot\theta\frac{d}{d\theta}-\frac{k_1}{\sin^2\theta} +k_2\right]\Theta(\theta)=0,\label{JP1}
\\&&
\left[\frac{d^2}{ d\phi^2} -\frac{\gamma^2-\frac{1}{4}}{4\sin^2\frac{\phi}{2}} -\frac{\delta^2-\frac{1}{4}}{4\cos^2\frac{\phi}{2}}-\frac{2(1-b\cos \phi)}{(b-\cos \phi)^2}+k_1 \right] Z(\phi)=0,\label{EP2}
\end{eqnarray}
where $k_1$, $k_2$ are the associated separation constants.

We now turn to (\ref{EP2}), which can be converted, by setting $z=\cos\phi$, $Z(z)=(z+1)^{\frac{1}{4}(\delta+2)}(z-1)^{\frac{1}{4}(\gamma+2)}(z-b)^{-1}f(z)$, to 
\begin{eqnarray}
&&(z^2-1)\frac{d^2f(z)}{dz^2}+\left\{\gamma-\delta+(\gamma+\delta+2)z-\frac{2(z^2-1)}{z-b}\right\}\frac{df(z)}{dz}\nonumber\\&& +\left\{\frac{1}{4}(\gamma+\delta+1)^2-k_1+\frac{\gamma-\delta+(\gamma+\delta-1)z}{(b-z)}\right\}f(z)=0.\label{Jb3}
\end{eqnarray}
Comparing (\ref{Jb3}) with exceptional Jacobi differential equation \cite{gom1},
\begin{eqnarray}
T^{(\eta,\xi)}(Y)=(n-1)(n+\eta+\xi)Y,\quad n\in \mathbb{N},
\end{eqnarray}
where
\begin{eqnarray}
&&T^{(\eta,\xi)}(Y)=(X^2-1)Y''+2A\left(\frac{1-B X}{B-X}\right)\{(X-C)Y'-Y\},
\nonumber\\&&
A=\frac{1}{2}(\xi-\eta), \quad B=\frac{\xi+\eta}{\xi-\eta}, \quad C=B+\frac{1}{A},
\end{eqnarray}
we obtain $\gamma=\xi$, $\delta=\eta$ and the separation constant
\begin{eqnarray}
&& k_1=\left(n+\frac{\gamma+\delta-1}{2}\right)^2.\label{Sp2}
\end{eqnarray}
Hence the solutions of (\ref{Jb3}) are given in terms of the exceptional Jacobi polynomials $\hat{P}^{(\delta,\gamma)}_n$ \cite{pos3, gom1} as 
\begin{eqnarray}
Z(\phi)\equiv F_n(\gamma,\delta)\frac{(\cos \phi+1)^{\frac{1}{4}(2\delta+1)}(\cos \phi-1)^{\frac{1}{4}(2\gamma+1)}}{(\cos\phi-b)}\hat{P}^{(\delta,\gamma)}_n(\cos\phi),
\end{eqnarray}
These EOP  are related to their standard Jacobi polynomials $P^{(\delta,\gamma)}_n$ \cite{and1} via 
\begin{eqnarray}
\hat{P}^{(\delta,\gamma)}_n=-\frac{1}{2}(\cos\phi-b)P^{(\delta,\gamma)}_{n-1}+\frac{bP^{(\delta,\gamma)}_{n-1}-P^{(\delta,\gamma)}_{n-2}}{\delta+\gamma+2n-2}.
\end{eqnarray}
Using (\ref{Sp2}) in the angular part (\ref{JP1}), we have
\begin{eqnarray}
\left[\frac{d^2}{ d\theta^2}+\cot\theta\frac{d}{d\theta} -\frac{(n+\frac{\gamma+\delta-1}{2})^2}{\sin^2\theta} +k_2\right]\Theta(\theta)=0.\label{JP2}
\end{eqnarray}
Then (\ref{JP2}) can be converted, by setting $z=\cos\theta$, to 
\begin{eqnarray}
&&\left[(1-z^2)\frac{d^2}{dz^2}-2z\frac{d}{dz}+k_2-\frac{(n+\frac{\gamma+\delta-1}{2})^2}{1-z^2}\right] f(z)=0.\label{Jb5}
\end{eqnarray}
Comparing (\ref{Jb5}) with associated Legendre differential equation
\begin{eqnarray}
(1-x^2)y''-2xy'+\left[m(m+1)-\frac{\mu^2}{1-x^2}\right] y=0,
\end{eqnarray}
we obtain the constants
\begin{eqnarray}
k_2=m(m+1),\quad \mu=n+\frac{\gamma+\delta-1}{2}.\label{Sp4}
\end{eqnarray}
Hence the solutions of (\ref{JP1}) are given in terms of the Legendre polynomials $P^\mu_m$ \cite{and1} as 
\begin{eqnarray}
\Theta(\theta)\equiv F_m(\mu)P^{\mu}_m(\cos\theta),
\end{eqnarray}
where $F_m(\mu)$ is the normalized constant and $m, \mu\in \mathbb{Z}$.

Using (\ref{Sp4}), the radial part (\ref{Ra1}) becomes
\begin{eqnarray}
\left[\frac{d^2}{dr^2}+\frac{2}{r}\frac{d}{dr}+\frac{\alpha}{r}+2E-\frac{m(m+1)}{r^2}\right]R(r)=0. \label{Rd1}
\end{eqnarray}
(\ref{Rd1}) can be converted, by setting $z=\varepsilon r$, $R(z)=z^m e^{-\frac{1}{2}z}f_1(z)$ and $\varepsilon^2=-8E$, to 
\begin{eqnarray}
\left[z\frac{d^2}{dz^2}+(2m+2-z)\frac{d}{dz}+\frac{\alpha}{\varepsilon}-m-1\right]f_1(z)=0.\label{Lg1}
\end{eqnarray}
Set 
\begin{eqnarray}
N=\frac{\alpha}{\varepsilon}-m-1.
\end{eqnarray}
Then (\ref{Lg1}) can be identified with the Laguerre differential equation. Hence the solutions of (\ref{Ra1}) are given in terms of the $N$-th order Laguerre polynomials $L^\beta_N$ \cite{and1} as
\begin{eqnarray}
R(r)\equiv e^{-\frac{\varepsilon r}{2}}(\varepsilon r)^{m}L^{2m+1}_{N}(\varepsilon r).
\end{eqnarray}
Hence the energy spectrum of the model (\ref{Nh1}), $E=\frac{-\varepsilon^2}{8}$, is given by
\begin{eqnarray}
E=-\frac{\alpha^2}{8\left(N+m+1\right)^2}, \quad N=1, 2, 3,\dots \label{energy}
\end{eqnarray}
Here $N$ is the principal quantum number.

\section{Algebraic calculation to the extended Kepler-Coulomb system}
We can rewrite the Hamiltonian of three parameter extended Kepler-Coulomb system in the standard way as a sequence of operators corresponding to separation in spherical coordinates (\ref{Nh1}),
\begin{eqnarray}
&&H=\frac{1}{2}\left[\frac{\partial^2}{\partial r^2}+\frac{2}{r}\frac{\partial}{\partial r}+\frac{\alpha}{r}+\frac{L_\theta}{r^2} \right],\label{Nh2}
\end{eqnarray}
where
\begin{eqnarray}
&&L_\theta=\frac{\partial^2}{\partial\theta^2}+\cot\theta\frac{\partial}{\partial\theta}+ \frac{L_\phi}{\sin^2\theta}, \\&&
L_\phi= \frac{\partial^2}{\partial\phi^2} -\frac{\gamma^2-\frac{1}{4}}{4\sin^2\frac{\phi}{2}} -\frac{\delta^2-\frac{1}{4}}{4\cos^2\frac{\phi}{2}}-\frac{2(1-b\cos \phi)}{(b-\cos\phi)^2}.
\end{eqnarray}
Making a slight change in the definition of these operators, 
\begin{eqnarray} H_\theta =1-4L_\theta, \qquad H_\phi=-L_\phi,\end{eqnarray}
leads to the following system of eigenvalue equations, from the previous section, 
\begin{eqnarray} H\Psi=E\Psi,\qquad 
H_{\theta} \Psi=\rho^2 \Psi,\qquad
 H_{\phi}\Psi = \mu^2 \Psi. \end{eqnarray}
 Moreover, these three operators mutually commute, i.e.  $[H_\theta, H]=[H_\phi, H]=[H_\theta, H_\phi]=0$.
The wave functions found in the previous section are then
\begin{eqnarray}
\Psi(r,\theta,\phi)=\psi_N^{\rho}\Theta_{\frac{\rho-1}{2}}^\mu\hat{\mathcal{Z}}_n ,
\end{eqnarray}
\begin{eqnarray}
\text{where}&&\psi_N^{\rho}=e^{-\frac{\varepsilon r}{2}}(\varepsilon r)^{\frac{\rho-1}{2}}L^{\rho}_{N}(\varepsilon r),\quad
\Theta_{\frac{\rho-1}{2}}^{\mu}=P^{\mu}_{\frac{\rho-1}{2}}(\cos\theta),
\nonumber\\&&
\hat{\mathcal{Z}}_n=(\cos\phi-b)^{-1}\sin^{\delta+\frac{1}{2}}\frac{\phi}{2}\cos^{\gamma+\frac{1}{2}}\frac{\phi}{2}\hat{P}_n^{(\delta,\gamma)}(\cos\phi).
 \end{eqnarray}
Here $\varepsilon=2\alpha/(2N+\rho+1)$ and 
\[ \rho= 2m+1, \qquad m=0,1,2 \ldots, \]
\[\mu=n+\frac{\gamma+\delta-1}{2}, \qquad n=1,2,3 \ldots\] As in the previous section, the relation among $E$, $N$ and $\rho$ is the quantization condition (\ref{energy})
\begin{eqnarray}
E=-\frac{\alpha^2}{2\left(2N+\rho + 1\right)^2}.\label{Eg1}
\end{eqnarray}
In the following we will construct additional integrals of motion to prove the superintegrability of the Hamiltonian (\ref{Nh2}).

\subsection{Ladder and shift operators for the associated Laguerre and Legendre polynomials}
We now search for recurrence operators which preserve the energy $E$. Equation (\ref{Eg1}) shows that $E$ is preserved under either
\begin{eqnarray*}
N\rightarrow N+1,\quad \rho\rightarrow \rho-2 \quad \text{or}\quad N\rightarrow N-1,\quad \rho\rightarrow \rho+2.
\end{eqnarray*}
as well as arbitrary shifts in $n$ (equivalently $\mu$).
We now construct the ladder operators from the associated Laguerre functions, as in \cite{pos3, kal7},
\begin{eqnarray}
L_N^{-}=(\rho+1)\frac{\partial}{\partial r}+\alpha-\frac{1}{2r}(\rho^2-1),
\quad
R_N^{+}=(-\rho+1)\frac{\partial}{\partial r}+\alpha-\frac{1}{2r}(\rho^2-1),
\end{eqnarray}
whose action on the corresponding wave functions are given by
\begin{eqnarray}
L_N^{-}\psi_N^{\rho}=-\frac{2\alpha}{2N+\rho+1}\psi_{N-1}^{\rho+2}, \quad \quad  
R_N^{+}\psi_N^{\rho}=-\frac{2\alpha(N+1)(N+\rho)}{2N+\rho+1}\psi_{N+1}^{\rho-2}.
 \end{eqnarray}
We can also construct lowering and rising differential operators of the $\theta$ related part of separated solution for the associated Legendre functions 
\begin{eqnarray}
L_{\rho}^{-}=(1-z^2)\frac{\partial}{\partial z}+\frac{\rho-1}{2}z,
\quad
R_{\rho}^{+}=(1-z^2)\frac{\partial}{\partial z}-\frac{\rho+1}{2}z,
\end{eqnarray}
where $z=\cos\theta$, and obtain their action on the corresponding wave functions 
\begin{eqnarray}
L_{\rho}^{-}\Theta_{\frac{\rho-1}{2}}^{\mu}=(\mu + \frac{\rho-1}{2})\Theta_{\frac{\rho-3}{2}}^{\mu}, 
\quad
R_{\rho}^{+}\Theta_{\frac{\rho-1}{2}}^{\mu}=(\mu - \frac{\rho+1}{2})\Theta_{\frac{\rho+1}{2}}^{\mu}.
 \end{eqnarray}
Both of these pairs of ladder operators are obtained by taking the standard ladder operators of the special functions \cite{and1} and conjugating by the ground state. 
 
\subsection{Ladder and shift operators for the exceptional Jacobi polynomials and associated Legendre polynomials}

Ladder operators for the exceptional Jacobi polynomials can be constructed from ladder operators for the Jacobi polynomials \cite{and1}
\begin{eqnarray}
&&\mathcal{L}_n^{-}=\frac{1}{2}(2n+\gamma+\delta)(1-y^2)\frac{\partial}{\partial y}-\frac{1}{2}n\{\gamma-\delta-(2n+\gamma+\delta)y\},
\nonumber\\&&
\mathcal{R}_n^{+}=-\frac{1}{2}(2n+\gamma+\delta+2)(1-y^2)\frac{\partial}{\partial y}+\frac{1}{2}(n+\gamma+\delta+1)\nonumber\\&&\qquad\quad\times\{\gamma-\delta+(2n+\gamma+\delta+2)y\}.
\end{eqnarray}
Their action is as
\begin{eqnarray}
&&\mathcal{L}^{-}_{n} P_n^{(\delta,\gamma)}(y)=(n+\gamma)(n+\delta) P_{n-1}^{(\delta,\gamma)}(y),
\nonumber\\&&
\mathcal{R}^{+}_{n}P_n^{(\delta,\gamma)}(y)=(n+1)(n+\gamma+\delta+1) P_{n+1}^{(\delta,\gamma)}(y),\quad y=\cos \phi.
\end{eqnarray}
To extend these operators to the EOP case, we make use of forward and backward operators  \cite{pos3, ho4} for the exceptional Jacobi polynomials
\begin{eqnarray}
&&\mathcal{F}=(y-1)(y+\frac{\gamma+\delta}{\delta-\gamma})\frac{\partial}{\partial y}+\delta(y+\frac{2+\gamma+\delta}{\delta-\gamma}),
\nonumber\\&&
\mathcal{B}=\frac{\gamma-\delta}{\gamma+\delta-(\gamma-\delta)y}\{(1+y)\frac{\partial}{\partial y}+\delta\}, \quad y=\cos \phi,
\end{eqnarray}
whose action are 
\begin{eqnarray}
\mathcal{F} P_n^{(\delta+1,\gamma-1)}(y)=-2(n+\delta-1)\hat{P}_{n+1}^{(\delta,\gamma)}(y),
\\
\mathcal{B}\hat{P}_{n}^{(\delta,\gamma)}(y)=\frac{1}{2}(n+\gamma)  P_{n-1}^{(\delta +1,\gamma-1)}(y).
\end{eqnarray}
We can then define, as in  \cite{pos3}, the corresponding ladder operators for the exceptional Jacobi polynomials via
\begin{eqnarray}
L_n=\mathcal{F}\circ \mathcal{L}^{-}_n\circ \mathcal{B}, \quad R_n=\mathcal{F}\circ \mathcal{R}^{+}_n\circ \mathcal{B}.
\end{eqnarray}
The final step is to conjugate these  ladder operators by the ground state $y_0=(y+1)^{\frac{1}{4}(\delta+2)}(y-1)^{\frac{1}{4}(\gamma+2)}(y-b)^{-1}$ for the angular component of the eigenfunction 
\begin{eqnarray}
L^{-}_n=y_0 L_n y_0^{-}, \quad R^{+}_n=y_0 R_n y_0^{-},\label{ep1}
\end{eqnarray}
so that their action on the $\phi$-components of the wave function are as follows,
\begin{eqnarray}
&&L^{-}_n \hat{\mathcal{Z}}_{n}=-(n+\delta)(n+\gamma)(n+\delta-2)(n+\gamma-2)\hat{\mathcal{Z}}_{n-1},
\nonumber\\&&
R^{+}_n\hat{\mathcal{Z}}_{n}=-n(n+\delta)(n+\gamma)(n+\delta+\gamma-1)\hat{\mathcal{Z}}_{n+1}.
\end{eqnarray}
While these operators shift the parameter $n$ (equivalently $\mu$) in the $\phi$-factor $\hat{\mathcal{Z}}_{n}$, we much account for this shift in the $\Theta^{\mu}_{\frac{\rho-1}{2}}(z)$ component as well. To do so, 
we now construct a pair of operators from associated Legendre polynomials that can lower and rise $\mu$ while fixing $\rho$,
\begin{eqnarray}
L_{\mu}^{-}=\sqrt{1-z^2}\frac{\partial}{\partial z}-\frac{\mu z}{\sqrt{1-z^2}},
\quad
R_{\mu}^{+}=\sqrt{1-z^2}\frac{\partial}{\partial z}+\frac{\mu z}{\sqrt{1-z^2}},
\end{eqnarray}
where $z=\cos\theta$, and their action on the corresponding wave functions are given by
\begin{eqnarray}
L_{\mu}^{-}\Theta^{\mu}_{\frac{\rho-1}{2}}(z)=(\frac{\rho-1}{2}+\mu)(\frac{\rho+1}{2}-\mu)\Theta^{\mu-1}_{\frac{\rho-1}{2}}(z),
\quad
R_{\mu}^{+}\Theta^{\mu}_{\frac{\rho-1}{2}}(z)=-\Theta^{\mu+1}_{\frac{\rho-1}{2}}(z).
\end{eqnarray}

\subsection{Integrals of motion and subalgebra structures}
Let us now consider the following suitable combinations of the operators
\begin{eqnarray}
D_1^{-}=L^{-}_N R^{+}_\rho, \quad D_1^{+}=R^{+}_N L^{-}_\rho,
\quad
D_2^{-}=R^{+}_\mu L^{-}_n, \quad D_2^{+}=L^{-}_\mu R^{+}_n.
\end{eqnarray}
The action of the operators $D^{\pm}_{i}$, $i=1, 2$ fixes our complete basis of eigenfunctions, thus providing  higher order integrals of the motion. Their explicitly action on the eigenfunctions are  given by
\begin{eqnarray}
&D_1^{-}\Psi(r, \theta, \phi)&=\frac{(\rho-2\mu+1)\alpha}{(2N+\rho+1)}\psi_{N-1}^{\rho+2}\Theta_{\frac{\rho+1}{2}}^\mu\hat{\mathcal{Z}}_n,
\nonumber\\&
 D_1^{+}\Psi(r, \theta, \phi)&=-\frac{\alpha(N+1)(N+\rho)(\rho+2\mu -1)}{2N+\rho+1}\psi_{N+1}^{\rho-2}\Theta_{\frac{\rho-3}{2}}^\mu\hat{\mathcal{Z}}_n,
\\
& D_2^{-}\Psi(r, \theta, \phi)&=(n+\delta)(n+\gamma)(n+\delta-2)(n+\gamma-2) \psi_{N}^{\rho}\Theta_{\frac{\rho-1}{2}}^{\mu+1}\hat{\mathcal{Z}}_{n-1},\nonumber\\
&D_2^{+}\Psi(r, \theta, \phi)&=\frac{-1}{4}n(n+\delta)(n+\gamma)(n+\delta+\gamma-1)(\rho+2\mu-1)\nonumber\\&&\quad\times(\rho-2\mu+1) \psi_{N}^{\rho}\Theta_{\frac{\rho-1}{2}}^{\mu-1}\hat{\mathcal{Z}}_{n+1}.
\end{eqnarray} 
The following commutation relations of the operators can be easily verified via the action on the eigenfunctions  (\ref{Nh2}),
\begin{eqnarray}
&&[D_1^{-}, H]=0 = [D_1^{+}, H],\quad
[D_1^{-}, H_\phi]=0=[D_1^{+}, H_\phi],
\nonumber\\&&
[D_2^{-}, H]=0=[D_2^{+}, H],
\quad
[D_2^{-}, H_\theta]=0=[D_2^{+}, H_\theta].
\end{eqnarray}
For the convenience we present a diagram representation of the above commutation relations
\begin{eqnarray}
\begin{xy}
(0,0)*+{D^{-}_{1}}="m";(20,0)*+{L_{\phi}}="f"; (40,0)*+{D^{+}_{1}}="n"; (20,20)*+{H}="r";   (60,0)*+{D^{-}_{2}}="j";(80,0)*+{L_{\theta}}="k"; (100,0)*+{D^{+}_{2}}="l"; (80,20)*+{H}="g"; 
"f";"r"**\dir{--};
"j";"k"**\dir{--}; 
"l";"k"**\dir{--};
"l";"g"**\dir{--};
"g";"j"**\dir{--};
"g";"k"**\dir{--};
"f";"n"**\dir{--}; 
"m";"f"**\dir{--};
"n";"r"**\dir{--};
"r";"m"**\dir{--};
\end{xy}
\end{eqnarray}
Moreover, we obtain
\begin{eqnarray}
&&[H_\theta, D_1^{-}]=\frac14(\rho+1)D_1^{-},\quad\quad \quad \quad [H_\phi, D_2^{-}]=(2\mu+1) D_2^{-},
\nonumber\\&&
[H_\theta, D_1^{+}]=\frac{-1}{4}(\rho-1) D_1^{+}, \quad\quad\quad\quad [H_\phi, D_2^{+}]= -(2\mu-1)D_2^{+}.
\end{eqnarray}
Let us now define the higher order operators $D^{\mp}_{i}D^{\pm}_{i}$, $i=1, 2$. We can also obtain the action of the operators $D^{\mp}_{i}D^{\pm}_{i}$, $i=1, 2$ on the wave functions. It follows from construction that  they present algebraically independent sets of differential operators and hence the system is superintegrable. The system also evidences  a common feature of superintegrable systems in that is admits two higher-order subalgebras. A direct computation of the action of the operators $D_1^{\pm}$ on the basis leads to 
\begin{eqnarray}
[H_\theta, D_1^{-}]=\frac14(\sqrt{H_\theta}+1)D_1^{-}, \quad [H_\theta, D_1^{+}]=-\frac{-1}{4}(\sqrt{H_\theta}-1) D_1^{+},\label{d1}
\end{eqnarray}
\begin{eqnarray}
&D_1^{-}D_1^{+}& =\frac{1}{4}[\sqrt{-\alpha^2}-\sqrt{2H}\sqrt{H_\theta}+\sqrt{2H}][\sqrt{-\alpha^2}+\sqrt{2H}\sqrt{H_\theta}-\sqrt{2H}]\nonumber\\&&\times [\sqrt{H_\theta}+2\sqrt{-L_\phi}-1][\sqrt{H_\theta}-2\sqrt{H_\phi}-1],
\nonumber\\&
D_1^{+}D_1^{-} &=\frac{1}{4}[\sqrt{-\alpha^2}-\sqrt{2H}\sqrt{H_\theta}-\sqrt{2H}][\sqrt{-\alpha^2}+\sqrt{2H}\sqrt{H_\theta}+\sqrt{2H}]\nonumber\\&&\times [\sqrt{H_\theta}+2\sqrt{H_\phi}+1][\sqrt{H_\theta}-2\sqrt{H_\phi}+1].\label{d2}
\end{eqnarray}
Similarly for the $D_2^{\pm}$ operators
\begin{eqnarray}
[H_\phi, D_2^{-}]=(2\sqrt{H_\phi}+1) D_2^{-}, \quad [ H_\phi, D_2^{+}]= -(2\sqrt{H_\phi}-1)D_2^{+},\label{d3}
\end{eqnarray}
\begin{eqnarray}
&D_2^{-}D_2^{+}&  =-\frac{1}{1024}[1 + \sqrt{H_\theta}+2 \sqrt{H_\phi}][-1 +\sqrt{H_\theta} - 2\sqrt{H_\phi}]\nonumber\\&&\times [1 + 2\sqrt{H_\phi} - \gamma - \delta][-1 + 2\sqrt{H_\phi} + \gamma - \delta] [1 + 2\sqrt{H_\phi} + \gamma - \delta]\nonumber\\&&\times[3 + 2\sqrt{H_\phi} + \gamma - \delta][-1 + 2\sqrt{H_\phi} - \gamma + \delta][1 + 2\sqrt{H_\phi} - \gamma + \delta]\nonumber\\&&\times [3 + 2\sqrt{H_\phi} - \gamma + \delta][-1 + 2\sqrt{H_\phi} + \gamma + \delta],\nonumber\\
&D_2^{+}D_2^{-} &=-\frac{1}{1024}[-1 + \sqrt{H_\theta} {+}2 \sqrt{H_\phi}][1 + \sqrt{H_\theta} {-} 2 \sqrt{H_\phi}]\nonumber\\&&\times [-1 + 2 \sqrt{H_\phi} - \gamma - \delta][-3 + 2 \sqrt{H_\phi} + \gamma - \delta][-1 +2\sqrt{H_\phi} + \gamma - \delta]  \nonumber\\&&\times [1 + 2\sqrt{H_\phi} + \gamma - \delta][-3 + 2 \sqrt{H_\phi} - \gamma + \delta][-1 + 2\sqrt{H_\phi} - \gamma + \delta] \nonumber\\&&\times [1 + 2\sqrt{H_\phi} - \gamma + \delta][-3 + 2\sqrt{H_\phi} + \gamma + \delta].\label{d4}
\end{eqnarray}
So the above higher order algebraic structure is the full symmetry algebra for the superintegrable system (\ref{Nh2}).

\subsection{Higher rank polynomial algebras}
In this subsection we will redefine the operators in sense of \cite{kal11} and show that they form well-defined higher rank polynomial algebra. 
We now define the following operators as 
\begin{eqnarray}
&&J_1=\frac{D^{-}_1-D^{+}_1}{\rho}, \quad  J_2=D^{-}_1+D^{+}_1,
\nonumber\\&&
K_1=\frac{D^{+}_2-D^{-}_2}{2\mu}, \quad  K_2=D^{-}_2+D^{+}_2.
\end{eqnarray}
It is easily verified that 
\begin{eqnarray}
&&[J_1, H]=0 = [J_2, H],\quad\quad [J_1, H_\phi]=0=[J_2, H_\phi],
\nonumber\\&&
[K_1, H]=0 = [K_2, H],\quad [K_1, H_\theta]=0=[K_2, H_\theta].
\end{eqnarray}
The commutation relations also can be represented by the following diagrams
\begin{eqnarray}
\begin{xy}
(0,0)*+{J_{1}}="m";(20,0)*+{H_{\phi}}="f"; (40,0)*+{J_{2}}="n"; (20,20)*+{H}="r";   (60,0)*+{K_{1}}="j";(80,0)*+{H_{\theta}}="k"; (100,0)*+{K_{2}}="l"; (80,20)*+{H}="g"; 
"f";"r"**\dir{--};
"j";"k"**\dir{--}; 
"l";"k"**\dir{--};
"l";"g"**\dir{--};
"g";"j"**\dir{--};
"g";"k"**\dir{--};
"f";"n"**\dir{--}; 
"m";"f"**\dir{--};
"n";"r"**\dir{--};
"r";"m"**\dir{--};
\end{xy}
\end{eqnarray}
We obtain the following, still quantum-number dependent,  commutation relations
\begin{eqnarray}
&&[H_\theta, J_1]=\frac{1}{4}\left(J_1+J_2\right), \quad [ H_\theta, J_2]=\frac{1}{4}\left(\rho^2 J_1+J_2\right),
\nonumber\\&&
[H_\phi, K_1]=K_1+K_2, \quad [ H_\phi, K_2]=(2n+\delta+\gamma-1)^2 K_1+K_2.
\end{eqnarray}
These can be expressed back in terms of the algebra generators as  
\begin{eqnarray}
&&[H_\theta, J_1]=\frac{1}{4}\left(J_1+J_2\right), \quad [ H_\theta, J_2]=\frac{1}{4}\left(H_\theta J_1+J_2\right),
\nonumber\\&&
[H_\phi, K_1]=K_1+K_2, \quad [ H_\phi, K_2]=4 H_\phi K_1+K_2.
\end{eqnarray}
The last set of algebra relations to recover are the commutators $[J_1, J_2]$ and $[K_1, K_2].$ A first step is to see the following relations from the action on the eigenfunctions
\begin{eqnarray}
&&[J_1, J_2]=\frac{2}{\sqrt{H_\theta}}[D_1^{-}, D_1^{+}],
\quad
[K_1, K_2]=\frac{1}{\sqrt{H_\phi}}[D_2^{+}, D_2^{-}].
\end{eqnarray}
Moreover, $[J_1,K_1]=0=[J_2,K_2]$ as $[D_1^{\pm},D_2^{\pm}]=0.$
We can rewrite the expressions (\ref{d2}) as 
\begin{eqnarray}
&D_1^{-}D_1^{+}& =P_1(H, H_\theta, H_\phi)+P_2(H, H_\theta, H_\phi)\sqrt{H_\theta},
\nonumber\\&
D_1^{+}D_1^{-} &=P_1(H, H_\theta, H_\phi)-P_2(H, H_\theta, H_\phi)\sqrt{H_\theta},
\end{eqnarray}
where
\begin{eqnarray}
&P_1(H, H_\theta, H_\phi)&=-\frac{1}{4}[16 H  - 64 H H_\theta  + 32 H H_\theta^2 + 16 H H_\phi  - 32 H H_\theta H_\phi \nonumber\\&& \quad +  2 \alpha^2  - 4 H_\theta \alpha^2 + 4 H_\phi \alpha^2],
\nonumber\\&
P_2(H,H_\theta, H_\phi)& =\frac{1}{4}[ 16 H  - 32 H  H_\theta + 16 H  H_\phi  + 2 \alpha^2 ].
\end{eqnarray}
Hence we have 
\begin{eqnarray}
&&[D^{-}_1, D^{+}_1]= 2P_2(H,H_\theta, H_\phi)\sqrt{H_\theta},
\\&&
 \{D^{-}_1, D^{+}_1\}= 2P_1(H,H_\theta, H_\phi). 
\end{eqnarray}
Also, the expression (\ref{d4}) can be written as
\begin{eqnarray}
&D_2^{-}D_2^{+}& =P_3(H_\theta, H_\phi)+P_4(H_\theta, H_\phi)\sqrt{H_\phi},
\nonumber\\&
D_2^{+}D_2^{-} &=P_3(H_\theta, H_\phi)-P_4(H_\theta, H_\phi)\sqrt{H_\phi},
\end{eqnarray}
where 
\begin{eqnarray}
&P_3(H_\theta, H_\phi)&=-\frac{1}{1024}[((\gamma+\delta-1)^2 - 4 H_\phi][\{ (\gamma-\delta)^2-9\}(H_\theta-1)\nonumber\\&&\qquad - 4 \{(\gamma-\delta)^2 + H_\theta-22\}H_\phi + 16 H_\phi^2][(\gamma-\delta)^4 +(4 H_\phi-1)^2 \nonumber\\&&\qquad - 2 (\gamma-\delta)^2 (4 H_\phi+1)],
\nonumber\\&
P_4(H_\theta, H_\phi)& =\frac{1}{256} [(\gamma+\delta-1)^2 - 4 H_\phi] [(\gamma-\delta)^4 + (1 - 4 H_\phi)^2 - 2 (\gamma-\delta)^2 (1 + 4 H_\phi)]\nonumber\\&&\qquad \times [(\gamma-\delta)^2 + 3 H_\theta - 4 (3 + 4 H_\phi)].
\end{eqnarray}
Hence we also have 
\begin{eqnarray}
&&[D^{-}_2, D^{+}_2]= 2P_4(H_\theta, H_\phi)\sqrt{H_\phi},
\\&&
 \{D^{-}_2, D^{+}_2\}= 2P_3(H_\theta, H_\phi). 
\end{eqnarray}
Thus, we realize the final set of algebra relations as 
\begin{eqnarray}
&&[J_1, J_2]=4P_2(H,H_\theta, H_\phi),
\quad
[K_1, K_2]=2P_4(H_\theta, H_\phi).
\end{eqnarray}

Thus, we have shown that the operators $H$, $H_{\theta}$, $H_\phi$, $K_1$, $K_2,$ $D_1$ and $D_2$ close to form a polynomial algebra. Finally, we mention that it is possible to show that these operators are well-defined and can be expressed without recourse to the action on the wave-functions. This is accomplished via the usual observation that the operators constructed are polynomial in $\rho^2$ and $\mu^2$ and so these can be replaced with the appropriate operators and the algebra relations will still hold. 

\subsection{Deformed oscillators, structures functions and spectrum}
In order to derive the spectrum using the algebraic structure, we realize the substructure (\ref{d1}) and (\ref{d2}) as well as the substructure (\ref{d3}) and (\ref{d4}), respectively, in terms of deformed oscillator algebra \cite{das1, das2} $\{\aleph, b^{\dagger}, b\}$ of the form
\begin{eqnarray}
[\aleph,b^{\dagger}]=b^{\dagger},\quad [\aleph,b]=-b,\quad bb^{\dagger}=\Phi (\aleph+1),\quad b^{\dagger} b=\Phi(\aleph),\label{kpfh}
\end{eqnarray}
where $\aleph $ is the number operator and $\Phi(x)$ is well behaved real function satisfying 
\begin{eqnarray}
\Phi(0)=0, \quad \Phi(x)>0, \quad \forall x>0.\label{kpbc}
\end{eqnarray}
We recall (\ref{d1}) and (\ref{d3}) in the following forms
\begin{eqnarray}
[\sqrt{H_\theta}, D_1^{\mp}]=\pm 2D_1^{\mp}, \quad
[\sqrt{H_\phi}, D_2^{\pm}]=\pm D_2^{\pm}.
\end{eqnarray}
Setting
\begin{eqnarray}
&&\sqrt{H_\theta}=2(\aleph_1+u_1), \quad b_1=D_1^{+}, \quad b_1^{\dagger}=D_1^{-},
\\&&\sqrt{H_\phi}=(\aleph_2+u_2), \quad b_2=D_2^{-}, \quad b_2^{\dagger}=D_2^{+},
\end{eqnarray}
where  $u_1>0$ and $u_2>0$ are some representation dependent constants, we obtain from (\ref{d2}) and (\ref{d4}),
\begin{eqnarray}
&&[\aleph_i,b^{\dagger}_i]=b^{\dagger}_i,\quad\quad [\aleph_i,b_i]=-b_i, \quad i=1,2, \nonumber\\
&& b_1b^{\dagger}_1=\Phi_1 (\aleph_1+1,\aleph_2+1, H, u_1, u_2),\quad b^{\dagger}_1 b_1=\Phi_1 (\aleph_1,\aleph_2,H, u_1, u_2),\nonumber\\
&& b_2b^{\dagger}_2=\Phi_2 (\aleph_1+1,\aleph_2+1, u_1, u_2),\quad b^{\dagger}_2 b_2=\Phi_1 (\aleph_1,\aleph_2, u_1, u_2).
\label{kpfh1}
\end{eqnarray}
The corresponding structure functions are given by
\begin{eqnarray}
\Phi_1(\aleph_1,\aleph_2, H, u_1, u_2)&=& b_1^{\dagger} b_1=D^{-}_1D^{+}_1 \nonumber\\
&=& P_1(\aleph_1,\aleph_2, H, u_1, u_2)+2 P_2(\aleph_1,\aleph_2, H, u_1, u_2)(\aleph_1+u_1)
\end{eqnarray}
\begin{eqnarray}
\Phi_2(\aleph_1,\aleph_2, u_1, u_2)&=& b_2^{\dagger} b_2=D^{+}_2D^{-}_2 \nonumber\\
&=& P_3(\aleph_1,\aleph_2, u_1, u_2)+2 P_4(\aleph_1,\aleph_2, u_1, u_2)(\aleph_2+u_2).
\end{eqnarray}
At this stage, the explicit expressions of the corresponding structure functions are as follows
\begin{eqnarray}
\Phi_1(\aleph_1,\aleph_2, E, u_1, u_2) &=&\frac{1}{4}\left[\sqrt{-\alpha^2}-2(\aleph_1+u_1)\sqrt{2E}+\sqrt{2E}\right]\nonumber\\
& &\left[\sqrt{-\alpha^2}+2(\aleph_1+u_1)\sqrt{2E}-\sqrt{2E}\right]\nonumber\\ 
& &[2(\aleph_1+u_1)+2(\aleph_2+u_2)-1]\nonumber\\
& &[2(\aleph_1+u_1)-2(\aleph_2+u_2)-1],
\end{eqnarray}
\begin{eqnarray}
\Phi_2(\aleph_1,\aleph_2, u_1, u_2) &=&-\frac{1}{1024}\left[-1 + 2(\aleph_1+u_1) + 2(\aleph_2+u_2)\right]\nonumber\\
& & [1 + 2(\aleph_1+u_1)- 2 (\aleph_2+u_2)]\, [-1 + 2 (\aleph_2+u_2) - \gamma - \delta]\nonumber\\
& &[-3 + 2 (\aleph_2+u_2) + \gamma - \delta] \,[-1 +2(\aleph_2+u_2) + \gamma - \delta] \nonumber\\
& & [1 + 2(\aleph_2+u_2) + \gamma - \delta] \,
 [-3 + 2 (\aleph_2+u_2) - \gamma + \delta]\nonumber\\
& & [-1 + 2(\aleph_2+u_2) - \gamma + \delta] \, 
 [1 + 2(\aleph_2+u_2) - \gamma + \delta]\nonumber\\
& & [-3 - 2(\aleph_2+u_2) + \gamma + \delta].
\end{eqnarray}
Note that only $\Phi_1$ contains the energy parameter $E$. To determine the energy spectrum, we need to construct the finite-dimensional unitary representations of (\ref{kpfh1}). We thus impose the following constraints on the structure functions:
\begin{eqnarray}
&&\Phi_1(p_1+1, p_2+1, E, u_1, u_2)=0,\quad \Phi_1(0,0, E, u_1, u_2)=0,\label{Cont1}\\
&&\Phi_2(p_1+1, p_2+1, u_1, u_2)=0,\quad \Phi_2(0,0, u_1, u_2)=0,\label{Cont2}
\end{eqnarray}
where $p_i, i=1,2$, are positive integers. These constraints give rise to finite-dimensional unitary presentations. We now solve the constraints (\ref{Cont1}) and (\ref{Cont2}) simultaneously. 
First of all, it can be readily verified that the only solution for the constraints ((\ref{Cont2}) is given by
\begin{eqnarray}
u_2=u_1+\frac{1}{2},\quad\quad p_1=p_2=p,\label{Soln-to-Phi2}
\end{eqnarray}
where $p$ is a positive integer. It follows that these constraints (\ref{Cont1}) and (\ref{Cont2}) lead to $(p+1)$-dimensional unitary representations of (\ref{kpfh1}).  Now we find solutions to the constraints (\ref{Cont1}) which satisfy (\ref{Soln-to-Phi2}). This will provide us the energy spectrum $E$ of the system as well as the allowed values of the parameters $u_1$ and $u_2$. After some computations, we obtain all allowed values of the energy $E$ and the parameters $u_1$, $u_2$ as follows.
\begin{eqnarray}
 E=-\frac{\alpha^2}{2(p+1)^2} ,\quad u_1=\frac{1}{2}+\frac{p+1}{2}, \quad u_2=1+\frac{p+1}{2}.
\end{eqnarray}
Here $p=0,1, \cdots,$ is any positive integer. Making the identification $p=2(N+m)+1$, the energy spectrum becomes (\ref{energy}).

\section{Conclusion}
In this chapter, we see the construction of a new, exactly-solvable system with wave functions comprised of products of Laguerre, Legendre and exceptional Jacobi polynomials. This system is a perturbation of a superintegrable system, the singular Coulomb system. We show that the system is superintegrable by constructing 7 integrals of motion and show that the algebra closes to form a polynomial algebra.  The construction of the higher order integrals of the motion is a systematic constructive approach from ladder operators which based on (exceptional) orthogonal polynomials. We also discuss the representations of this new algebra via the deformed oscillator method to derive spectra and degeneracies of unitary representations. 

It would be of interest to further investigate this system further, in particular to understand how the singular term behaves in the classical limit. Here we have normalized $\hbar=1$, if it is reintroduced the system will depend non-trivially on this parameter. Also of interest could be to understand the scattering states and how the exception perturbation affects those.


\chapter{Conclusion and further research}\label{conc}

Superintegrable systems are very exclusive class of dynamical systems in quantum mechanics and have rigorous understanding polynomial algebra structures and connections with many domains of  pure and applied mathematics. This thesis is concerned with $N$-dimensional quantum superintegrable systems with scalar potentials as well as vector potentials with monopole type interactions, and the algebraic derivations for spectra and degeneracies of these systems. 
The main results of this thesis are summarized as follows.

In chapters \ref{ch3}, \ref{ch4}, \ref{ch5}, \ref{ch7} and \ref{ch8}, we have applied the direct approach to new families of the $N$-dimensional superintegrable Kepler-Coulomb systems with non-central terms and double singular oscillators in the Euclidean space, and to new families of superintegrable Kepler and deformed Kepler systems interacting with Yang-Coulomb monopoles in the flat and curved Taub-NUT spaces. We have constructed the integrals of motion of the models, the corresponding polynomial algebras satisfied by these integrals and the Casimir operators of the polynomial algebras. It is shown that in 3D, 5D and arbitrary dimensions, the integrals close to the higher-rank algebra of the form $Q(3)\oplus\mathcal{L}$, where $\mathcal{L}=\mathcal{L}_1\oplus \mathcal{L}_2\oplus...$ is a direct sum of certain Lie algebras $\mathcal{L}_1, \mathcal{L}_2,\dots$, and $Q(3)$ is the quadratic algebra generated by three generators involving Casimir operators of the Lie algebras in its structure constants. 

In chapters \ref{ch6}, \ref{ch9} and \ref{ch10}, we have highlighted the applications of the constructive approach to the St\"{a}ckel equivalents of the $N$-dimensional superintegrable Kepler-Coulomb model with non-central terms and the double singular oscillators, the 3D extended superintegrable Kepler-Coulomb system from exceptional orthogonal polynomials in the flat Euclidean space, and the new families of the MIC-harmonic oscillators with monopole interactions in the flat and curved Taub-NUT spaces. We have constructed the ladder and shift operators based on the corresponding wave functions of the models. The suitable combinations of these operators generate higher-order integrals of motion and close to form the higher-rank polynomial algebras $P(3)\oplus\mathcal{L}$, where $P(3)$ is higher-order polynomial algebra involving Casimir operators of certain Lie algebras in its structure constants.  

The realizations of these symmetry polynomial algebras in terms of the deformed oscillators have enabled us to obtain the finite dimensional unitary representations and the structure functions. The most interesting advantage of the algebraic approach is that the structure functions are in factorized form which significantly simplify the computations of the process. The structure functions provide the energy spectra algebraically and a deeper understanding of the degeneracies for the superintegrable quantum models.  Let us remark that the recurrence approach was previously restricted to systems with scalar potentials, and this thesis is the first to extend and apply this approach to superintegrable systems with monopole interactions.

As one of the future research directions, it is interesting to study higher dimensional superintegrable models with monopole interactions \cite{men1, kri1, mar10}. It is expected that there would be exciting new symmetry polynomial algebras of such models to be found. It is also interesting to study the symmetry algebras of superintegrable systems with spins \cite{win2} and systems in more general curved spaces. 

Additionally, the application of the co-algebra technique to the general $N$-dimensional superintegrable models with scalar as well as vector potentials interactions is worth investigating \cite{bal4, pos1, rig1}. The deformed oscillator algebra approach to the Calogero type many body problems is another exciting research topic \cite{calo2, calo3, vei1}.

\bibliographystyle{abbrv}
\bibliography{MonographV6}


\appendix 


\end{document}